\documentclass[sigconf]{acmart}
\usepackage{amsmath,amsfonts}
\usepackage{graphicx}
\usepackage{textcomp}
\usepackage{fancyhdr}
\usepackage{graphicx}
\usepackage{textcomp}
\usepackage{xspace}
\usepackage{setspace}
\usepackage{enumitem}
\usepackage{comment}
\usepackage{listings}
\usepackage{caption}
\usepackage{algorithm}
\usepackage[noend]{algpseudocode}
\renewcommand{\algorithmiccomment}[1]{\bgroup\hfill\fontsize{5.8}{6}\selectfont$\triangleright$ {\textcolor{blue}{#1}}\egroup}
\usepackage{booktabs} 
\usepackage{graphicx}
\usepackage{textcomp}
\usepackage{balance}
\usepackage{tikz}
\usepackage[rightcaption]{sidecap} 
\usepackage{xcolor}
\usepackage[11pt]{moresize}
\usepackage{array,graphicx}
\usepackage{float}
\usepackage{svg}
\usepackage{pifont}
\usepackage{threeparttable}
\usepackage{graphics}
\usepackage{multirow}
\usepackage{tabularx}
\usepackage{verbatim}
\usepackage{booktabs}
\usepackage{graphicx, caption, subcaption}
\usepackage{dblfloatfix}
\usepackage{layout}
\usepackage[us,12hr]{datetime}
\usepackage{cleveref}
\usepackage{notoccite}
\usepackage{anyfontsize}
\usepackage{soul,color}
\soulregister\cite7
\soulregister\ref7
\soulregister\pageref7
\usepackage[acronym]{glossaries}

\usepackage[colorinlistoftodos]{todonotes}
\newcommand{\js}[1]{{\color{black}{#1}}}
\newcommand{\nh}[1]{{\color{black}{#1}}}
\newcommand{\juan}[1]{{\color{black}{#1}}}

\newcommand{\namePaper}{QRator\xspace} 
\renewcommand{\namePaper}{Memoria\xspace}

\renewcommand{\namePaper}{Prometheus\xspace}
\renewcommand{\namePaper}{Telemus\xspace}
\renewcommand{\namePaper}{Coeus\xspace}
\renewcommand{\namePaper}{Epiphron\xspace}
\renewcommand{\namePaper}{Odin\xspace} 
\renewcommand{\namePaper}{Sibyl\xspace} 
\newcommand{\ignore}[1]{}
\definecolor{dred}{rgb}{0.75, 0.00, 0.00}
\definecolor{dgreen}{rgb}{0.00, 0.5, 0.00}
\definecolor{ddgreen}{rgb}{0.00, 0.50, 0.00}
\definecolor{dpink}{rgb}{0.75, 0.0, 0.75}
\definecolor{dblack}{rgb}{0.00, 0.00, 0.00}
\definecolor{dblue}{rgb}{0.00, 0.00, 0.75}
\definecolor{gyell}{rgb}{0.5, 0.5, 0.0}
\definecolor{dbleudefrance}{rgb}{0.19, 0.55, 0.91}
\definecolor{darkgoldenrod}{rgb}{0.72, 0.53, 0.04}

\definecolor{magenta}{rgb}{160,0,255}
\newcommand{\gaganF}[1]{[{\color{magenta}GAGAN:#1}]}
\newcommand{\gh}[1]{{\color{black}#1}}

\newcommand{\ghpca}[1]{{\color{black}#1}}

\newcommand{\grakesh}[1]{{\color{black}#1}}
\newcommand*{\Scale}[2][4]{\scalebox{#1}{$#2$}}%

\newcommand*\circled[1]{\tikz[baseline=(char.base)]{
            \node[shape=circle,draw,inner sep=0pt,fill=black, text=white] (char) {#1};}}

\newcommand*\circleds[1]{\tikz[baseline=(char.base)]{
            \node[shape=circle,draw,inner sep=-0.5pt,fill=black, text=white] (char) {#1};}}

\newcommand{\rakeshisca}[1]{{\color{black}#1}}
\newcommand{\gagan}[1]{{\color{black}#1}}
\newcommand{\gf}[1]{{\color{black}#1}}
\newcommand{\go}[1]{{\color{black}#1}}

\newcommand{\gup}[1]{{\color{black}#1}}

\newcommand{\rakesh}[1]{{\color{black}#1}}

\newcommand{\jgll}[1]{{\color{ddgreen}\textbf{\textit{JGL: #1}}}}

\newcommand{\juangg}[1]{{\color{black}{#1}}}
\newcommand{\juanggg}[1]{{\color{black}{#1}}}
\newcommand{\juancr}[1]{{\color{black}{#1}}}
\newcommand{\jgl}[1]{}

\newcommand{\hlssd}{\textsf{H\&L}\xspace}
\newcommand{\hmssd}{\textsf{H\&M}\xspace}
\newcommand{\hps}{\textsf{HPS}\xspace}
\newcommand{\cde}{\textsf{CDE}\xspace}
\newcommand{\arc}{\textsf{Archivist}\xspace}
\newcommand{\slow}{\textsf{Slow-Only}\xspace}
\newcommand{\fast}{\textsf{Fast-Only}\xspace}
\newcommand{\oracle}{\textsf{Oracle}\xspace}
\newcommand{\kleio}{\textsf{RNN-HSS}\xspace}
\newcommand{\sibyldef}{\textsf{Sibyl$_{Def}$}\xspace}
\newcommand{\sibylopt}{\textsf{Sibyl$_{Opt}$}\xspace}
\newcommand{\etal}{\textit{et al.}}

\newcommand{\thold}{\textsf{H\&M\&L}}
\newcommand{\thnew}{\textsf{H\&M\&L$_{SSD}$}}

\crefformat{section}{\S#2#1#3} 

	\makeatletter
	\g@addto@macro{\normalsize}{%
	  \setlength{\abovedisplayskip}{1pt plus 1pt minus 1pt}
	  \setlength{\belowdisplayskip}{1pt plus 1pt minus 1pt}
	  \setlength{\abovedisplayshortskip}{0pt}
	  \setlength{\belowdisplayshortskip}{0pt}
	  \setlength{\intextsep}{1pt plus 1pt minus 1pt}
	  \setlength{\textfloatsep}{1pt plus 1pt minus 1pt}
	  \setlength{\skip\footins}{4pt plus 1pt minus 1pt}}
	\makeatother
\definecolor{NavyBlue}{rgb}{0.19, 0.55, 0.91}
\newcommand{\comm}[1]{{\color{NavyBlue}#1}}

\DeclareMathOperator{\argmaxH}{argmax}   

\newacronym{hss}{HSS}{Hybrid Storage System}

\newcommand{\ho}[2][white]{{%
    \colorlet{foo}{#1}%
    \sethlcolor{foo}\hl{#2}}%
}
\newcommand{\hy}[2][white]{{%
    \colorlet{foo}{#1}%
    \sethlcolor{foo}\hl{#2}}%
}

\definecolor{schrift}{RGB}{0,73,174}

\makeatletter

\newcommand{\disableAcronymHyperlink}{%
  \def\AC@hyperlink##1##2{##2}%
  \def\AC@hyperref[##1]##2{##2}%
  \def\AC@hypertarget##1##2{##2}%
  \def\AC@phantomsection{}%
}
\makeatother

\definecolor{dollarbill}{rgb}{0.52, 0.99, 0.4}

\newcommand{\head}[1]{{\noindent\textbf{#1.}\xspace}} 
\newcommand{\fig}[1]{{Figure~#1}\xspace} 
\newcommand{\tab}[1]{{Table~#1}\xspace} 
\newcolumntype{?}{!{\vrule width 1pt}} 
\newcolumntype{;}{!{\vrule width 0.5pt}}
\newcolumntype{P}[1]{>{\centering\arraybackslash}p{#1}}

\newcommand\hm{\textsf{hm\_1}\xspace}
\newcommand\proxy{\textsf{prxy\_0}\xspace}

\definecolor{byzantine}{rgb}{0.84, 0.2, 0.64}
\definecolor{amber}{rgb}{1.0, 0.75, 0.0}
\definecolor{canaryyellow}{rgb}{1.0, 0.94, 0.0}
\definecolor{amethyst}{rgb}{0.6, 0.4, 0.8}

\newcommand\mix{\textsf{mix\_1}\xspace}
\newcommand{\gca}[1]{{\color{black}#1}}
\newcommand{\gsa}[1]{{\color{black}#1}}
\newcommand{\gor}[1]{{\color{black}#1}}
\newcommand{\gon}[1]{{\color{black}#1}}
\newcommand{\gonn}[1]{{\color{black}#1}}
\newcommand{\gont}[1]{{\color{black}#1}}
\newcommand{\gonff}[1]{{\color{black}#1}}
\newcommand{\gonz}[1]{{\color{black}#1}}
\newcommand{\gonx}[1]{{\color{black}#1}}

\newcommand{\gonf}[1]{{\color{black}#1}}
\newcommand{\rcam}[1]{{\color{black}#1}}
\newcommand{\rcamfix}[1]{{\color{black}#1}}
\newcommand{\rncam}[1]{{\color{black}#1}}
\newcommand{\rnlast}[1]{{\color{black}#1}}
\newcommand{\rbc}[1]{{\color{black}#1}}

\newcommand{\gonzz}[1]{{\color{black}#1}}

\newcommand\scalemath[2]{\scalebox{#1}{\mbox{\ensuremath{\displaystyle #2}}}}

\usepackage{algpseudocode}
\algrenewcommand\algorithmicrequire{\textbf{Initialize:}}
\algrenewcommand\algorithmicensure{\textbf{Output:}}
\AtBeginDocument{%
  \providecommand\BibTeX{{%
    \normalfont B\kern-0.5em{\scshape i\kern-0.25em b}\kern-0.8em\TeX}}}
    
\usepackage{xspace}

\usepackage{algorithm}
\usepackage[noend]{algpseudocode}

\usepackage{tikz}

\newif\ifcameraready
\camerareadyfalse

\definecolor{amber}{rgb}{1.0, 0.49, 0.0}
\definecolor{darkgreen}{rgb}{0.0, 0.2, 0.13}
\definecolor{darkbyzantium}{rgb}{0.36, 0.22, 0.33}
\definecolor{darkseagreen}{rgb}{0.56, 0.74, 0.56}
\definecolor{darkspringgreen}{rgb}{0.09, 0.45, 0.27}
\definecolor{dollarbill}{rgb}{0.52, 0.73, 0.4}

\definecolor{darkcyan}{rgb}{0.0, 0.55, 0.55}
\definecolor{forestgreen}{rgb}{0.0, 0.27, 0.13}
\definecolor{azure}{rgb}{0.0, 0.5, 1.0}
\definecolor{amber}{rgb}{1.0, 0.49, 0.0}

\definecolor{darkpink}{rgb}{0.88, 0.28, 0.54}

\definecolor{mypink}{RGB}{224,8,95}
\definecolor{myblue}{RGB}{31,116,186}
\definecolor{mygreen}{RGB}{35,155,51}
\definecolor{myorange}{RGB}{201,108, 32}
\definecolor{mypurple}{RGB}{122,32,201}

\definecolor{dblue}{rgb}{0.00, 0.00, 0.55}
\definecolor{ddblue}{rgb}{0.00, 0.00, 0.90}

\definecolor{magenta}{rgb}{255,0,255}
\definecolor{red}{rgb}{255,0,0}

\usepackage{siunitx}
\usepackage[us,12hr]{datetime}

\usepackage{enumitem}

\copyrightyear{2022} 
\acmYear{2022} 
\setcopyright{acmlicensed}\acmConference[ISCA '22]{The 49th Annual International Symposium on Computer Architecture}{June 18--22, 2022}{New York, NY, USA}
\acmBooktitle{The 49th Annual International Symposium on Computer Architecture (ISCA '22), June 18--22, 2022, New York, NY, USA}
\acmPrice{15.00}
\acmDOI{10.1145/3470496.3527442}
\acmISBN{978-1-4503-8610-4/22/06}

\begin{document}

\rtitle{\namePaper: Adaptive and Extensible  Data Placement in \\ Hybrid Storage Systems Using Online Reinforcement Learning}
\title{\fontsize{17}{15}\selectfont\namePaper: Adaptive and Extensible  Data Placement in \\ Hybrid Storage Systems Using Online Reinforcement Learning} 
\rtitle{\namePaper: Adaptive and Extensible  Data Placement in \\ Hybrid Storage Systems using Online Reinforcement Learning}
\title{\namePaper: Adaptive and Extensible  Data Placement in \\ Hybrid Storage Systems Using Online Reinforcement Learning} 


\author{Gagandeep Singh$^{1}$ \hspace{0.5cm}  Rakesh Nadig$^1$ \hspace{0.5cm}  Jisung Park$^1$ \hspace{0.5cm}  Rahul Bera$^1$ \hspace{0.5cm}  Nastaran Hajinazar$^1$  \\ David Novo$^3$ \hspace{0.5cm}  Juan G{\'o}mez-Luna$^1$ \hspace{0.5cm}    Sander Stuijk$^2$ \hspace{0.5cm}  Henk Corporaal$^2$ \hspace{0.5cm}  Onur Mutlu$^1$ \\ \vspace{0.2cm}
 $^1$ETH Z{\"u}rich \hspace{1cm}  $^2$Eindhoven University of Technology \hspace{1cm} $^3$LIRMM, Univ. Montpellier, CNRS \vspace{0.4cm} }









\renewcommand{\shortauthors}{G. Singh, et al.}
\renewcommand{\authors}{Gagandeep Singh, Rakesh Nadig, Jisung Park, Rahul Bera, Nastaran Hajinazar, David Novo, Juan G{\'o}mez-Luna, Sander Stuijk, Henk Corporaal, Onur Mutlu}

\begin{abstract}

\go{Hybrid storage systems (HSS) use multiple different storage devices to \gonn{provide} high and scalable storage capacity at high performance. Data placement across different devices is critical to maximize the benefits \gonn{of} such a hybrid system. Recent research proposes various techniques that aim to accurately identify performance-critical data to \gca{place} it in a ``best-fit'' storage device. Unfortunately, most of these techniques are 
\rakeshisca{rigid}, which (1) limits their \gon{adaptivity} to perform well \gon{for} a wide range of \gon{workloads} and storage device configurations, and (2) makes it difficult for designers to extend these techniques \gon{to} different storage system configuration\gon{s} (e.g., \gon{with \gonn{a} different number \gonn{or different types} of} storage devices) than the configuration \rbc{they are} designed for.  }
\go{Our goal} is to design a new data placement technique for hybrid storage \gon{systems} that \gon{overcomes these issues and provides: (1) \emph{adaptivity}\gonn{,} by}  
\emph{continuously learn\gon{ing}} from and \go{adapt\gon{ing} to} the \gon{workload} and the storage device characteristics, and (2) \emph{easy \gon{extensibility}} to  a wide range of \go{\gon{workloads} and} HSS configurations.

We introduce \emph{\namePaper}, \gon{the first technique that uses} 
reinforcement learning \gca{for data placement in hybrid storage systems}. \namePaper observes different features \gon{of} the \gon{running workload}  \gon{as well as the}  storage devices to make system-aware data placement decisions. For every decision \gon{it makes}, \namePaper receives a reward from the system that it uses to evaluate the long-term \gon{performance} impact of its decision and continuously optimizes its data placement policy online. 

We implement \namePaper on  \emph{real} \gonn{systems with various} \go{HSS configurations, including dual- and tri-hybrid storage systems}, and extensively compare it against \gon{four previously proposed data placement techniques (both heuristic- and machine learning-based) } over a wide range of \go{\gon{workloads}}.
Our results show that  
 \namePaper~\gon{ provides \gonzz{21.6\%/19.9\%} performance improvement in a performance-oriented\gonn{/cost-oriented} HSS configuration 
 compared to the best previous data placement technique.} 
Our evaluation using an HSS configuration with three different storage devices shows that \namePaper outperforms \gonn{the state-of-the-art} \gonzz{data placement} 
policy by 23.9\%-48.2\%\gonn{,} while significantly reducing the system architect's burden 
in designing a data placement mechanism that can simultaneously incorporate three storage devices.
We show that \namePaper  achieves 80\% of the performance of an {oracle} policy that has \gon{complete} knowledge~of~future access patterns while incurring \gon{a \gonn{very modest} storage}  overhead of only \gonzz{124.4} KiB. 
\end{abstract}

\vspace{10pt}



\maketitle
\vspace{-0.2cm}
\section{Introduction}
\label{sec:introduction}
Hybrid storage systems (HSS) take advantage of both \gon{fast-yet-small} storage devices and large-yet-slow storage devices to deliver high storage capacity at low latency~\cite{
meza2013case,bailey2013exploring,smullen2010accelerating,lu2012pram,tarihi2015hybrid,xiao2016hs,wang2017larger,lu2016design,luo2015design,srinivasan2010flashcache,reinsel2013breaking,lee2014mining,felter2011reliability,bu2012optimization,canim2010ssd,bisson2007reducing,saxena2012flashtier,krish2016efficient,zhao2016towards,lin2011hot,chen2015duplication,niu2018hybrid, oh2015enabling,liu2013molar,tai2015sla,huang2016improving,kgil2006flashcache,kgil2008improving,oh2012caching,yang2013hec,ou2014edm,appuswamy2013cache,cheng2015amc,chai2015wec,dai2015etd,ye2015regional,chang2015profit,saxena2014design,li2014enabling,zong2014faststor,do2011turbocharging,lee2015effective,baek2016fully,liu2010raf,liang2016elastic,yadgar2011management,zhang2012multi,klonatos2011azor}. The key challenge in designing a high-performance and cost-effective hybrid storage system is to accurately identify the performance-critical\gon{ity} of  application \gon{data} and place \gon{data} in the ``best-fit'' storage device~\cite{niu2018hybrid}. 


{Past works~\cite{matsui2017design,sun2013high,heuristics_hyrbid_hystor_sc_2011,vasilakis2020hybrid2,lv2013probabilistic,li2014greendm,guerra2011cost,elnably2012efficient,heuristics_usenix_2014,doudali2019kleio,ren2019archivist,cheng2019optimizing,raghavan2014tiera,salkhordeh2015operating,hui2012hash,xue2014storage,zhang2010automated,zhao2010fdtm,shi2013optimal,wu2012data,ma2014providing,iliadis2015exaplan,wu2009managing,wu2010exploiting,park2011hot}  propose many different data placement techniques to improve the performance of an HSS. We identify \gon{two major} shortcomings \gon{of} prior proposals that significantly limit their performance: \gonn{lack of (1) adaptivity \gonzz{to workload changes and the storage device characteristics,} and (2) extensibility.}

\head{(\gon{1a}) Lack of adaptivity \gca{to workload changes}} \gon{To guide data placement, past techniques consider only a limited number of \gon{workload} characteristics} ~\cite{matsui2017design,sun2013high,heuristics_hyrbid_hystor_sc_2011,vasilakis2020hybrid2,lv2013probabilistic,li2014greendm,guerra2011cost,elnably2012efficient,heuristics_usenix_2014,lv2013hotness,montgomery2014extent}.
Designers statically tune the {parameters values} for all considered \gon{workloads} at design time based on
\rcam{empirical analysis and designer} experience, and expect those {statically-fixed values} to be equally effective \rnlast{for} a wide range of \rcam{dynamic }\gon{workload \rcam{demands}} and system configurations seen in the real world. As a result, such data placement techniques \gon{cannot easily} adapt 
\rakeshisca{to} a wide range of \gonn{dynamic} \gon{workload} demands and significant\gon{ly} \gon{underperform}  when compared to an oracle technique that has \gon{complete} knowledge of future \gon{storage} access patterns (up to $41.1\%$ lower performance, \gonn{ref.} \cref{sec:motivation_limitations}).


\head{(\gon{1b}) \gon{Lack of adaptivity to changes in device types and configurations}} Most prior \gon{HSS} data placement techniques \gonn{(e.g.,~\cite{matsui2017design,sun2013high,heuristics_hyrbid_hystor_sc_2011,vasilakis2020hybrid2,lv2013probabilistic,li2014greendm,guerra2011cost,elnably2012efficient,heuristics_usenix_2014,doudali2019kleio,ren2019archivist})}
\rakeshisca{do not adapt well to changes in the}
underlying storage device characteristics (e.g., \gonn{changes in the level of} asymmetry in the \mbox{read/write} latenc\gonzz{ies},  \gon{or the number and types of storage devices}
). As a result, \rcam{existing techniques} cannot effectively take  into account the cost of data \gca{movement} between  storage devices while making data placement decisions. This lack of 
\rcam{adaptivity} leads to highly inefficient data placement policies, especially in HSSs with significantly-different device access latencies \gon{than what prior techniques were designed for} (as show\rcam{n} in \cref{sec:motivation_limitations}). 

\head{(2) Lack of extensibility} A large number of prior \gon{data placement} techniques \rcam{(e.g., ~\cite{matsui2017design,sun2013high,heuristics_hyrbid_hystor_sc_2011,lv2013probabilistic,li2014greendm,guerra2011cost,elnably2012efficient,heuristics_usenix_2014})} are typically designed for an HSS that consists of only two storage devices. As modern HSSs \gon{already} incorporat\gon{e} more than two types of storage devices\rakeshisca{~\cite{ren2019archivist, matsui2017tri, matsui2017design}}, system architects need to put significant effort into extending prior techniques to accommodate more than two devices. We observe that a \rcam{state-of-the-art} heuristic-based data placement technique optimized for an HSS with two storage devices~\cite{matsui2017tri} often leads to \gf{suboptimal} performance in an HSS with three storage devices (up to \gonzz{48.2}\% lower performance, \gonn{ref.} \cref{subsec:trihybrid}).  

\textbf{Our goal} is to develop a new, efficient, and high-performance data placement mechanism for hybrid storage systems that \gonn{provides} (1) \gon{\emph{adaptivity}}\gonn{,}  
by \emph{continuously learning} from and {adapting to} the \gon{workload and  storage device characteristics,} and  (2) \emph{\gon{easy extensibility}} to a wide range of \gon{\gon{workloads} and HSS} configurations.
}

\textbf{Key ideas.} To this end, we propose \namePaper, 
a reinforcement \linebreak learning-based data placement technique for hybrid storage systems.\footnote{In Greek mythology, \namePaper is an oracle who makes accurate prophecies\rcam{~\cite{wiki:Sibyl}}.} Reinforcement learning (RL)~\cite{sutton_2018} is a goal-oriented decision-making process in which an autonomous agent learns to \gon{take} optimal actions that maximize a reward function by interacting with an environment.    {The key idea of \namePaper is to design the data placement module in hybrid storage systems as a reinforcement learning agent that \emph{autonomously learns} and adapts to the best-fit data placement policy for the \gon{running} \gon{workload} and \gon{the current} hybrid storage \gonn{system} configuration.} 
For every storage \gon{page} access,  \namePaper observes
different features from the running \gon{workload} and the underlying storage system (e.g., access \gonz{count} of the current request, \gon{remaining capacity in the fast storage, etc.}). \rbc{It uses the features as  \textbf{\emph{state}} information to \gon{take} a data placement \textbf{\emph{action}} (i.e., which device to place the page into). \gont{For every action,} \namePaper receives 
a \gon{delayed} \textbf{\emph{reward}}} from the system in terms of per-\gup{request} latency. 
\gca{This} reward encapsulates the internal device characteristics  of \juanggg{an} HSS (such as read/write latenc\rcamfix{ies}, latency of garbage collection, \rcamfix{queuing delays,} \gont{error handling latencies,} and write buffer state). \namePaper uses this reward to estimate
the long-term impact of its \gon{action (i.e., data placement} decision\gon{)} on the overall application performance \gup{and continuously optimizes its data placement policy online} \gon{to maximize the long-term benefit \gonn{(i.e., reward)} of its actions}.

\rbc{\textbf{Benefits.}} 
\rbc{Formulating the data placement module as an RL agent} \rbc{enables a human designer} to specify only \emph{what} performance objective \rbc{the} data placement module should target, rather than designing and implementing a new data placement \gon{policy} that requires \gca{explicit specification of} \emph{how}  to achieve the performance objective. The use of RL not only \gon{enables} the data placement module to \emph{autonomously} learn the ``best-fit'' data placement strategy for a wide range of \gon{workloads} and hybrid storage \gon{system} configurations but also significantly reduces the burden of a human designer \gon{in finding a good data placement policy}. 

\rbc{\textbf{Challenges.}} While RL provides a promising alternative to existing data placement techniques, we identify two main  challenges in applying RL to data placement in an HSS.

\head{(1) Problem formulation}  The RL agent's effectiveness depends on how the data placement problem is cast as a reinforcement learning-based task. Two key issues arise  \gonn{when} formulating \gon{HSS} data placement as an RL problem\rnlast{:}  (1) \gonn{taking into account} the \gonn{latency} asymmetry \gonn{within and across} storage devices, and (2) deciding which actions to reward and penalize (\gca{also known as the} \emph{credit assignment problem}~\gca{\cite{minsky1961steps}}). 
First, we need to make the agent aware of the asymmetry in read and write latencies of \gon{each} storage \gon{device} and \gonx{the} differences in latencies across \gon{multiple} storage devices. \gup{Real-world storage devices could have \gon{dynamic} latency variations due to their complex hardware and software components (e.g., internal  caching, garbage collection, \gon{error handling, multi-level cell reading,} etc.)~\cite{cai2015data,grupp2009characterizing, jung2012nandflashsim, cai2014neighbor, cai2017error, cui2017dlv, Cai2018, park2021reducing}. 
Second, \gonn{if the fast storage is running \gonzz{out} of free space, there might be evictions \gup{in the background} \gon{from} the fast storage to the slow storage.
}
\gon{As a result, when} we reward the agent, \gon{not only there is a \gon{variable} \gonn{and delayed} reward,
~but} it is \gon{also} hard to properly \gon{assign} \emph{credit} or \emph{blame} \gon{to} different decisions.

\head{(2) Implementation overhead} A \gon{workload} could have \gon{hundreds of} thousands \rcam{ of} pages \gon{of \gonn{storage} data}, making it challenging to efficiently handle the large data footprint with a low design overhead \gon{for the learning agent.} 

To address the first challenge, we use two \gon{main} techniques. First, we design a \textit{reward} structure in terms of \emph{request latency}, \gon{which} allows \namePaper~to learn the \gon{workload and storage} device characteristics 
\nh{when} \gonn{continuously and frequently} interacting with a hybrid storage system. We  \gca{add} a negative penalty \gca{to the reward structure} in case of eviction, \gonn{which helps  \gca{with handling} the credit assignment problem and \gont{encourages the} agent \gont{to} place only performance-critical pages in the fast storage.} {Second, we perform thorough hyper-parameter tuning \gon{to find parameter values that work well for a wide variety of workloads.}} 
To address the second challenge, we use two main techniques. First, we divide states into a small number of bins that reduce \gon{the} state space, which directly \gon{affects}  the implementation overhead. {Second, instead of adopting a traditional table-based RL approach  (e.g., \cite{ipek2008self,pythia}) to store \gonn{the} agent's state-action information (collected by interacting with \gon{an HSS})\gonn{, which} can \gonn{easily} introduce significant performance overhead\gonn{s in the presence of}  a large \gont{state/action} space}, we use a {simple} feed-forward neural network~\cite{zell1994simulation} with only two hidden layers of 20 and 30 nodes, respectively. 


\textbf{Key results.}} \gup{We  evaluate \namePaper~\gf{using \gon{two different dual-HSS} configurations and two different tri-HSS configurations.} We use fourteen diverse \gon{storage} traces from \rakeshisca{Microsoft Research Cambridge (MSRC)} \cite{MSR} collected on real enterprise servers. \gon{We  evaluate \namePaper on workloads from  FileBench~\cite{tarasov2016filebench} on which it has never been trained}. \gon{We compare \namePaper to four state-of-the-art data placement techniques.} We demonstrate four key results. \gon{ First, \namePaper~\gonn{ provides \gonzz{21.6\%/19.9\%} performance improvement in a performance-oriented\gonn{/ cost-oriented} HSS configuration 
 compared to the best previous data placement technique.}} 
Second,  \namePaper outperforms the best-performing supervised learning-based technique  \gon{ on workloads it has never been trained on} by 46.1\% and 54.6\%, on average, } \gon{in} performance-oriented and cost-oriented HSS \gca{configurations, respectively}. 
Third,  \namePaper provides 23.9\%-48.2\% higher performance in tri-hybrid storage systems than a \gonn{state-of-the-art} heuristic-based \gon{data placement} technique \gonn{demonstrating that} \namePaper \gon{~is} \gonn{easily} extensible and alleviates the designer's burden in \gon{finding} sophisticated data placement mechanisms \gon{for new and complex HSS configurations}. Fourth, \namePaper's performance benefits come with a low \gagan{storage} implementation overhead of only \gonzz{124.4} KiB. 

\vspace{0.2em}
\noindent This work makes the following  \textbf{major contributions}:
\begin{itemize}[leftmargin=*, noitemsep, topsep=0pt]
 \item  We show on real \gonn{hybrid storage systems (HSSs)}  that prior state-of-the-art \gont{HSS data placement mechanisms} fall 
  short of the oracle placement due to: lack of  \gon{(1) adapt\gonn{i}vity to workload changes and \gonn{storage device characteristics}, and (2) extensibility. } 
  \item We propose \namePaper
  , a new self-\gon{optimizing}  mechanism 
  \gon{that uses} reinforcement learning \gon{to make data placement decisions in hybrid storage systems}. \namePaper{} 
  \go{dynamically}
  \emph{learns}\gont{,} \gonn{ using both multiple workload features and system-level feedback information\gont{, how} to continuously adapt its policy to improve \rnlast{its} long-term performance \rnlast{for} a workload.}
    
    \item We conduct an in-depth evaluation of \namePaper on \gon{real \gonn{systems with various} HSS configurations\gont{,}} showing that it outperforms \gonn{four} state-of-the-art 
    techniques over a wide variety of applications with a low implementation overhead.
    \item We provide an \rbc{in-depth} explanation of \namePaper's actions that show that \namePaper performs dynamic data placement decisions by learning \gonn{ changes in the level of asymmetry in the read/write latenc\gonzz{ies} and the number and types of storage devices.} 
    \item \gca{ We \gon{freely} open-source \namePaper to aid 
 {future research \gon{in} data placement for storage systems~\cite{sibylLink}}.} 
    
    \end{itemize}
 
\vspace{-0.2cm}
\section{Background\label{sec:background}}
\vspace{-0.1cm}
\subsection{Hybrid Storage Systems (HSSs)\label{subsec:bg_hybrid}}

\begin{figure}[h]
  \centering
   \includegraphics[width=0.6\linewidth]{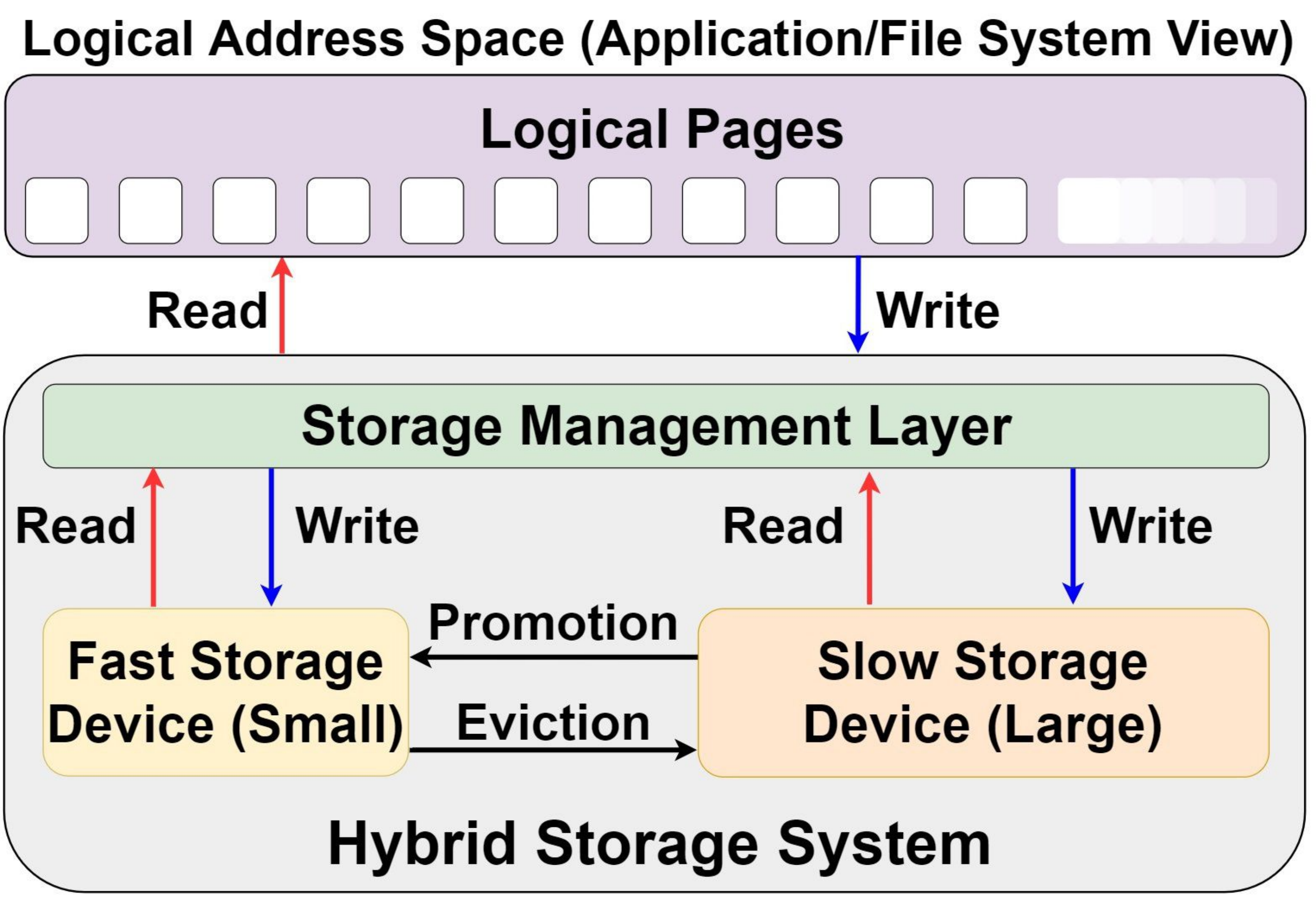}
  \caption{Overview of a hybrid storage system} 
  \label{fig:hybrid}
  \end{figure}

\fig{\ref{fig:hybrid}} depicts a typical HSS \gca{consisting} of a fast-yet-small \gon{storage} device \rcam{(e.g., \cite{inteloptane,samsung2017znand})} and a large-yet-slow \gon{storage} device \rcam{(e.g., \cite{intels4510, intelqlc, seagate, adatasu630})}. \gon{Traditional hybrid storage systems~\cite{heuristics_hyrbid_hystor_sc_2011,feng2014hdstore,micheloni2018hybrid} were designed with a smaller NAND flash-based SSD and a larger HDD. 
Nowadays, hybrid storage systems come with emerging {NVM} devices \rcam{(e.g.,~\cite{kim2014evaluating, tsuchida201064mb, kawahara20128, choi201220nm})} coupled with slower {high-density} NAND flash devices~\cite{matsui2017design, oh2013hybrid, lee2010high, okamoto2015application}. } The storage management layer can be implemented either as system software running on the host system or as \gca{the} firmware of a hybrid storage device (e.g., flash translation layer (FTL) in flash-based SSDs \rcam{\cite{cai2017error,gal2005algorithms}}), depending on the configuration of the HSS. In this work, we \gon{demonstrate and implement our ideas in \gonzz{the} storage management layer \gonzz{of} the operating system (OS)\gonzz{,} but they can be easily implemented in firmware as well.} The storage management layer {in the OS} orchestrates host I/O requests across heterogeneous {devices}, which are connected via an NVM Express (NVMe)~\cite{nvme} or SATA~\cite{sata} interface. {The storage management layer} provides the operating system with a unified logical \rcam{address} space (like the multiple device driver (md) kernel module in Linux~\cite{linux}).\rcam{~As illustrated in \fig{\ref{fig:hybrid}}, the unified logical address space is divided into a number of logical pages (e.g., 4 KiB pages). The pages in the logical address space are assigned to an application. The storage management layer translates \rcamfix{a} read/write performed on \rcamfix{a} logical page into a read/write operation on a target storage device based on the data placement policy. In addition, the storage management layer manages data migration between the storage devices in an HSS. When data currently stored in the slow storage device is moved to the fast storage device, it is called \emph{promotion}. Promotion is usually performed when \rcamfix{a page} in the slow storage device is accessed frequently. Data is moved from the fast storage device to the slow storage device during an \emph{eviction}. Eviction typically occurs when the data in the fast storage device is infrequently accessed or when the fast storage device becomes full.}

\gon{The performance of {a} hybrid storage system highly depend{s} on the ability of the storage management layer (\fig{\ref{fig:hybrid}}) {to \gonn{effectively} manage diverse devices and}
workloads~\cite{matsui2017design, ren2019archivist}.
This \gonn{diversity} presents a {challenge} 
for system {architects} 
\gonn{when they} design an intelligent data placement policy. 
{A desirable policy has} to effectively utilize the low latency characteristics of the fast device {while making} optimal use of its small capacity 
and \gon{should provide easy extensibility to a wide range of workloads and HSS configurations. }}

\vspace{-0.5cm} 
\section{Motivation}\label{sec:motivation_limitations}


{ To assess the effectiveness of existing \gon{HSS} data placement techniques under diverse workloads and hybrid storage configurations, we evaluate {state-of-the-art } heuristic-based (\emph{\cde}~\cite{matsui2017design} and \emph{\hps}~\cite{meswani2015heterogeneous}) and supervised learning-based  (\emph{\arc}~\cite{ren2019archivist}) techniques. We also implement an RNN-based data placement technique (\kleio), adapted from hybrid main memory~\cite{doudali2019kleio}. \gon{To evaluate the effect of underlying storage device technologies, we use {three  different storage devices}: high-end {(\textsf{H})}~\cite{inteloptane}, middle-end {(\textsf{M})}~\cite{intels4510}, and low-end {(\textsf{L})}~\cite{seagate}, \gon{configured into two different hybrid storage configurations: a performance-oriented HSS (\hmssd) and a cost-oriented HSS (\hlssd).  
Table~\ref{tab:devices} provides details \gonn{of our system and} devices.}   \gup{We restrict the fast storage \gon{capacity} to 10\% of the working set size \gonn{of a workload}, which ensures \gon{eviction} of data from  fast storage to slow storage when  fast storage capacity is full.}}}

{\cde}~\cite{matsui2017design} \gon{allocates} hot or random write requests in the faster storage, {whereas} cold and sequential write requests are evicted to the slower device. 
\hps~\cite{meswani2015heterogeneous} uses the access \gonz{count} \gca{of pages} to periodically migrate cold pages to the slower storage device. 
\rakeshisca{\arc~\cite{ren2019archivist} \gca{uses a neural network classifier to predict the target device for data placement.} 
\emph{\kleio}, adapted from~\cite{doudali2019kleio}, is a supervised learning-based mechanism {that} exploits recurrent neural networks (RNN) to \gca{predict the hotness of a page} \gon{and place hot pages in fast storage}.} 
We compare the above policies with three extreme \gon{baselines}:
(1) \slow{}, {where all data reside\rcamfix{s in} \gon{the} slow \gon{storage \rcamfix{device} (i.e., there is no fast storage \rcamfix{device})}}, (2) \fast, {where all data resides in \gon{the}  fast \gon{storage} \rcamfix{device}}, {and (3) \textsf{Oracle}}
\cite{meswani2015heterogeneous}, which exploits \gon{complete} {knowledge of} future I/O-access patterns  
{to perform data placement and to }\js{select victim data blocks for eviction from} the fast device. 

We identify \gon{two major}  shortcomings \gon{of the \gonn{state-of-the-art} baseline data placement techniques}: \gonzz{lack of (1) adaptivity to workload changes and the storage device characteristics, and (2) extensibility.}

\begin{figure}[h]
    \centering
 \includegraphics[width=1\linewidth,trim={0.2cm 0cm 0.2cm 0cm},clip]{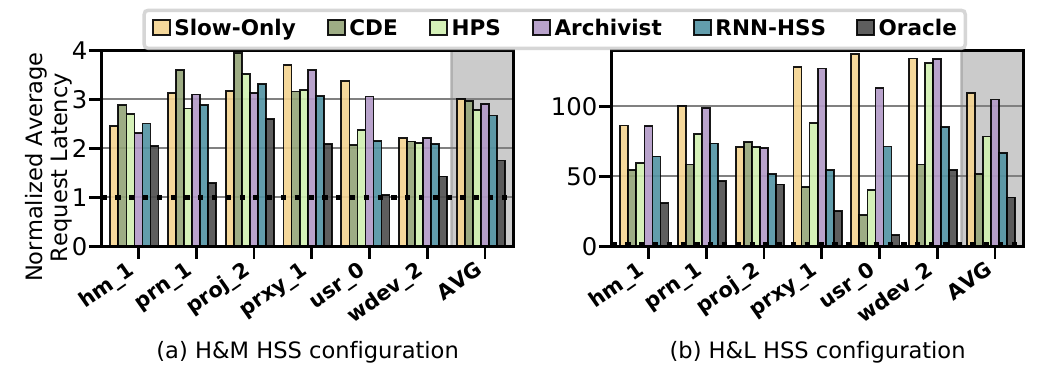}
 \vspace{-0.55cm}
    \caption{Average request latency normalized to \fast~policy} 
    \vspace{-0.1cm}
    \label{fig:motivation_iops}
\end{figure}

\head{(1a) Lack of adaptivity \gca{to workload changes}} 
\fig{\ref{fig:motivation_iops}} shows \gonzz{the} \gon{average request} latency of \gon{all} policies, \gon{normalized to \fast,} under two different hybrid storage configurations. \gon{We make the following \gonn{three} observations.}
First, \gon{all the baseline}  techniques are only effective under specific workloads, showing significantly \gonn{lower} performance \gonn{than} \textsf{Oracle} in most workloads.  \cde, \hps, \arc, and \kleio~achieve comparable performance to \textsf{Oracle} {for specific workloads} \gagan{(e.g.,  \textsf{hm\_1} for \hps~in \textsf{H\&M},  \textsf{usr\_0} for \cde~in \textsf{H\&L}, \gon{\textsf{hm\_1} for \arc~in \textsf{H\&M}, and \kleio in \textsf{proj\_2} for \cde~in \textsf{H\&L} }). Second, the baselines show a {large} \gon{average} performance \gon{loss} of 41.1\% (32.6\%), 37.2\% (55.5\%), \gonn{39.7\% (66.7\%), and 34.4\% (47.6\%)} \gon{compared} to \textsf{Oracle}'s performance, under the \textsf{H\&M} (\textsf{H\&L}) hybrid {storage} configuration, respectively.} 
\gonn{Third, in \hmssd, the baseline techniques provide a performance improvement of only 1.4\%, 7.4\%, 3.5\%, and 11.3\% compared to \slow.  }

\begin{SCfigure}[][h]
  \centering
 \includegraphics[width=0.7\linewidth,trim={0cm 0.02cm 0.2cm 0cm},clip]{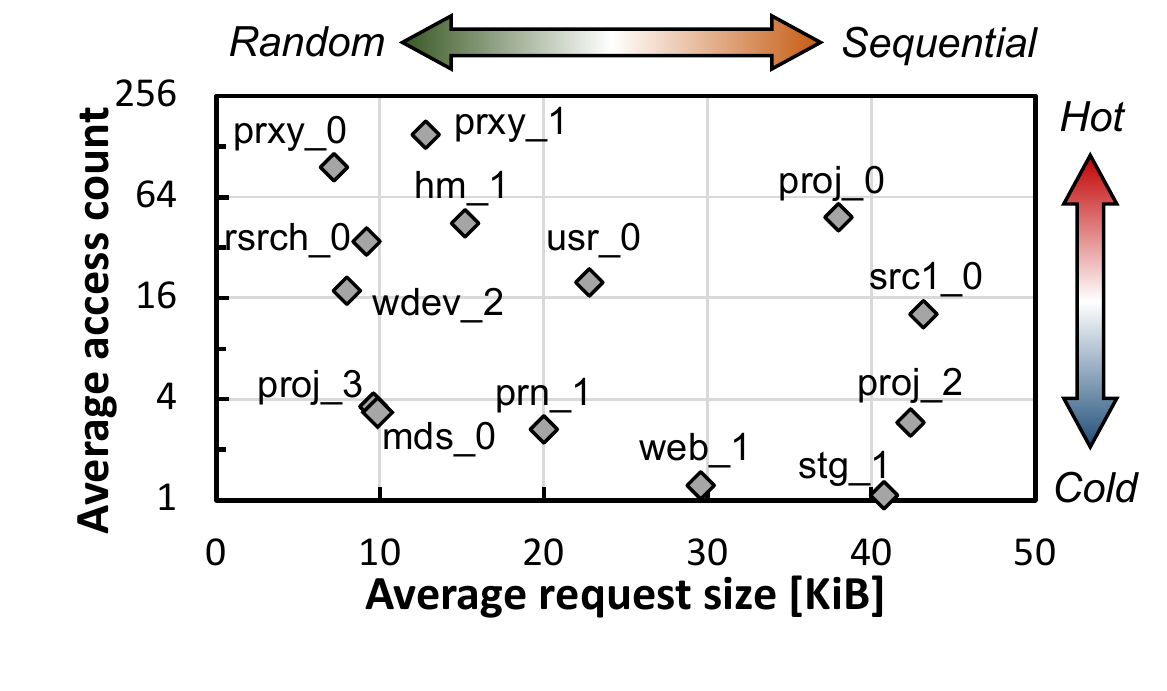}
 \vspace{-23pt}
 \caption{\gonx{Randomness and hotness} characteristics of real-world MSRC \gonx{workloads}~\cite{MSR}}
\label{fig:apps}
\end{SCfigure}

{

\gonn{We conclude that \gon{all four baselines} consider only a limited number of \gon{workload} characteristics to construct a data placement technique, which leads to a significant performance gap compared to the \oracle policy. Thus, there is no single  policy that} 
\juangg{works} well for all \gon{the workloads}. 

\gon{To further analyze the characteristics  of our evaluated workloads, 
we} plot the \gca{average} hotness (y-axis) and randomness (x-axis) in \fig{\ref{fig:apps}}. 
\gagan{We provide details on these workloads in Table~\ref{tab:workload}.} \gonn{  In these workload traces,} each \gon{storage} request is labeled with a timestamp that indicates the time \gonzz{when} the request was issued from the processor core. Therefore, the time interval between two consecutive I/O requests represents the time the core has spent computing.  {We quantify a workload’s hotness (or coldness) using the average access 
\rcamfix{count}, \gca{which is} the average \rcamfix{of the access counts of all pages} in a workload; the higher (lower) the average access \rcamfix{count}, the hotter (colder) the workload. We quantify a workload’s randomness using the average request size in the workload; the higher (lower) the average request size, the more sequential (random) the workload. } 
\gon{From \fig{\ref{fig:apps}}\rcamfix{,} we make \rcamfix{the}  following two observations. First, }the average hotness and randomness {vary widely between \gon{workloads}}. \gon{Second, we observe that each of our evaluated \gon{workload}\rcamfix{s}  exhibits highly dynamic behavior throughout its execution. For example, in Figure~\mbox{\ref{fig:dynamic}}, we show the execution \gonn{timeline} of \texttt{rsrch\_0}, \gonn{which \gonx{depicts} the accessed address\gonx{es} and request sizes.} 
\gon{We conclude that} an efficient policy needs to incorporate continuous adaptation to highly dynamic changes in workload behavior.}}  


{
\begin{SCfigure}[][h]
  \centering
  \hspace{0.2cm}
 \includegraphics[width=0.6\linewidth,trim={0cm 0cm 0.2cm 0cm},clip]{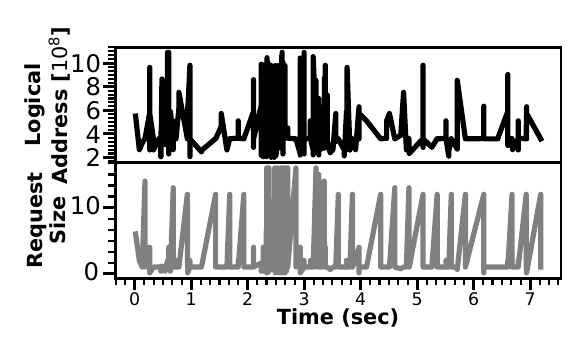}
  \caption{\gor{\gonz{Timeline of} accessed logical addresses and request sizes during \gont{the} execution of \texttt{rsrch\_0} workload}}
\label{fig:dynamic}
\vspace{-0.2cm}
\end{SCfigure}
}

\vspace{0.1cm}
\head{\gon{(1b) Lack of adaptivity to changes in device types and configurations}} 
\juangg{\gon{There are a wide variety and number of storage devices 
~\cite{samsung2017znand, seagate, inteloptane, cheong2018flash,micron3dxpoint, hady2017platform, intelp4610, intelqlc, intels4510, adatasu630, kim2014evaluating, tsuchida201064mb, kawahara20128, choi201220nm, matsui2017design, oh2013hybrid, lee2010high, okamoto2015application} that can be used  to configure \gca{an HSS}.} The \gca{underlying storage} technology used in \gca{an HSS}  significantly influences \gca{the} effectiveness \gca{of a data placement policy}.} \gon{We demonstrate this with an example  observation from \fig{\ref{fig:motivation_iops}}}. In the \textsf{H\&M} configuration (\fig{\ref{fig:motivation_iops}(a)}), we \gon{observe} that  for certain workloads (\textsf{hm\_1} and \textsf{prn\_1}), both \cde  and \hps provide rather 
\juangg{low} performance even compared to \slow. Similarly, \arc and  \kleio provide lower performance for \textsf{hm\_1} and \textsf{proj\_2} in \hmssd compared to \slow.   
While in the \textsf{H\&L} configuration (\fig{\ref{fig:motivation_iops}(b)}), we  observe that \cde, \hps, \arc, and \kleio~
\juangg{result in} lower latency than \slow for the respective workloads. 
\gonn{Thus, we conclude that both heuristic-based and learning-based data placement policies lead to poor performance due to their inability to holistically take into account the device characteristics. The high diversity in \gagan{device} characteristics makes it very difficult for a system architect to design a generic  data-placement technique that is suitable for all \acrshort{hss} configurations. }

\head{(2) Lack of extensibility} \gon{As} modern \gon{HSSs already incorporate more than two types of storage  devices~\cite{ren2019archivist, matsui2017tri, matsui2017design, meza2013case}, system architects need  to put significant effort into extending prior \gonn{data placement} techniques to
accommodate more than two devices.} \rakeshisca{In \cref{subsec:trihybrid}, we evaluate the effectiveness of a state-of-the-art heuristic-based policy~\cite{matsui2017tri} \gon{for different tri-HSS configurations, \gonn{comprising} of three different storage devices.} This policy is based on the \cde~\cite{matsui2017design} policy that divides pages into hot, cold, and frozen data and allocates these pages to \textsf{H}, \textsf{M}, and \textsf{L} devices, respectively. \gon{ A system architect needs to statically define the hotness values and explicitly handle the eviction and promotion between the three devices during  design-time.} 
Through  our evaluation in \cref{subsec:trihybrid}, we conclude that such a heuristic-based policy (1) lacks extensibility, \rcam{thereby increasing}  
 the system architect's effort, and (2) leads to lower performance when compared to an RL-based solution (up to 48.2\% lower).} 

 \gon{Our} empirical study  shows \js{that \textbf{the \gonn{state-of-the-art} {heuristic- and learning-based} data placement techniques are rigid and far from 
optimal,}} which strongly motivates us to develop a new data placement technique \gonn{to achieve significantly higher performance than existing policies}.
The new technique should provide \mbox{(1) \emph{adaptivity}} to better capture the features \gonn{and dynamic changes} in I/O-access patterns and storage device characteristics, 
and \mbox{(2) \emph{easy}} \emph{extensibility} to a wide range of \gon{workloads} and HSS configurations. \gonn{Our goal is to develop such a technique using reinforcement learning.}  

\vspace{-0.2cm}
\section{{Reinforcement Learning}}
\vspace{-0.1cm}
\subsection{{Background}}
\label{background:rl}
\begin{figure}[h]
  \centering
    \includegraphics[width=0.7\linewidth,trim={2.2cm 7.6cm 3cm 1.4cm},clip]{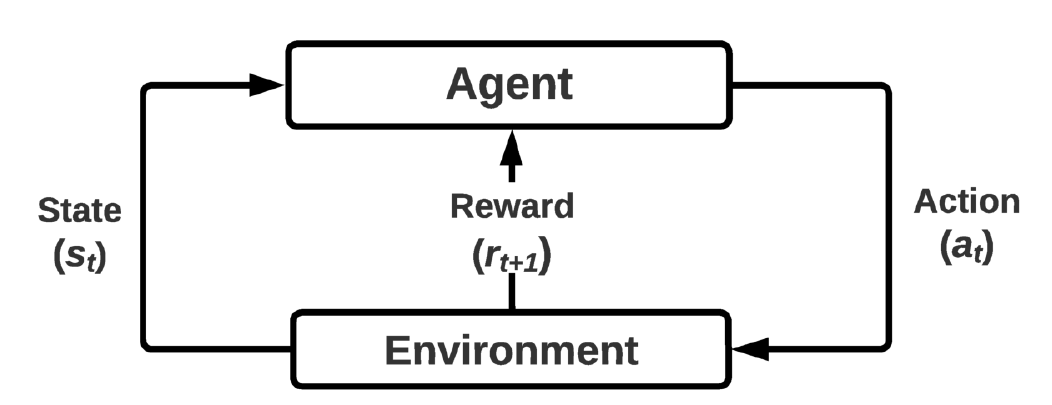}
  \caption{Main components of \gon{general} RL}
\label{fig:rl_basic}
 \end{figure}

Reinforcement learning (RL)~\cite{sutton_2018} is a class of machine learning (ML) algorithms that involve an \emph{agent} learning to achieve an objective by interacting with its \emph{environment}{, as shown in \fig{\ref{fig:rl_basic}}}. The agent starts from an initial representation of its environment in the form of an initial state\footnote{{State is a representation of an environment using different features.} } $s_0$ $\in$ $S$, where $S$ is the set of all  possible states. 
Then, at each \emph{time step} $t$, the agent performs an \emph{action} $a_t$ $\in$ $A$ in state $s_t$ ($A$ represents the set of possible actions) and moves to the next state $s_{t+1}$. The agent receives a numerical reward $r_{t+1}$ $\in$ $R$, \gon{which could be \textit{immediate} or \textit{delayed} in time},  for action $a_t$ that \gca{changes} the environment state from $s_t$ to $s_{t+1}$. The sequence of states and actions starting from an initial state to the final state is called an \emph{episode}. 
The agent \gca{\gonx{makes} decisions and receives corresponding rewards} 
while trying to maximize the \textit{accumulated} reward, \gont{\rcamfix{as opposed to} \gonz{maximizing} the reward for \gonz{only} \emph{each} action}. In this way, the agent can optimize for the long-term impact of its decisions. 

\gon{
The policy $\pi$ governs an agent's action in a state.  The agent’s goal is to find the optimal policy that maximizes the cumulative reward\footnote{The total cumulative reward is also known as the \emph{return}~\cite{sutton_2018}.} collected from the environment
over time.  
The agent finds an optimal policy $\pi^{*}$ by calculating the optimal action-value function \gont{($Q^{*}$), also known as the \textbf{Q-value}} of the state-action pair, where $Q(S,A)$ represents the expected cumulative reward by taking an action A in a given state S.  

\gonz{Traditional RL methods (\rbc{e.g.,}~\cite{watkins1989learning,qlearning_ML_1992,rummery1994line,ipek2008self,pythia}) use a tabular approach with a lookup table to store the \rbc{Q-values associated with each} state-action pair. 
These approaches can lead to high storage and computation overhead for environments with a large number of states and actions. 
 To overcome this issue, \rbc{\emph{value function approximation} was proposed.~\cite{silver2016mastering,silver2017mastering,mnih2013playing,liang2016deep}}. Value function approximation replaces \gonzz{the} lookup table with \gonzz{a} supervised-learning model~\cite{sutton1999policy,baird1995residual,liang2016deep,mnih2013playing,silver2016mastering,silver2017mastering}, which provides the capability to generalize over a large number of state-action pairs with a low storage and computation overhead. } }

\vspace{0.1cm}
\vspace{-0.2cm}
\subsection{Why \gca{Is RL}  a Good Fit \gon{for Data Placement in \rcamfix{Hybrid Storage Systems}?}}

We choose RL \gon{for data placement in HSS} due to the following advantages
compared to heuristic-based (e.g.,~\cite{matsui2017design,
meswani2015heterogeneous}) and supervised learning-based (e.g.,~\cite{ren2019archivist}) techniques.

\head{(1) \gon{Adaptivity}} As discussed in \cref{sec:introduction} and \cref{sec:motivation_limitations}, a data placement technique should have the ability to adapt to changing workload demands and underlying device characteristics. This \gon{adaptivity}  requirement of data placement makes RL a good fit \gca{to model} data placement. 
The RL agent works autonomously \gon{in an HSS} using the provided \gca{state} features and reward to \gon{\gonx{make} data placement decisions} without any human intervention.

\head{(2) Online learning} Unlike an \emph{offline} learning-based approach, an RL agent uses an \emph{online} learning approach. Online learning allows an RL agent to  continuously adapt  its decision-making policy using  system-level feedback and specialize to \juangg{the} current workload and system configuration. 

\gont{
\head{(3) Extensibility} RL provides the ability to easily extend a mechanism with \gonzz{a} \gonff{small} effort required to implement the extension. As shown in \cref{subsec:trihybrid}, unlike heuristic-based mechanisms,  RL can be easily extended to different \gonff{types and} number of storage devices\gonff{. Such extensibility} reduces the system architect's burden in designing sophisticated data placement mechanisms.} 

\gonff{\head{(4) Design Ease} \gonz{With RL, the designer of the HSS} does not need to specify \gonz{a data placement} policy. They need to specify \emph{what} to optimize (via reward function) but not \emph{how} to optimize it. }

\gon{
\head{(5) Implementa\gont{tion Ease}} RL provides ease of implementation \gonff{that requires a} \gonz{small}  computation overhead. 
{As shown} in \cref{sec:results}, \textit{function approximation}-based~RL~techniques can generalize over all the possible state-action pairs by using a simple feed-forward neural network to provide high performance at low implementation overhead {\gont{(compared to} sophisticated RNN-based mechanisms}).}

\vspace{-0.1cm}
\section{\namePaper: RL Formulation}
\label{sec:rl_formulation}

\begin{figure}[h]
  \centering
 \includegraphics[width=0.85\linewidth,trim={1.12cm 3.5cm 1.8cm 5.5cm},clip]{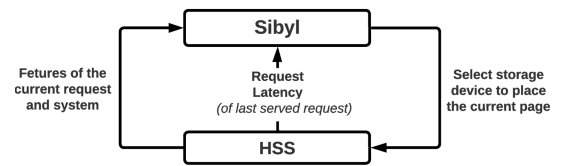}
  \caption{Formulating data placement as an RL problem}
\label{fig:rl_formulation}
 \end{figure}

\fig{\ref{fig:rl_formulation}} shows our formulation of  data placement as an RL problem. We design \namePaper as an RL agent that learns to perform accurate and system-aware data placement decisions by interacting with the hybrid storage system. With every storage request, \namePaper observes \gup{multiple \gon{workload}} \gont{and system-level} features \gca{as \gonzz{a} \emph{state}} to make a data placement decision. After every \emph{action}, \namePaper receives a \emph{reward} \gca{in terms of \gont{the served} request latency} that takes into account the data placement decision and internal storage system state. \namePaper's goal is to find an optimal data placement policy that 
\gont{maximizes} overall performance for \gon{the running} workload and \rcam{the current} system configuration.
\gon{To reach its performance goal, \namePaper needs to minimize the \gont{average request} latency \gon{of the running workload} \gont{by maximizing} the use of the fast storage device while avoiding the eviction penalty \gont{due to non-performance critical pages}. }

\head{Reward} 
\gon{After every data placement decision} at time-step\footnote{\gor{In HSS, a time-step is defined as a new storage request.
}}  $t$, {\namePaper gets a reward 
from the environment} at time-step $t+1$ that acts as a feedback to \namePaper's previous action. 
{To achieve \namePaper's performance goal, we craft \gon{the} reward \gon{function} \textit{R} as follows:
\begin{equation}
R =\begin{cases}
       {\scalemath{1}{ \frac{1}{L_t} }}& 
        \begin{gathered}
            \textit{if no eviction \gon{of a page from the}} \\[-\jot]
            \textit{\gon{fast storage to the slow storage}}
        \end{gathered}\\
        max(0,\frac{1}{L_t}-\textit{$R_p$}) & \text{\textit{in case of eviction} }
    \end{cases}
\end{equation}
{where $L_t$ and \textit{$R_p$} represent \rcamfix{the} \gonff{last} \gont{served} \gca{request latency}  and eviction penalty, respectively. \gont{If the fast storage is running \gonff{out} of free space, there might be evictions in the background from the fast
storage to the slow storage. Therefore, we add an eviction penalty (\textit{$R_p$}) to  guide \namePaper to  place only performance\gonzz{-}critical pages in the fast storage. 
} 
\gon{We empirically select \gont{\textit{$R_p$} to be equal to} 0.001$\times$$L_e$ ($L_e$ is the time spent in evicting pages from the fast storage to the slow storage)}\rcamfix{, which}  prevents the agent from aggressively \gonff{placing} all requests \gonff{into} the fast storage device.}}

\gont{$L_t$} \gonff{(request latency)}  is the time taken to service \gont{the last} read or write  I/O request \gont{from the OS}. Request latency can faithfully capture the status of the hybrid storage system, as it significantly varies depending on the request type, device type, and the internal state \gont{and \gonz{characteristics}} of the device (e.g., such as read/write latenc\rcamfix{ies}, \gonzz{the} latency of garbage collection, \rcamfix{queuing delays, \gonz{and error handling latencies})}.
\rcamfix{Intuitively, if} \gont{$L_t$} is low (high), i.e., if the agent serves a storage request from the fast (slow) device, the agent receives a high (low) reward. However, \gonff{if there is an eviction,} we \rcam{penalize} the agent \gonff{so as} \gup{ to \rcamfix{encourage} the agent \rcamfix{to} place only performance-critical pages in the fast storage device.} 
   We need the \rcamfix{eviction} penalty \gonff{to be} 
   large enough to \rcamfix{discourage} the agent \rcamfix{from evicting} and small enough not to deviate the learned policy too much on a \gonff{placement} decision \gonz{that leads} to higher latency.  

\head{State}
At each time-step $t$, the state features for a particular \mbox{read/write} request  are collected in an \emph{observation} vector. 
  \gon{We  perform feature selection~\cite{kira1992featureselection} to determine the best state features to include in \namePaper's \gont{observation} vector.} 

We use a limited number of features due to two reasons. First, a limited feature set allows us to reduce the implementation overhead of our mechanism \gont{(\gonff{see} \cref{sec:overhead_analysis})}. Second, we empirically observe that our RL agent is more sensitive to the reward structure than to the \gca{number} of features in the \gonff{observation vector}.  Specifically, using the request latency as a reward provides indirect feedback on the internal timing \rcam{characteristics} \emph{and} the current state \gon{(e.g., queu\rcamfix{e}ing delays, buffer dependencies, effects of garbage collection, read/write latenc\gonzz{ies},  write buffer state, \gonzz{and} \rcamfix{error handling latencies})} of the hybrid storage system. Our observation aligns with a recent study~\cite{silver2021reward} that argues that the reward is the most critical component \gca{in RL to find an optimal decision-making policy}.

In our implementation of \namePaper{}, the observation vector is a 6-dimensional tuple:
\vspace{-0.1cm}
\begin{equation}
\gonx{
O_t = (size_t, type_t, intr_t, cnt_t, cap_t, curr_t).}
\end{equation}

\noindent
Table~\ref{tab:state} 
{lists our \js{six}} selected features. \gon{We \rcamfix{quantize the representation of each state} into a small number of bins \gonff{to} reduce \gonff{the} \gont{storage overhead of state representation}.} These features can be captured in the block layer of the storage system \gup{and stored in a separate metadata table (\cref{sec:overhead_analysis})}.  
$size_t$ represents the size \gon{of the current request} in terms of the number of pages associated with \gon{it}. It {indicates whether the incoming request} is sequential or random. 
{$type_t$ \gon{(request type)} differentiates between read and write requests, \gon{important} for data placement decisions since storage devices have asymmetric read and write latencies.} 
$intr_t$ (access interval)  and $cnt_t$ (access \gonz{count}) represent the temporal and spatial reuse \gon{characteristics} of \gont{the currently requested page}\gca{, respectively}.  \gup{Access interval  
\rcamfix{is defined as} the number of page accesses between two  references to the same page. 
\gf{ Access \gonz{count} 
\rcamfix{is defined as} \gonff{the} total \gon{number of} accesses to \gon{the page}.} \gonz{These metrics} provide
insight into \gonzz{the} \gonz{dynamic} \gon{behavior of the currently requested page.}
\footnote{We did not use the reuse distance as a locality metric due to its high computation overhead during online profiling~\cite{zhong2009program}.}
} 
$cap_t$ is a global counter \gonz{that tracks} the remaining capacity in the fast storage \gon{device, which }is an important feature since our agent's goal is to maximize the use of the limited fast storage capacity \gca{while avoiding evictions} \gon{from the fast storage device}.  \gup{By including this feature, the agent can learn}
to \gonff{avoid the eviction penalty (i.e., learn to } restrain {itself} from \gonff{placing in fast storage} non-performance critical pages that would lead to evictions). 
 \gon{$curr_t$ \gont{is the} current placement of the requested page. Since every data placement decision affects the decision for future requests, $curr_t$ guides \namePaper to perform past-aware decisions.}


{
\begin{table}[h]
 \caption{State features \gon{used by} \namePaper}
    \label{tab:state}
\centering
\ssmall
 \renewcommand{\arraystretch}{0.9}
\setlength{\tabcolsep}{2pt}
  \resizebox{1\linewidth}{!}{%
\begin{tabular}{l||l|c|c}
\hline
\textbf{Feature}             & \textbf{Description} & \textbf{\# of bins} & 
\textbf{Encoding (bits)}\\ 
\hline
\gonx{$size_t$}                      &  Size of the current request  (in pages) &  8  & 8 \\
\gonx{$type_t$}                      &  Type of the \gont{current request (read/write) }  &2 & 4  \\
\gonx{$intr_t$}                      &  \gh{Access} interval \gon{of the requested page}&64 & 8 \\ 
\gonx{$cnt_t$}                      & Access \gonz{count} of the \gon{requested} page &64 &8   \\
\gonx{$cap_t$}   
&  Remaining capacity in the fast storage device  &8&8    \\
\gonx{$curr_t$}                      &   Current {placement} \gon{of the requested page} \gonff{(fast/slow)} &2 & 4\\
\hline
\end{tabular}
}
\end{table}

}

\head{Action} 
{At each time-step $t$, in a given state, \namePaper selects an action \gup{($a_t$ in Figure~\ref{fig:rl_formulation})} from \gon{all} possible actions. \gca{In a hybrid storage system with two devices,  possible actions are: } placing data in \gon{(1) the fast storage device}} 
or (2) the slow storage device. \gont{This is easily extensible to $N$ storage devices, where $N\ge3$.} 

\vspace{-0.2cm}
 
\section{\namePaper: Design}
\label{sec:mechanism}
  \begin{figure*}[t]
\centering
\begin{subfigure}[t]{.53\linewidth}
  \centering
  \includegraphics[width=\linewidth,trim={0.8cm 2.4cm 1cm 1.2cm},clip]{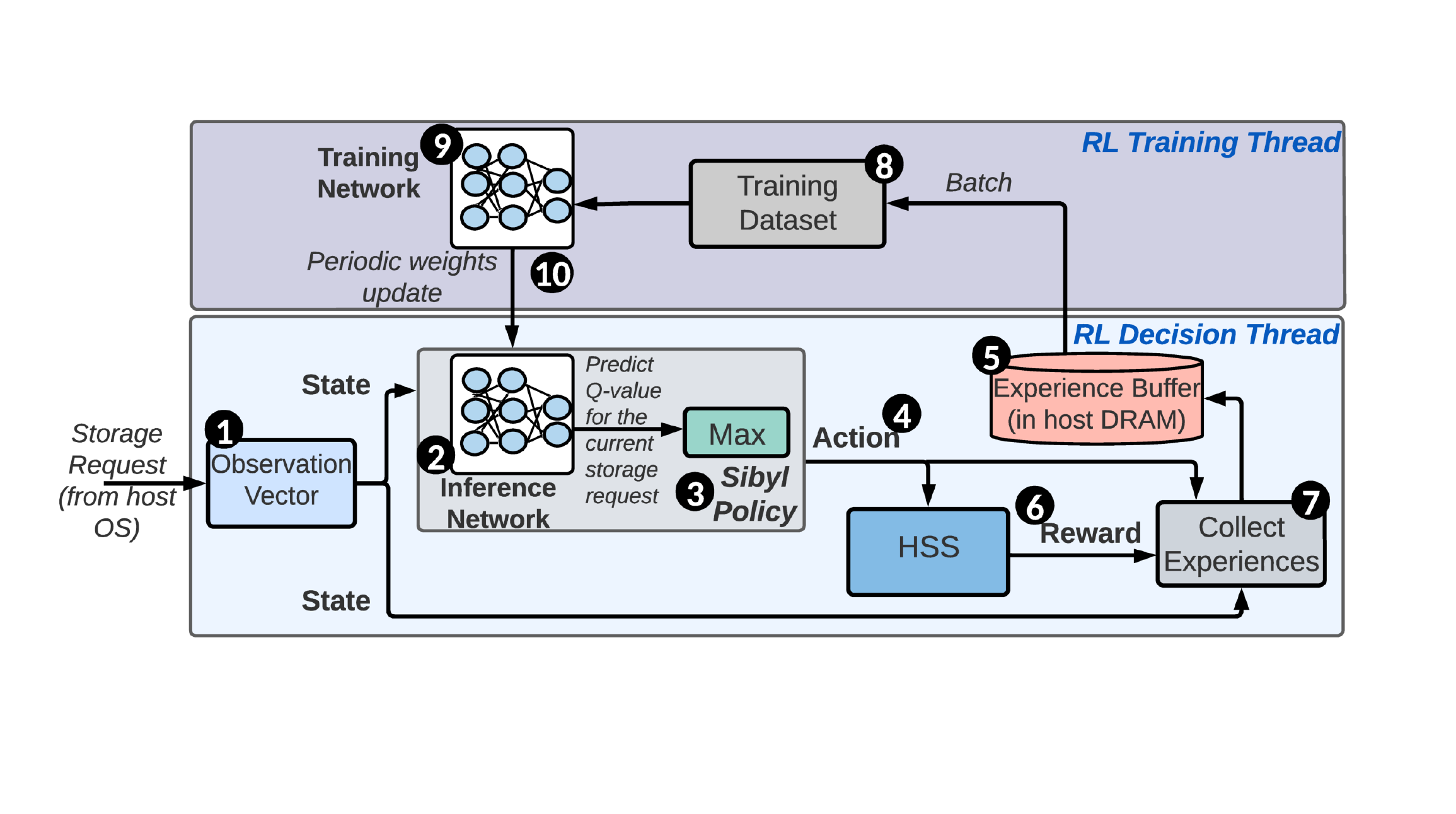}
  \vspace{-0.7cm}
  \caption{
  \label{fig:curator}}
\end{subfigure}%
\hspace{0.3cm}
\begin{subfigure}[t]{.25\linewidth}
  \centering
  \includegraphics[width=\linewidth,trim={0cm 1.4cm 0cm 0.2cm},clip]{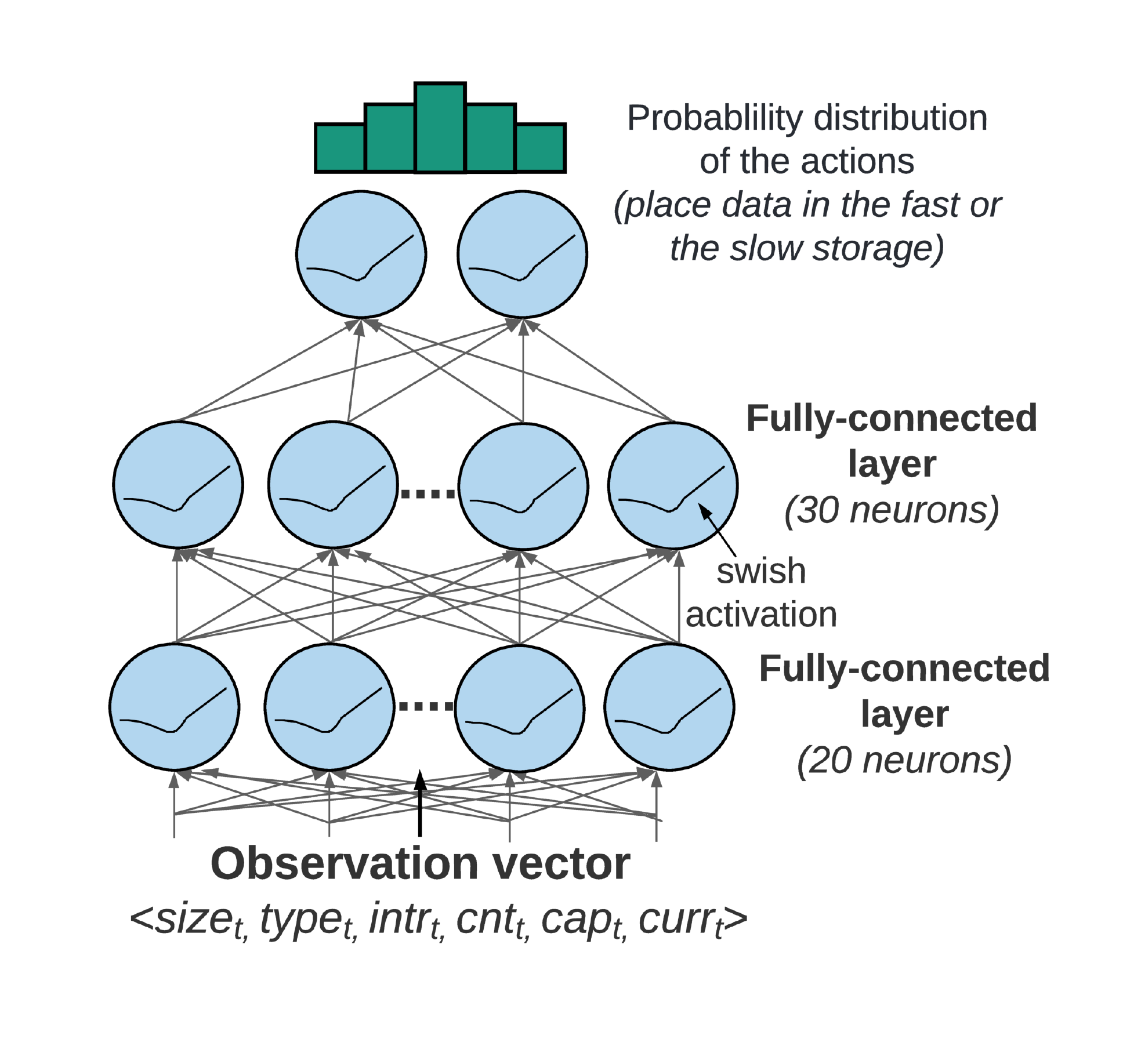}
   \vspace{-0.7cm}
   \caption{
  \label{fig:model}}
\end{subfigure}
\caption[Two numerical solutions]{(a) Overview of \namePaper (b) \gon{\gonz{Training}} network design using as input the \gonx{state} features from Table~\ref{tab:state}. \gon{The inference network \gonz{is identical except it is used only for inference}} }
\label{fig:qrator}
\vspace{-0.5cm}
\end{figure*}

 \rbc{We implement \namePaper in \gon{the storage management layer of the host system's \gonz{operating system}}}. 
 Figure~\ref{fig:qrator}(a) shows a high-level overview of \namePaper. \gonz{Sibyl} is \rbc{composed of} two parts, \gonz{each implemented as a separate thread},  that run in parallel 
{\rbc{(1)} the \emph{RL decision thread}, \gont{where Sibyl decides the data placement \gonx{\circled{4}} of the current storage request  while collecting  \gonz{information \circled{7} about its decisions \gonx{\circled{4}} and their effects} \gonx{\circled{6}} in an \emph{experience buffer} \circled{5}, and} 
\rbc{(2)} the \emph{RL training thread}}, \gonff{where} \namePaper uses the collected \textit{experiences}\footnote{Experience is a representation of a transition from one time step to another, in terms of $\langle State, Action, Reward, Next State \rangle$.} \gonx{\circled{8}}  to \gca{ update its decision-making policy online \gonx{\circled{9}}.} \gont{Sibyl continuously learns from its past decisions \gonz{and their impact}. Our two-threaded implementation avoids that the learning (i.e., training) interrupts \gonz{or delays} data placement decisions for incoming requests. To enable the parallel execution of the two threads, we duplicate the neural network \gonz{that is} used to \gonz{make} data placement \gonz{decisions}. While one network \gonz{(called the \emph{inference network} \circled{2})} is deployed \gonx{(i.e., makes decisions)} the second network \gonz{(called the} \gonz{training} network \circled{9}), is  trained in the background. \gonx{The} inference network is used \emph{only} for inference, while \gonx{the}
training network is used \emph{only} for training.} \gonz{Therefore, \gonx{\namePaper does} \textit{not} perform a separate training step for the inference network and} \gonx{instead} periodically cop\gonx{ies} the  \gonz{\gonz{training}} network weights to the \gonz{inference} network \circleds{10}.

\gont{For every new storage request to the HSS, Sibyl uses the state information \circled{1} to \gonz{make} a data placement decision \gonx{\circled{4}}. 
The inference network  predicts the Q-value for each available action given the state information. Sibyl policy \circled{3} selects the action with the maximum Q-value \gonx{or, with a low probability, a random action for exploration} and performs the data placement. 
}

\vspace{-0.2cm}
 \subsection{\namePaper Data Placement Algorithm}
\label{subsec:mechanism_rl_algo}

Algorithm~\ref{algo:rl_algo} describes \gon{how} \namePaper~\gon{performs} {data placement}  \grakesh{for} an HSS. \gon{Initially, the experience buffer is allocated to hold $e_{EB}$ entries \gonz{(line~\ref{algo:intial_eb})}, and the \gonz{training} and the inference network weights are initialized to \gonz{random values } \gonz{(lines~\ref{algo:intial_exp} and \ref{algo:intial_inf})}.} \gon{When a storage request is received \gonz{(line~\ref{algo:intial_req})}}, Sibyl policy (\circled{3}  Figure~\ref{fig:qrator}(a)) either (1) randomly selects an action  with $\epsilon$ probability (lines~\ref{algo:exploration1}-\ref{algo:exploration2}) to perform exploration in an HSS environment,  or \gon{(2) selects \gonx{the action that maximizes the Q-value, based on information stored in the inference network}} 
(lines~\ref{algo:exploitation1}-\ref{algo:exploitation2}). After performing the \gon{selected} action \gont{(line~\ref{algo:act})}, \namePaper collects its reward\gonx{, whose value depends on whether an eviction is needed from fast storage} (lines~\ref{algo:reward_if}-\ref{algo:reward_evict}). The \gont{generated experience \gonx{is} stored} in the experience buffer (line~\ref{algo:store_exp}). \gonz{Once the experience buffer has $e_{EB}$ entries (line~\ref{algo:eb_full}), \gonx{\namePaper} trains the training network.} \gont{During training,} the training network samples a batch of experiences  from the \gor{ experience buffer} (line~\ref{algo:line:sample_experiences}) and updates its weights using stochastic gradient descent (SGD)~\cite{bottou2003stochastic} (line~\ref{algo:line:bellman_update}). \gonx{\namePaper} does \textit{not} perform a separate training step for the inference network.  \gon{Instead}, the training network weights are copied to the inference network (line~\ref{algo:weight_trans}), which removes the training
of the inference network from the critical path \gonx{of decision-making}.  
\setlength{\textfloatsep}{0.1cm}
\setlength{\floatsep}{0.00cm}
\begin{algorithm}[h]
\scriptsize
\setstretch{0.7}
 \caption{\namePaper's \gont{reinforcement learning-based} data placement \gon{algorithm}}
 \label{algo:rl_algo}
 \begin{algorithmic}[1]
 \State \textbf{Intialize:} \texttt{the experience buffer \textit{EB} to capacity \textit{$e_{EB}$}\label{algo:intial_eb}}
 \State \textbf{Intialize:}  \texttt{{the} {{training} network}  with random weights $\theta$\label{algo:intial_exp}}
 \State \textbf{Intialize:}  \texttt{{the} {inference} {network}  with random weights \textit{$\hat{\theta}$}\label{algo:intial_inf}}
 \State \textbf{Intialize:} \texttt{the observation vector $O_t$=$O(s_1)$ with storage request $s_1$=\{$req_{t}$\}, and host and  storage features\label{algo:intial_req}}


        \ForAll{\texttt{storage requests}     }
            \If{\texttt{(rand() $<\epsilon$)}} \Comment{\comm{with probability $\epsilon$, perform exploration}\label{algo:exploration1}}
                \State    \texttt{ random action $a_t$} \label{algo:exploration2}
            \Else \Comment{\comm{{with probability 1-$\epsilon$}{,} perform exploitation}} \label{algo:exploitation1}
                \State  \texttt{ $a_t=argmax_a Q_t(a)$}  \label{algo:exploitation2}
                \Comment{\comm{select action with the highest $Q_t$ value {from inference network}}}
            \EndIf
           \State \texttt{execute  $a_t$} \label{algo:act}
             \Comment{\comm{{place} the requested page {to}  fast or  slow storage}}
           \If{ no eviction}\label{algo:reward_if} 
                \State \texttt{$r_t \gets$  $\frac{1}{L_t}$\label{algo:reward_no_evict}} 
                \Comment{\comm{{reward, given} no eviction of a page from fast to slow storage}}
            \Else
              \State \texttt{$r_t \gets$ max(0,$\frac{1}{L_t}$-\textit{$R_p$})\label{algo:reward_evict}} 
              \Comment{\comm{{reward with} an eviction penalty in case of an eviction}}
              \EndIf
           \State \texttt{store {experience} $(O_t,a_t,r_t,O(t+1))$ in \textit{EB} \label{algo:store_exp}}
            \If{{\texttt{({num} requests \gonx{in \textit{EB}}==$e_{EB}$)}}\label{algo:eb_full}}   \Comment{\comm{train  {training} network when EB is full}}
               \State \texttt{sample random batch{es} of  experiences  from \textit{EB}, which are in format $(O_j,a_j,r_j,O(j+1))$} \label{algo:line:sample_experiences}\;
               \Comment{\comm{where $O_j$ represents an observation  at a time instant j from \textit{EB}}}
            \State \texttt{Perform stochastic gradient descent}
               \Comment{ \comm{update the {training} network weights \label{algo:line:bellman_update} }}
                  \State \texttt{$\hat{\theta} \gets \theta$ \label{algo:weight_trans}} \Comment{ \comm{{copy} the {training} network weights to the inference network }}
              \EndIf 

        \EndFor
\end{algorithmic}
   \end{algorithm}


\subsection{\gont{Detailed Design of \namePaper}}
\label{subsec:detail_design}

 \subsubsection{RL Decision Thread} In this thread, \namePaper~\gonx{makes} data placement decisions while storing \textit{experiences} in an experience buffer. 
 Sibyl extracts the observation vector \circled{1} from the attributes of the incoming request and the current system state (e.g., access \gonz{count}, remaining capacity in the fast storage) and uses \gonx{the} inference network \circled{2} to predict the Q-values for each \gonx{possible} action with 
 the given state vector.
 \gonz{While making data placement decisions, Sibyl balances the \gonx{random} \textit{exploration} of the environment (to \gonx{find a better} policy \gonzz{without getting stuck at a suboptimal one}) with the \textit{exploitation} of its current  policy (to \gonx{maximize its reward based on the current inference network weights}).} 

\head{Sibyl policy} For every storage request, 
Sibyl policy selects the action that leads to \gonx{the highest long-term} reward \circled{6}.
 We use \rbc{a} Categorical Deep Q-Network (also known as C51)~\cite{C51} to update  $Q(s,a)$.
C51's objective is to learn the distribution of Q-values, 
whereas other variants of Deep Q-Networks~\cite{sutton1999policy,baird1995residual,liang2016deep,mnih2013playing,silver2016mastering,silver2017mastering} aim to approximate \gonx{a single value for $Q(s,a)$}. This distribution helps \rbc{Sibyl} to capture more information from the environment 
to make \gonz{better} \gont{data placement} decisions~\cite{harrold2022data}.  

\gont{For \gonx{tracking} the state, we divide each feature into a small number of bins to reduce the state space \gonx{(see \cref{sec:rl_formulation})}, which directly affects the implementation overhead of \namePaper.} We \gonx{select the number of bins (Table~\ref{tab:state}) based on empirical} sensitivity analysis. 
Our state representation uses a more relaxed encoding of 40 bits \gont{(than using only 20 bits for the \gonx{observation vector})} 
to allow for future extensions (e.g., features with more bins). 
Similarly,  we use a relaxed 4\gonzz{-bit} encoding for the action to allow extensibility to a different number of storage devices. \gont{For the reward structure, we use a half-precision floating-point (16-bit) representation. }

\head{Experience buffer}  
 \gonx{\namePaper} stores \emph{experiences} \gonx{it} collect\gonx{s} while interacting with the HSS in an  \emph{experience buffer}~\cite{dqn}. The experience buffer is allocated in  the host \gonz{main memory (DRAM)}. 
To minimize \gonx{its} design overhead, we 
deduplicate data in the stored experiences. 
  To improve the training quality, we \gonx{perform batch training where each batch consists of randomly sampled experiences}. \gonzz{This technique of randomly sampling experiences from the experience buffer} is called  \textit{experience replay}~\cite{dqn}. 

\begin{figure}[h]
\centering
  \includegraphics[width=0.5\linewidth,trim={0cm 0.1cm 0cm 0cm},clip]{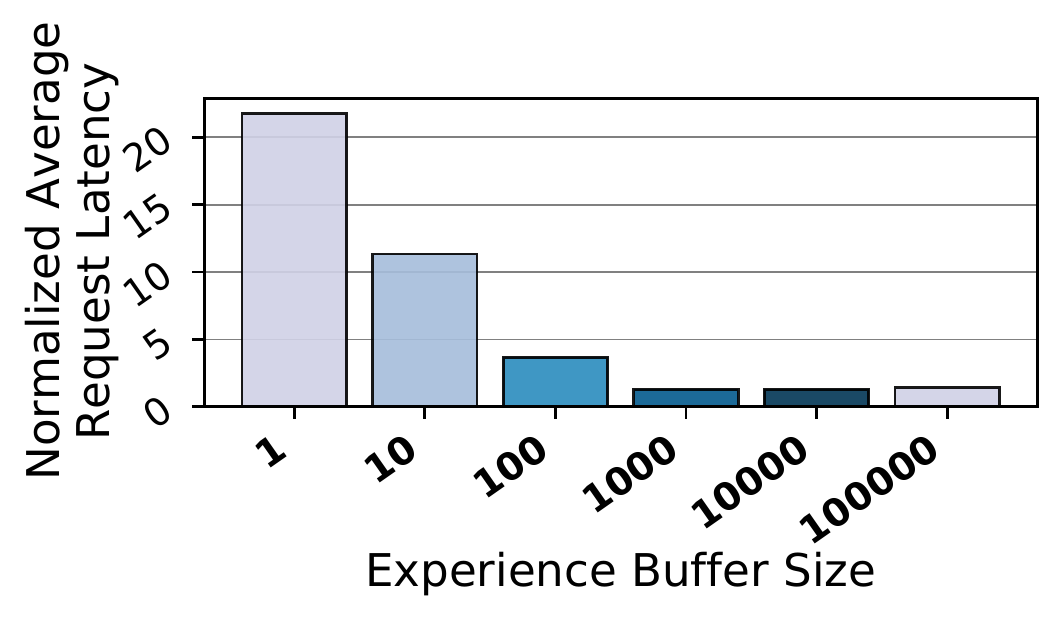}
  \vspace{-0.1cm}
\caption[Two numerical solutions]{Effect \gonx{of different experience buffer sizes on} the \gont{average request} latency (normalized to \fast) \label{fig:training_f}}
\end{figure}

Figure~\ref{fig:training_f} shows the effect \rbc{of} different \rbc{experience} buffer sizes \gonx{on \namePaper's performance} in the \hmssd configuration. 
We observe that \namePaper's performance saturates \gonx{at 1000 entries, which we select as \gonzz{the experience buffer} size}. 
{Since the size of our state representation is 40 bits, to store a single experience tuple, 
we  need 40-bit+\gonzz{4}-bit+16-bit+40-bit, i.e., 100 bits. 
\gont{In total, for 1000 experiences, the \gor{experience buffer} requires 100 KiB in the host DRAM.}}

\head{Exploration vs. exploitation}  An \gonx{RL} agent needs to \emph{explore} the environment to improve its policy 
\rbc{to maximize \gonx{its} long-term reward beyond  local maxima~\cite{sutton_2018}}.
At the same time, the agent \gonx{needs} to \emph{exploit} what it has already \emph{experienced} \gonx{so that it can take advantage of its learning so far}.   
\rbc{To balance exploration and exploitation,}
we use the $\epsilon$-greedy policy~\cite{tokic2011value}: the best-known action based on the agent's experience is selected with (1-$\epsilon$) probability, and \rbc{otherwise}, i.e., with $\epsilon$ probability, \rbc{another} action is chosen randomly. \gonx{Exploration allows \namePaper to experience states it may not otherwise get into~\cite{sutton_2018} and thus avoid missing higher long-term rewards.} 
To perform exploration, \namePaper randomly chooses to \gonx{place data} \gonx{to} the fast or the slow storage \juancr{device}, \gonx{so that it can} get more information about the HSS and the workload. Based on the received reward, \namePaper updates its \gonx{training network}. 
  \gonx{Such exploration} helps \namePaper to avoid making suboptimal data placement decisions \gonx{in the long run.}

\subsubsection{RL Training {Thread}}  
 \gont{This thread uses a} batch \rbc{of collected experiences} \circled{8} from the experience buffer \gonx{to train the training} network \circled{9}. 
The updated weights \gonx{of the training network} are transferred to the inference network after every 1000 requests \circleds{10}. 

\head{\gonz{Training} and inference \rbc{networks}} 
The training an\gonzz{d} inference network allows \gonzz{the} parallel execution of \gonx{decision} and training threads.
We use an identical neural network structure for the training and inference networks.
\gup{A \rbc{deep neural network can be} prohibitive due to the long time \gonx{it requires for} train\gonx{ing} and converge\gonx{nce}, preventing \namePaper to  adapt to new state-action pairs \gonx{in a timely manner}. Based on experiments, we find that a simple feed-forward network~\cite{bebis1994feed} with only two hidden layers~\cite{de1993backpropagation} \gonx{provides} good \gonzz{performance} 
for \gonx{\namePaper's} data placement task.} Figure~\ref{fig:qrator}(b) shows the structure of our \rbc{\gonz{training}} network.\footnote{\rbc{The inference network is identical in shape to the \gonz{training} network.}} The network takes the observation vector $O_t$ as \gonx{its} input and produces a probability distribution of \rbc{Q-values} as \gonx{its} output. Before feeding the data to the network, we preprocess the data by normalizing and casting the data to low precision data types, which allows us to reduce memory in the experience buffer. 
Next, we apply two fully-connected \juangg{hidden} layers of 20 and 30 neurons, respectively. 
{We select these neurons based on our extensive design space exploration with different numbers of hidden layers and neurons per layer.} After the two hidden layers, we have an output layer of 2 neurons, one for each action. \gonx{Sibyl policy \circled{3} selects the action with the maximum Q-value.} All fully-connected layers use the swish {activation} function~\cite{ramachandran2017searching},  a non-monotonic function that
 outperforms ReLU~\cite{agarap2018relu}. 
 
 \gca{During the training of the \gonz{training} network, \gonzz{the} inference network's weights are fixed.}  \gont{ After every 1000 requests, \gonzz{the} weights of the \gonz{training} network are copied to the inference network, 
which removes the training of the inference network from the critical path. We set the number of requests to 1000 based on our empirical evaluation of the experience buffer size \gonx{(Figure~\ref{fig:training_f})}. Each training step is composed of \gont{8 batches of experiences from an experience buffer of 1000 experiences with a batch size of 128}. We perform the training on the host CPU rather than on \gonx{a dedicated hardware accelerator} because (1) the network size is small and the weights perfectly fit in on-chip caches of the CPU in our evaluated system, \gonzz{and}  (2) to avoid continuous weight transfer overhead between the host CPU and the \gonx{accelerator} over the external interface. }

\sloppy\head{Hyper-parameter tuning}
\label{subsec:lhs}
\gont{We improve \namePaper's accuracy by tuning its hyper-parameters.  Hyper-parameters are sets of RL algorithm variables that can be tuned to optimize the accuracy of the RL agent~\cite{paine2020hyperparameter,NAPEL}.} 
{For hyper-parameter tuning, we perform 
cross-validation~\cite{arlot2010survey} 
using different hyper-parameter values. During cross-validation, we randomly select one workload for hyper-parameter tuning and  \gonx{use the} other \gont{thirteen} workloads for validation. On the selected workload, we use different hyper-parameter configurations that we choose using \gonzz{the} \mbox{\textit{design of experiments}} (DoE)~\cite{montgomery2017design}. }DoE allows us  to minimize the number of experiments needed
{ to find the \gonx{best} hyper-parameter \grakesh{values}} without sacrificing the quality of the information gathered by the experiments. \gont{Unlike traditional supervised learning methods, we do \emph{not} train Sibyl \emph{offline} using a training dataset before deploying it for data placement. \gonx{All training happens online in \namePaper.} For every evaluated workload, Sibyl starts with no prior knowledge and gradually learns to make data placement decisions online by interacting with the hybrid storage system.  \namePaper  needs only one-time \gonx{offline} hyper-parameter tuning.}

Table~\ref{tab:hyperparameter} shows the hyper-parameters considered \gonx{in} \namePaper's design \gonx{as well as their chosen values after the tuning process.  
The discount  factor ($\gamma$) \gonx{determines the balance between} the immediate and future rewards. \gonx{At $\gamma$=0 ($\gamma$=1), \namePaper gives importance only to the immediate (long-term) reward.} The learning rate ($\alpha$) determines the rate at which neural network weights are updated. \gonzz{A lower $\alpha$ makes small updates to the neural network weights, which could take more training iterations to converge to an optimal policy. While a higher $\alpha$ results in large updates to the neural network weights, which could cause the model to converge too quickly to a suboptimal solution.} 
The exploration rate ($\epsilon$) balances exploration and exploitation for \gonx{\namePaper}. We also explore different batch sizes \gonx{(i.e., the number of samples processed in each training iteration)} and experience buffer sizes to train our \gonz{training} network.} 
\begin{table}[h]
\ssmall
\begin{center}
 \caption{Hyper-parameters considered \gon{for} tuning }
    \label{tab:hyperparameter}
     \renewcommand{\arraystretch}{0.8}
\setlength{\tabcolsep}{10pt}
  \resizebox{1\linewidth}{!}{%
\begin{tabular}{l||l|l}
\hline
\textbf{Hyper-parameter}             & \textbf{Design Space} & \textbf{Chosen Value} \\  \hline

Discount factor   ($\gamma$)                           & 0-1  &0.9   \\
Learning rate     ($\alpha$)                  & $1e^{-5}-1e^{0}$ & $1e^{-4}$     \\
Exploration rate  ($\epsilon$)                           & 0-1  &0.001   \\
Batch size                          & 64-256   & 128     \\
\gonx{Experience buffer size ($e_{EB}$)}                         & 10-10000 & 1000      \\
\hline
\end{tabular}
}
\end{center}
\end{table}

\section{Evaluation Methodology} \label{sec:evaluation}
\head{Evaluation setup}
{
We evaluate  \namePaper~\gon{using}  real \gonf{systems with various HSS}   configurations.
The \gonf{HSS devices} \gon{appear} as a single flat {block device \rcamfix{that exposes} one contiguous logical block address space \gon{to the OS}, as depicted in \fig{\ref{fig:hybrid}}}. { We implement a lightweight custom block driver interface that manages the I/O requests to storage devices.  \tab{\ref{tab:devices}} {provides our system details, including the characteristics of the \gon{three}  storage devices \gon{we use}.}  
}  {To analyze the {sensitivity of our approach to} different device characteristics, we evaluate two different hybrid storage configurations (1) \em{performance-oriented HSS}: high-end device (\textsf{H})~\cite{inteloptane} and middle-end device (\textsf{M})~\cite{intels4510}, and (2) \em{cost-oriented HSS}: high-end device  (\textsf{H})~\cite{inteloptane} and low-end device (\textsf{L}) ~\cite{seagate}. 
\gca{We also evaluate two tri-hybrid \gon{HSS} configurations \gon{consisting of (1)}  \thold~and  \gon{(2)} \thnew~ devices.  }
{We \gon{run the} Linux Mint 20.1  operating system~\cite{linuxmint} with the  Ext3
file system~\cite{tweedie1998journaling}}. We use \gon{the} TF-Agents API~\cite{TFagents} to develop \namePaper. \gon{We evaluate \namePaper using two different metrics: \mbox{(1) \emph{average}} \emph{request latency}\rcamfix{, i.e.,} average \rcamfix{of the latencies of all storage \mbox{read/write} requests in a workload}, \gonzz{and} (2) \emph{request throughput (IOPS)}\rcamfix{, i.e., throughput of all storage requests in a workload in terms of completed I/O operations per second}. 
} 
}}
\begin{table}[h]
\begin{center}
\small
\setstretch{0.9}
\vspace{5pt}
\caption{{Host system and \gca{storage devices used in} hybrid storage configurations }}
 \label{tab:devices}
  \renewcommand{\arraystretch}{1} 
\setlength{\tabcolsep}{2pt} 
 \resizebox{1.0\columnwidth}{!}{%
\begin{tabular}{lll}
\hline
\multicolumn{1}{l|}{\textbf{Host System}}                  & 
\multicolumn{2}{l}{\begin{tabular}[c]{@{}l@{}}AMD Ryzen 7 2700G~\gon{\cite{amdryzen}}, 8-core\gon{s}@3.5 GHz, \\ 8$\times$64/32 KiB L1-I/D, 4 MiB L2, 8 MiB L3,  \\16 GiB RDIMM DDR4 2666 MHz
\end{tabular}}  
\\ \hline
\multicolumn{1}{l|}{\textbf{Storage \rcamfix{Devices}}}                  & \multicolumn{2}{l}{\textbf{Characteristics}}                                                                                                                                        \\ \hline
\multicolumn{1}{l|}{\textsf{H}: Intel Optane SSD P4800X~\cite{inteloptane}}     & \multicolumn{2}{l}{\begin{tabular}[c]{@{}l@{}}375 GB, PCIe 3.0 NVMe, SLC, R/W: 2.4/2 GB/s,\\ random R/W: 550000/500000 IOPS\end{tabular}} \\ \hline

\multicolumn{1}{l|}{\textsf{M}: Intel SSD D3-S4510~\cite{intels4510}} & \multicolumn{2}{l}{\begin{tabular}[c]{@{}l@{}}1.92 TB, SATA TLC (3D), R/W: 550/510 MB/s, \\ random R/W: 895000/21000 IOPS\end{tabular}}           \\ \hline
\multicolumn{1}{l|}{\textsf{L}: Seagate \ghpca{HDD} ST1000DM010~\cite{seagate} } & \multicolumn{2}{l}{\begin{tabular}[c]{@{}l@{}}1 TB, SATA 6Gb/s 7200 RPM \\ \rakesh{Max. Sustained Transfer Rate: 210 MB/s}\end{tabular}}   \\      \hline
\multicolumn{1}{l|}{\textsf{L$_{SSD}$}: ADATA SU630 SSD ~\cite{adatasu630} } & \multicolumn{2}{l}{\begin{tabular}[c]{@{}l@{}}960 GB, SATA 6 Gb/s, TLC,   \\ Max R/W: 520/450 MB/s\end{tabular}}      
\\ \hline                          \multicolumn{1}{l|}{\textbf{HSS \gon{Configurations}}}              & \multicolumn{1}{l|}{\textbf{Fast \ghpca{Device}}  }                                                            & \textbf{Slow \ghpca{Device}}                                                             \\ \hline
\multicolumn{1}{l|}{\textsf{H\&M} (Performance-oriented) }                       & \multicolumn{1}{l|}{high-end (\textsf{H})}                                                                   & middle-end (\textsf{M})                                                                \\
\multicolumn{1}{l|}{\textsf{H\&L} (Cost-oriented) }                          & \multicolumn{1}{l|}{high-end (\textsf{H})}                                                                   & low-end (\textsf{L})                                                                   \\ \hline                                           
\end{tabular}
}
\end{center}
\end{table}

{{\head{Baselines}} \label{subsec:eval:baselines} {We compare \namePaper against \rcamfix{two } state-of-the-art heuristic-based \gon{HSS} data placement techniques, (1) cold data eviction (\cde) \cite{matsui2017design} and (2) history-based \gon{page selection} (\hps)~\cite{meswani2015heterogeneous}, \gon{(3)} a state-of-the-art supervised learning-based technique (\arc)~\cite{ren2019archivist}, \gon{and (4)} a recurrent neural network (RNN)-based data placement technique (\kleio), adapted from \rakeshisca{Kleio~\cite{doudali2019kleio}, a data placement technique for hybrid memory systems}. \gup{\kleio 
provides a state-of-the-art ML-based data placement \gon{baseline}.} 
We compare the above policies with three extreme \gon{baselines}:
(1) \slow{}, {where all data resides in \gon{the} slow \gon{storage (i.e., there is no fast storage)}}, (2) \fast, {where all data resides in \gon{the}  fast \gon{storage}}, {and (3) \textsf{Oracle}}
\cite{meswani2015heterogeneous}, which exploits \gon{complete} {knowledge of} future I/O-access patterns  
{to perform data placement and to }\js{select victim data blocks for eviction from} the fast device.

{\head{Workloads}} 
We use fourteen different \js{block-I/O traces} from {the} MSRC benchmark suite~\cite{MSR} that are collected from real enterprise server workloads. 
{We carefully select the fourteen traces to have {distinct} I/O-access patterns, as shown in \tab{\ref{tab:workload}}, in order to study a diverse set of workloads with different \rcamfix{randomness} and  \rcamfix{hotness} \gon{properties} (see Figure~\ref{fig:apps}). 
We quantify a workload’s randomness using the average request size of the workload; the higher (lower) the average request size, the more sequential (random) the workload. \gon{The average access \rcamfix{\gonz{count}} provides the average \rcamfix{of the access counts of all pages in a workload;}
the higher (lower) the average access \rcamfix{\gonz{count}}, the hotter (colder) the workload.  
\rcamfix{\tab{\ref{tab:workload}} also  shows}  \rcamfix{the number of} unique requests in a workload.} To demonstrate {\namePaper}'s ability to generalize and provide performance gains across \gon{\emph{unseen traces}}, \gon{i.e., traces} that are \emph{not} used to tune the \gon{hyper-parameters of} \namePaper, we evaluate {\namePaper} using \gon{four} additional workloads from FileBench~\cite{tarasov2016filebench}.} 

\begin{table}[h]
\caption{Characteristics of 14 evaluated workloads}
 \label{tab:workload}
 \centering
 \tiny
 \setstretch{0.85}
   \renewcommand{\arraystretch}{0.85} 
\setlength{\tabcolsep}{2pt} 
\resizebox{0.9\linewidth}{!}{%
\begin{tabular}{@{}lrrrrr@{}}
\hline
 \textbf{Workload}        & \multicolumn{1}{l}{\textbf{Write}} & \multicolumn{1}{l}{\textbf{Read}} & \multicolumn{1}{r}{\textbf{Avg. request}} & \multicolumn{1}{r}{\textbf{Avg. access}} & \textbf{No. of unique}  \\ 
                &         \%                  &    \%                      & \multicolumn{1}{r}{\textbf{size}}          & \multicolumn{1}{r}{\textbf{\gonz{count}}} & \textbf{requests} \\ 
         
         \hline
hm\_1    & 4.7\%                     & 95.3\%                   & 15.2                                  & 44.5                                  & 6265                     \\

mds\_0    & 88.1\%                     & 11.9\%                   & 9.6                                  & 3.5                                  &        31933              \\
prn\_1   & 24.7\%                    & 75.3\%                   & 20.0                                  & 2.6                                  & 6891                     \\
proj\_0  & 87.5\%                    & 12.5\%                   & 38.0                                  & 48.3                                 & 1381                     \\
proj\_2  & 12.4\%                    & 87.6\%                   & 42.4                                 & 2.9                                  & 27967                    \\
proj\_3  & 5.2\%                     & 94.8\%                   & 9.6                                  & 3.6                                  & 19397                    \\
prxy\_0  & 96.9\%                    & 3.1\%                    & 7.2                                  & 95.7                                 & 525                      \\
prxy\_1  & 34.5\%                    & 65.5\%                   & 12.8                                  & 150.1                                 & 6845                     \\
rsrch\_0 & 90.7\%                    & 9.3\%                    & 9.2                                 & 34.7                               & 5504                     \\
src1\_0  & 43.6\%                    & 56.4\%                   & 43.2                                 & 12.7                                  & 13640                    \\
stg\_1   & 36.3\%                    & 63.7\%                   & 40.8                                 & 1.1                                  & 3787                     \\
usr\_0   & 59.6\%                    & 40.4\%                   & 22.8                                  & 19.7                                  & 2138                     \\
wdev\_2  & 99.9\%                    & 0.1\%                    & 8.0                                  & 17.7                                 & 4270                     \\
web\_1   & 45.9\%                    & 54.1\%                   & 29.6                                  & 1.2                                  & 6095                     \\ \hline
\end{tabular}
}
\end{table}


 \begin{figure*}[t]
\centering
\begin{subfigure}[h]{.45\textwidth}
  \centering
  \includegraphics[width=\linewidth,trim={0.2cm 0.2cm 0cm 0.2cm},clip]{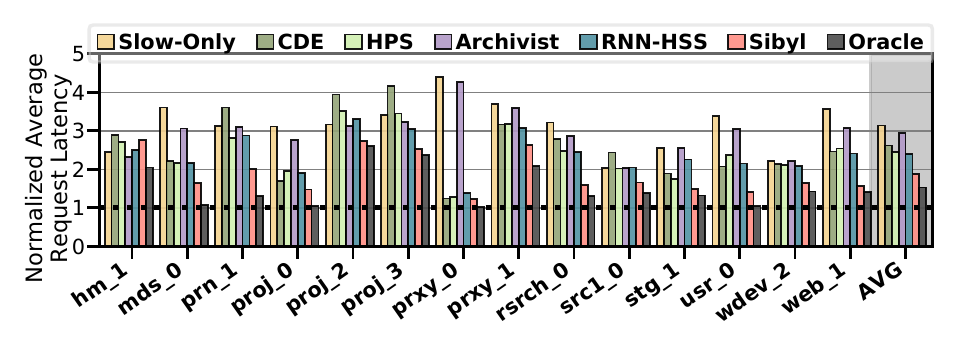}
  \vspace{-0.6cm}
   \caption{\textsf{H\&M} \gon{HSS} configuration\label{fig:perf_1}}
\end{subfigure}%
\begin{subfigure}[h]{.45\textwidth}
  \centering
  \includegraphics[width=\linewidth,trim={0.2cm 0.2cm 0.2cm 0.2cm},clip]{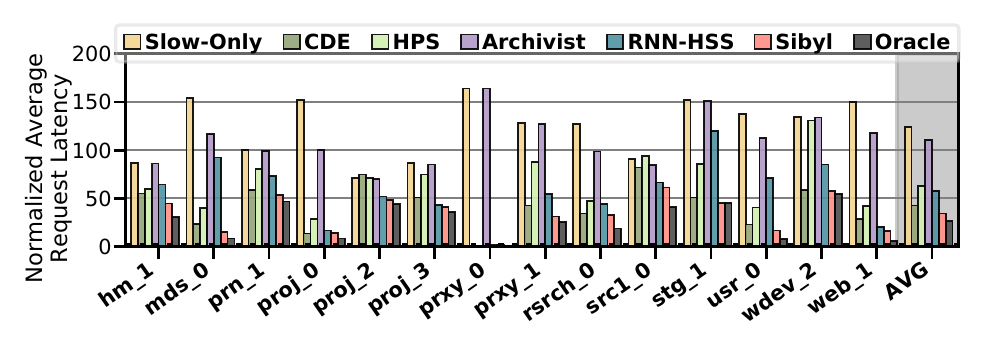}
    \vspace{-0.6cm}
   \caption{\textsf{H\&L} \gon{HSS} configuration \label{fig:perf_2}}
\end{subfigure}
\caption[Two numerical solutions]{\gon{Average request latency} under two different hybrid storage configurations (normalized to \fast
)
\label{fig:perf}}
 \vspace{-0.1cm}
\end{figure*}
 
\begin{figure*}[h]
\vspace{-1em}
\centering
\begin{subfigure}[h]{.45\textwidth}
  \centering
  \includegraphics[width=\linewidth,trim={0.2cm 0.2cm 0cm 0.2cm},clip]{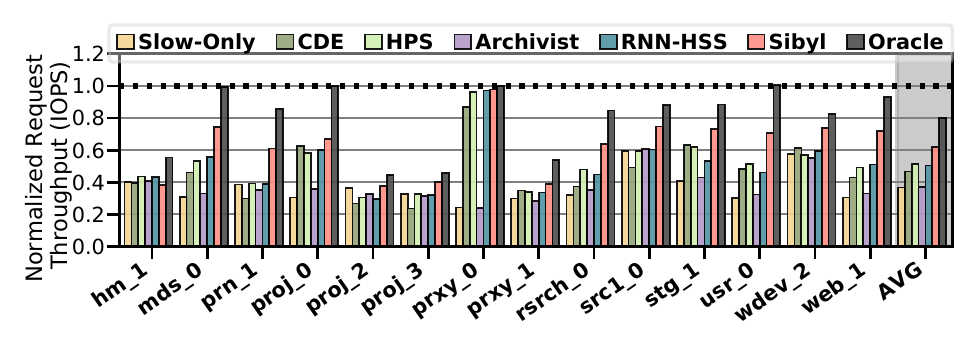}
  \vspace{-0.6cm}
\caption{\textsf{H\&M} \gon{HSS} configuration 
  \label{fig:iops_1}}
\end{subfigure}%
\begin{subfigure}[h]{.45\textwidth}
  \centering
  \vspace{-0.04cm}
  \includegraphics[width=\linewidth,trim={0.2cm 0.2cm 0.2cm 0.2cm}, clip]{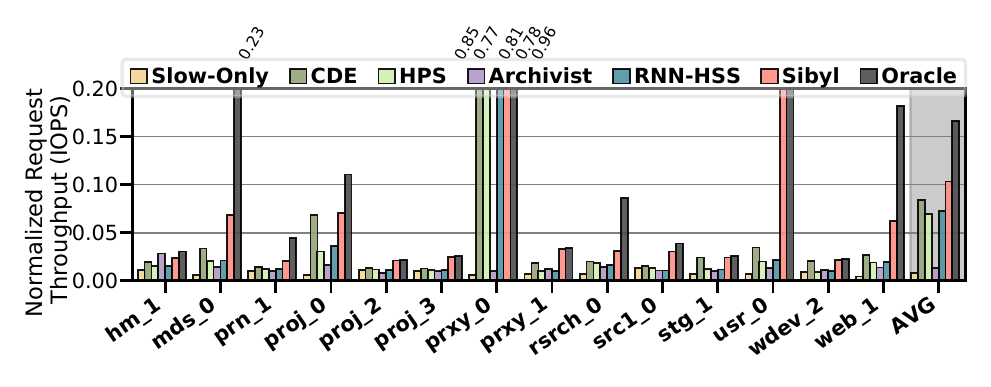}
  \vspace{-0.6cm}
  \caption{\textsf{H\&L} \gon{HSS} configuration 
  \label{fig:iops_2}}
\end{subfigure}
\caption{\gon{Request throughput (IOPS)} under two different hybrid storage configurations (normalized to \fast
)
\label{fig:iops}}
\vspace{-0.5cm}
\end{figure*}

\vspace{-0.1cm}
\section{Results}
\label{sec:results}
 \subsection{Performance Analysis}
 \label{subsection:Evaluation_perf}
Figure~\ref{fig:perf} compares the average \gon{request} latency of \namePaper against the baseline policies \gon{for \hmssd (Figure~\ref{fig:perf}(a)) and \hlssd (Figure~\ref{fig:perf}(b)) HSS configurations}. \js{All values are} normalized to \js{\fast}. 
We make \rcamfix{five major} observations. 
First, \namePaper consistently outperforms all the baselines for all the workloads in \hlssd and all but two workloads in \hmssd. \gon{In the} \js{\textsf{H\&M}} \gon{HSS} configuration (Figure~\ref{fig:perf}(a)), {where the latency difference between  two  devices  is  relatively  smaller than \textsf{H\&L},} \namePaper improves \gon{average} performance by {28.1\%, 23.2\%, 36.1\%, and 21.6\%} 
\gonzz{over} \cde, \hps, \arc, and \kleio, respectively. In \gon{the} \js{\textsf{H\&L}} \gon{HSS} configuration (Figure~\ref{fig:perf}(b)), where there is a large difference between the latencies of the two {storage devices}, \namePaper~improves performance by 19.9\%, 45.9\%, 68.8\%, and 34.1\% 
\gonzz{over} \cde, \hps, \arc and \kleio, respectively.  } 
\gor{
\rcamfix{We observe that the} larger the latency gap between HSS devices, the higher the expected benefits {of avoiding} the eviction penalty by placing only performance-critical pages in the fast storage.
Second, in \rcamfix{the} \textsf{H\&M} HSS configuration, \gon{\cde and \hps are ineffective for certain workloads (\textsf{hm\_1}, \textsf{prn\_1}, \textsf{proj\_2}, \textsf{proj\_3}, and \textsf{src1\_0}) even when compared to \slow.} 
\rcamfix{In contrast,} ~\namePaper~\gon{consistently \rcamfix{and significantly} outperforms \slow for all workloads because it} can learn the small latency difference between \js{the two storage devices in \textsf{H\&M}} and dynamically \gonzz{adapts its} data placement decisions, which is \rcamfix{difficult} \js{for \cde and \hps due to their inability to holistically take into account the underlying device characteristics.} \gsa{Third, \namePaper provides slightly lower performance than \gon{other} baselines in only two workloads: \slow, \hps, \arc, and \kleio for \hm and \cde and \hps for \proxy in the \hmssd HSS configuration. We observe that such workloads are write-intensive and have many random requests (in terms of \gon{both access pattern}  and request size). Therefore,  such  workload\gon{s} would benefit from  more frequent retraining of Sibyl’s \gonzz{training} network. We experimentally  show in \cref{subsection:mixed_workload} that using a lower learning rate \gonzz{during} the training of the \gonzz{training} network helps to improve \rcamfix{\namePaper's} performance for such workloads. \gon{ Fourth, \namePaper achieves, on average, 80\% of the performance of the \oracle, \rcamfix{which} has complete knowledge of future access patterns, \gont{across \hmssd and \hlssd}. Fifth, \kleio provides higher performance than heuristic-based policies (2.1\% and 8.9\% than \cde and \hps, respectively, in \hmssd and 9.8\% than \hps in \hlssd), \rcamfix{but \namePaper outperforms it by \gonz{27.9\%}}.
Unlike \namePaper, \rcamfix{the two machine learning-based policies,} \arc and \kleio, do \emph{not} consider any system-level feedback, which leads to \rcamfix{their} suboptimal performance. }}

Figure~\ref{fig:iops} \gon{compares the request throughput (IOPS)} of {\namePaper} against other {baseline} policies. We make two observations. 
\rcamfix{First, in the \textsf{H\&M} ({\textsf{H\&L}}) HSS configuration (Figure~\ref{fig:iops}), {\namePaper} improves throughput by 32.6\% (22.8\%), 21.9\% (49.1\%), 54.2\% (86.9\%), and 22.7\% (41.9\%) \gonzz{over} {\cde}, {\hps}, {\arc}, and {\kleio}, respectively.}
Second, \namePaper provides slightly lower performance than \slow, \cde, \hps, \arc, and \kleio for \rcamfix{only} \hm in \hmssd HSS configuration. We draw similar observations \rcamfix{for throughput results} as \rcamfix{we did for latency results} (Figure~\ref{fig:perf}) because as \rcamfix{\namePaper} consider\rcamfix{s} the request size in state features and \gonz{request} latency in the reward, \rcamfix{it} {also} indirectly captures throughput (size/latency).

\gon{ We conclude that \namePaper 
consistently provides higher performance than \rcamfix{all five} baselines and \rcamfix{significantly} improves both average request latency and request throughput.}}

\subsection{{Performance on Unseen Workloads}}
\label{subsection:perf_unseen}
To demonstrate {\namePaper}'s ability to generalize and provide performance gains across \textit{unseen} \rcamfix{workloads} that are \textit{not} used to tune the \rcamfix{hyper-parameters of the} data placement policy of \namePaper, we evaluate {\namePaper} using \gonzz{four} additional  workloads from FileBench~\cite{tarasov2016filebench}. \gon{No data placement policy \rcamfix{we evaluate}, including \namePaper, \rcamfix{is} tuned on these workloads. } Figure~\ref{fig:unseen} shows the performance of \rcamfix{these} unseen \rcamfix{workloads}. 
We observe the following observations. First, in \textsf{H\&M} (\textsf{H\&L}) \gon{HSS} configuration, {\namePaper}~outperforms {\kleio} and {\arc}  by $46.1\%$ ($54.6\%$) and $8.5\%$ ($44.1\%$), respectively. Second, \mbox{\namePaper} may misplace some pages during the \rcamfix{online} adaptation period, but it provides significant performance benefit\rcamfix{s} over existing ML-based data placement techniques.} 
We conclude that {\namePaper} \gon{
provides high performance benefits on unseen \rcamfix{workloads} for which it has not been tuned.}

\begin{figure}[h]
    \centering
 \includegraphics[width=0.9\linewidth]{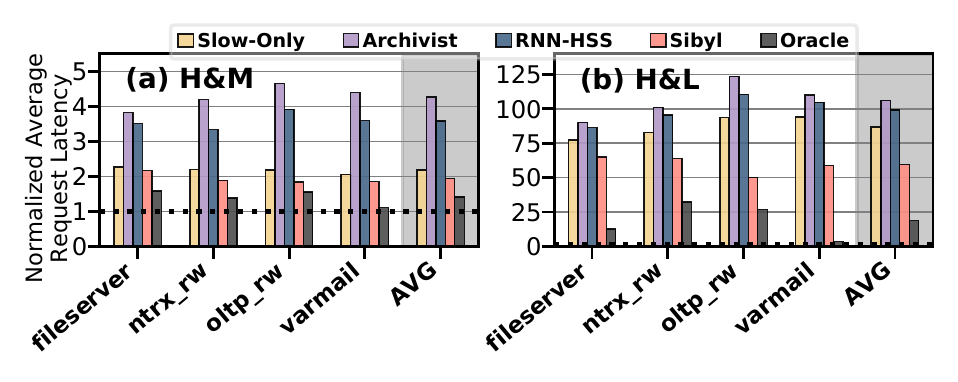}  \vspace{-0.2cm}
    \caption{{\gon{Average request} latency on unseen workloads (normalized to \fast) under two HSS configurations}}
    \vspace{-0.2cm}
    \label{fig:unseen}
\end{figure}

\subsection{Performance on Mixed Workloads}
\label{subsection:mixed_workload}
We evaluate mixing  two or more workloads at the same time while \rcamfix{randomly varying} their relative start times.  Table~\ref{tab:mixed_workloads} describes the characteristics of these mixed workloads. These workloads are truly independent of each other, potentially creating more evictions from the fast storage device than a single workload.  Such a scenario \mbox{(1) leads} to unpredictable execution where requests arrive at different, unpredictable timesteps, (2) mimics distributed workloads, and (3) further tests the \gon{ability} of Sibyl to dynamically adapt its decision-making policy.

\begin{figure}[h]
    \centering
 \includegraphics[width=1\linewidth]{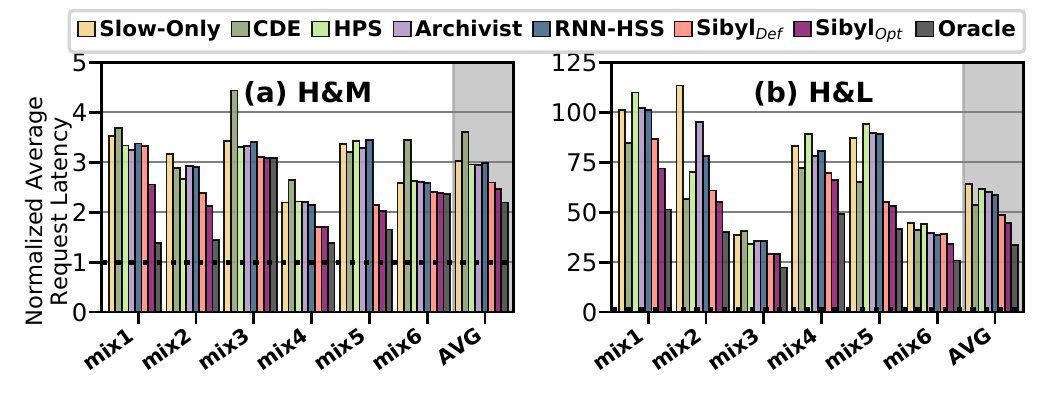}    
\vspace{-0.4cm} 
 \caption{{\gon{Average request} latency on mixed workloads (normalized to \fast) \rcamfix{and two HSS configurations}}}
  
    \label{fig:mixedWorkloads}
\end{figure}

Figure~\ref{fig:mixedWorkloads} shows \gonzz{average request latency} for mixed workloads. We use two different settings for \namePaper: (a) \sibyldef, where we use our default hyper-parameters (\cref{subsec:lhs}), and (b) \sibylopt, where we optimize the hyper-parameters \rcamfix{for} these mixed workloads and use a lower learning rate ($\alpha$) of $1e^{-5}$. A lower learning rate performs small\rcamfix{er} updates to the \gonz{training}  network’s weights in each training \rncam{iteration}, thus requiring more training to converge to an optim\gon{al} solution.
\gsa{{
\begin{table}[h]
 \caption{Characteristics of mixed workloads}
    \label{tab:mixed_workloads}
\centering
\normalsize
 \setstretch{0.8}
   \renewcommand{\arraystretch}{0.8} 
\setlength{\tabcolsep}{2pt}
  \resizebox{0.8\linewidth}{!}{%
\begin{tabular}{c||c|l}
\hline
\textbf{\rcamfix{Mix}}             & \textbf{\rcamfix{Workloads}} & \textbf{Description} \\ 
\hline
\hline
\textbf{mix1}                     &  prxy\_0~\cite{MSR} and ntrx\_rw~\cite{tarasov2016filebench} & \begin{tabular}{l}    
 Both prxy\_0 and \\ ntrx\_rw are write-intensive
\end{tabular}
    \\
\hline
\textbf{mix2}                      &  rsrch\_0~\cite{MSR} and oltp\_rw~\cite{tarasov2016filebench}  &
\begin{tabular}{l} 
rsrch\_0 is  write-intensive and \\ oltp\_rw is read-intensive
\end{tabular}
\\
\hline
\textbf{mix3}                      &  proj\_3~\cite{MSR} and YCSB\_C~\cite{cooper2010benchmarking}& 
\begin{tabular}{l}
Both proj\_3 and \\ YCSB\_C are read-intensive
\end{tabular}
\\
\hline
\textbf{mix4}                      & src1\_0~\cite{MSR} and fileserver~\cite{tarasov2016filebench} &
\begin{tabular}{l}
Both src1\_0 and \\ fileserver have nearly equal \\ numbers of reads and writes
\end{tabular} \\
\hline
\textbf{mix5}                     & 
\begin{tabular}{l}
prxy\_0~\cite{MSR}, oltp\_rw~\cite{tarasov2016filebench} and \\ fileserver~\cite{tarasov2016filebench}
\end{tabular}
&
\begin{tabular}{l}
prxy\_0 is write-intensive, \\ oltp\_rw is read-intensive, and \\fileserver has nearly equal \\numbers of reads and writes
\end{tabular}
\\
\hline
\textbf{mix6}                     &   
\begin{tabular}{l}
src1\_0~\cite{MSR}, YCSB\_C~\cite{cooper2010benchmarking} and \\ fileserver~\cite{tarasov2016filebench}  
\end{tabular}
&
\begin{tabular}{l}
src1\_0 and fileserver have\\nearly equal numbers \\of reads and writes while \\YCSB\_C is read-intensive
\end{tabular}
\\
\hline
\end{tabular}
}
\end{table}
}
}

We make two observations.   First, \sibyldef consistently outperforms \cde, \hps, \arc, and \kleio~ \gon{by} 27.9\%, 12.2\%, 12.1\%, and 12.9\%,  
 respectively, in \rcamfix{the} \textsf{H\&M} HSS \gon{configuration} and 9.4\%, 21.3\%, 19.4\%, and 17.1\%, respectively, in \textsf{H\&L} HSS \gon{configuration}. Second,  with a lower learning rate \rcamfix{and optimized hyper-parameters}, \sibylopt provides  5.2\% (9.3\%) higher average performance for \textsf{H\&M} (\textsf{H\&L}) \gon{HSS} configuration than \sibyldef.
 \gon{Third}, for \mix,  \hps provides comparable performance to \sibyldef in \textsf{H\&M},
 and \cde provides slightly  better performance  in \textsf{H\&L}.
 As discussed in \cref{subsection:Evaluation_perf}, \proxy  is write-intensive and has  random requests (with an average request size of 7.2) \gon{within} every 1000 requests, which is the experience buffer size to train the \gonz{training}  network. Such a workload requires more frequent retraining of Sibyl’s \gonz{training}  network to achieve higher performance. \gon{We conclude that \namePaper ~\gonz{can} \rcamfix{effectively} adapt its \rcamfix{data placement} policy \rcamfix{online} to highly dynamic workloads.}


\gon{
\subsection{Performance with Different Features}
{Figure~\ref{fig:ablation} compares the use of some of the most \rcamfix{useful} features for the \emph{state} of \namePaper in our \textsf{H\&L} \gon{HSS} configuration.  \textsf{All} represents using all the six features \gca{in Table~\ref{tab:state}}. \namePaper~\gca{autonomously} decides which features are important to maximize the performance of \gon{the running} workload. 

\begin{figure}[h]
  \centering
    \includegraphics[width=1\linewidth,trim={0cm 0.35cm 0cm 0cm},clip]{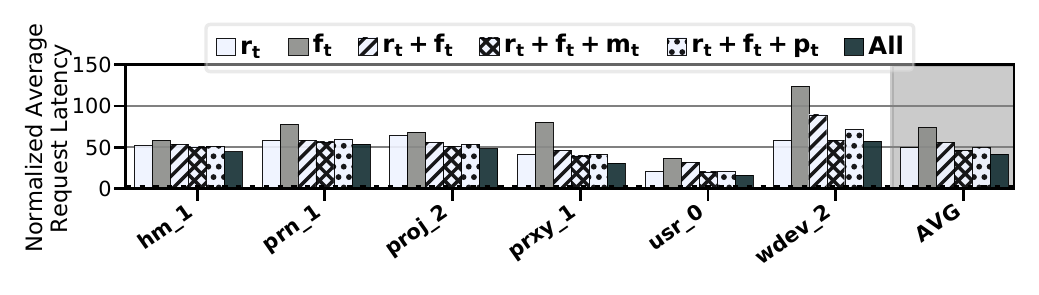}
  \caption{\rcamfix{Average request latency when using} different features (see Table~\ref{tab:state}) for the state space of \namePaper in the \textsf{H\&L} \gon{HSS configuration}  (normalized to \fast) }
\label{fig:ablation}
 \end{figure}

We make two {key} observations from Figure~\ref{fig:ablation}. First, \namePaper consistently achieves the lowest latency \gup{(up to 43.6\% lower)} by using all the features {mentioned in Table~\ref{tab:state}} (\textsf{All} in {Figure~\ref{fig:ablation}}). Second, by using the same features as in \rcamfix{baseline} heuristic-based polic\gonz{ies}, \namePaper is able to perform better data placement decisions. For example, \rcamfix{\textsf{$r_t$} and \textsf{$f_t$}  configurations of} \namePaper \rcamfix{~(in Figure~\ref{fig:ablation})} use \rcamfix{only one} feature, \rcamfix{just like} {\cde and \hps} do. 
\rcamfix{These two \namePaper configurations outperform {\cde and \hps}  policies by 4.9\% and 5.5\%, respectively (\textit{ref.} {Figure~\ref{fig:perf}(b)}). Using the same features as a heuristic-based policy, Sibyl autonomously finds a \rncam{higher-performance} dynamic policy that can maximize the reward function, which heuristic-based policies cannot possibly do.}
We conclude that {\namePaper}~uses a richer set of features 
that can capture multiple aspects of a storage request to \gon{\gonz{m}ake \gonz{better}}  {data} placement decisions than a heuristic-based policy. \gup{RL reduces 
\juangg{the design burden on system architects,} as \namePaper autonomously learns to use \gon{the provide\rcamfix{d}} features to achieve the highest cumulative reward.} In contrast, traditional heuristic-based policies use \rcamfix{features to make rigid data placement decisions without any system-level feedback,} \gonz{and thus they underperform \rncam{compared to \namePaper}.}
} 
}

\subsection{\gon{Performance with Different Hyper-Parameters}}

\begin{figure}[h]
\begin{subfigure}[t]{.45\linewidth}
  \includegraphics[width=\linewidth,trim={0cm 0cm 0cm 0cm},clip]{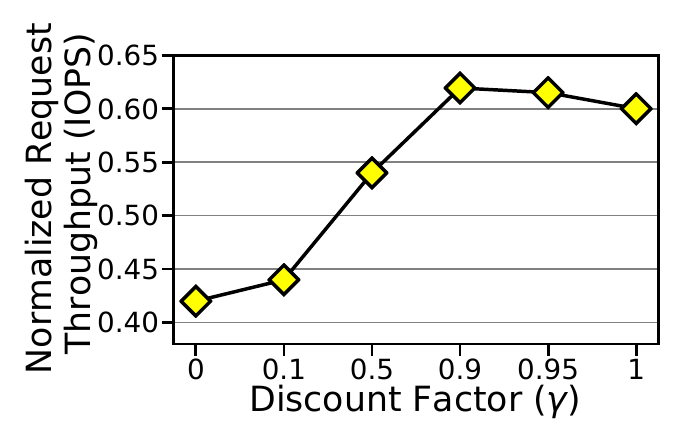}
   \vspace{-1.8cm}
  \caption{ \hspace{-2cm}
  \label{fig:discount}}
\end{subfigure}%
\begin{subfigure}[t]{.45\linewidth}
  \includegraphics[width=\linewidth,trim={0cm 0cm 0cm 0cm},clip]{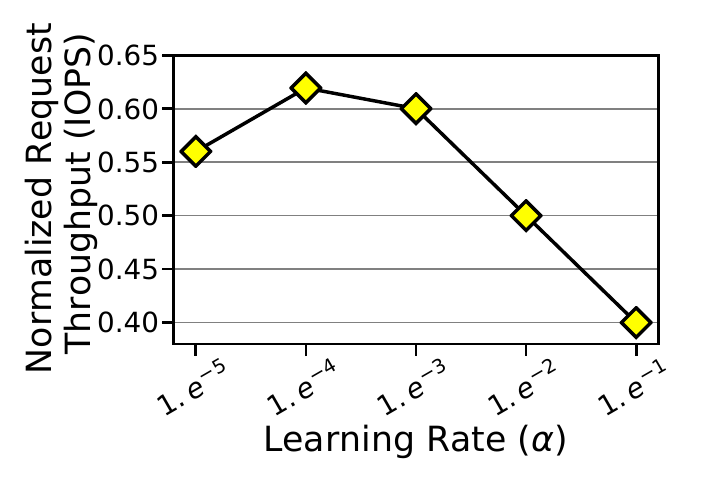}
  \vspace{-1.9cm}
   \caption{\hspace{-2cm}
  \label{fig:learning_rate}}
\end{subfigure}
\begin{subfigure}[t]{.45\linewidth}
  \includegraphics[width=\linewidth,trim={0cm 0cm 0cm 0cm},clip]{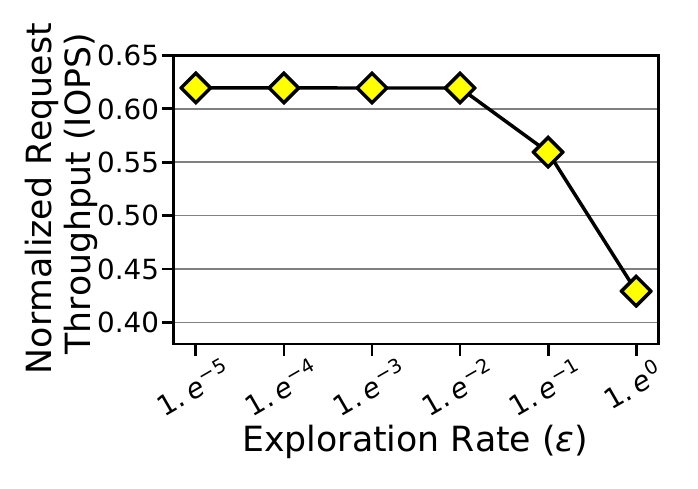} 
  \vspace{-1.9cm} 
   \caption{\hspace{-2cm}
  \label{fig:epsilon}}
\end{subfigure}
\caption{Sensitivity of \namePaper~\rcamfix{throughput} to: (a) the discount factor ($\gamma$), (b) the learning rate ($\alpha$), (c) the exploration rate ($\epsilon$), \rcamfix{averaged across 14 workloads (normalized to Fast-Only)} \label{fig:hyper_senstivity}}
\end{figure}

Figures~\ref{fig:hyper_senstivity}(a), \ref{fig:hyper_senstivity}(b), and \ref{fig:hyper_senstivity}(c) show the effect \gon{of}  three critical hyper-parameters (discount factor, learning rate, and exploration rate) \rcamfix{on \namePaper's throughput in \hmssd HSS configuration}. 
Figure~\ref{fig:hyper_senstivity}(a) \rcamfix{shows} that \namePaper's \rcamfix{throughput} drops sharply at $\gamma=0$.
At $\gamma=0$, \namePaper gives importance \gonz{only} to the immediate reward \gonz{and not} \rncam{at all to} the long-term reward, \gca{leading to} lower performance. We use $\gamma=0.9$, where \namePaper ~\rcamfix{is more forward-looking, giving enough weight to} long-term rewards.
\rcamfix{Figure~\ref{fig:hyper_senstivity}(b) shows} that at \rcamfix{a learning rate of} $\alpha=1e^{-4}$, \namePaper~\gon{provides} the best performance. \rcamfix{The learning rate determines the rate at which \gonz{training} network w\gonz{e}ights are updated.} \gonz{Both too slow and too fast} updates are detrimental \gonz{for}  adaptive learning and \gonz{stable \rncam{exploitation} of a learned policy}, \rncam{respectively}.
Third, Figure~\ref{fig:hyper_senstivity}(c)  shows  that the performance of \mbox{\namePaper} drops sharply if it performs exploration \rncam{too frequently} \rcamfix{(i.e., $\epsilon=1e^{-1}$)} and \rcamfix{thus} does not \gonz{sufficiently} exploit its learned policy. \namePaper achieves \rcamfix{the highest} performance improvement\rcamfix{s} for \rcamfix{$ 1e^{-5}\leq ~\epsilon \leq 1e^{-2}$}. 



{\subsection{{Sensitivity to Fast Storage Capacity}}
{Figure~\ref{fig:sensitivity} shows the \gon{average request latency} of \namePaper and baseline policies \gon{as we vary}  the available capacity in the fast  \gon{storage}. \gonzz{The} \rcamfix{\rncam{x}-axis denotes a range of fast storage device sizes available for data placement and represented in terms of percentages of the entire fast storage device capacity, \gonz{where 100\% represents \rncam{the size where} all pages of a workload can fit in the fast storage}}. 
\begin{figure}[h]
    \centering
 \includegraphics[width=1\linewidth]{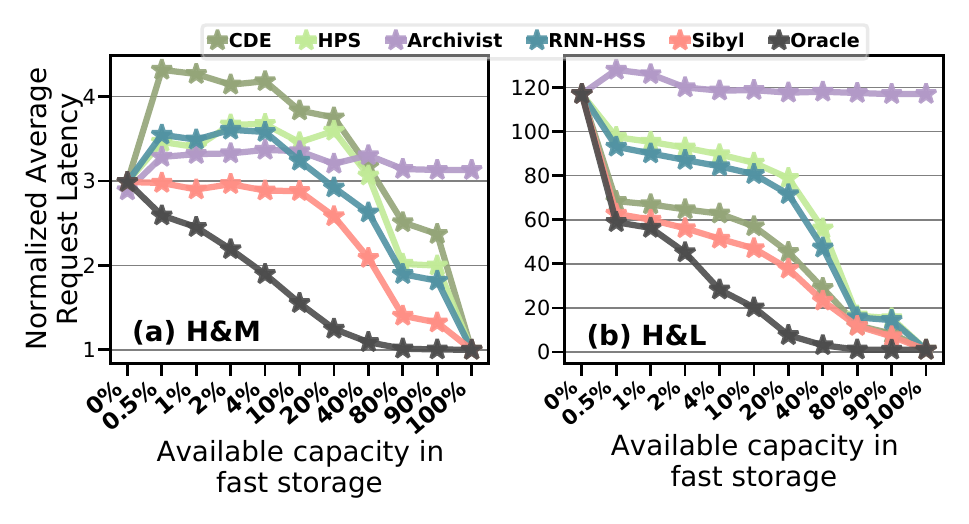}
\vspace{-0.5cm}
    \caption{\gon{Average request latency for various fast storage \gonz{device} sizes (normalized to \fast)}}
    \label{fig:sensitivity}
\end{figure} 
}

\gonzz{We make two observations.} First, \rcamfix{for all fast storage sizes}, \namePaper performs better than  the baseline heuristic- 
and supervised learning-based policies for both \textsf{H\&M} and \textsf{H\&L} HSS configurations. 
\rcamfix{Even when the fast storage size is as small as 1\%,}~\namePaper outperforms \cde, \hps, \arc, \kleio by 47.2\% (11.5\%), 17.3\% (58.9\%), 12.3\% (110.1\%), 21.7\% (50.2\%), respectively, in \hmssd (\hlssd)
}. Second, at a larger 
({smaller}) \rcamfix{fast storage device size}, the 
\rcamfix{performance approaches that of} \gca{the} \fast (\slow) policy, \gf{ except \rcamfix{for} \arc. \arc classifies pages \gca{as} hot or cold \gca{at} the beginning of an epoch and does not change its placement decision throughout the execution of that epoch. It does not perform any promotion or eviction of data. We observe that \arc often mispredicts the target device for a request and classifies the same number of requests for the fast and slow storage device under different fast storage sizes.

As we vary the \gonz{size} of \gonz{the} \rcamfix{fast storage device}, a dynamically adaptable data placement policy is required, which considers features from both the running workload and the underlying storage system. \gon{ We conclude that \namePaper can provide scalability by dynamically \rcamfix{and effectively} adapting \rcamfix{its policy} to the available storage size to achieve high performance. 
}} 

\vspace{-0.2cm}
\subsection{{Tri-Hybrid Storage Systems}}
\label{subsec:trihybrid}
\begin{figure}[h]
\centering
\begin{subfigure}[h]{.45\textwidth}
  \centering
  \includegraphics[width=\linewidth,trim={0.2cm 0.2cm 0cm 0.2cm},clip]{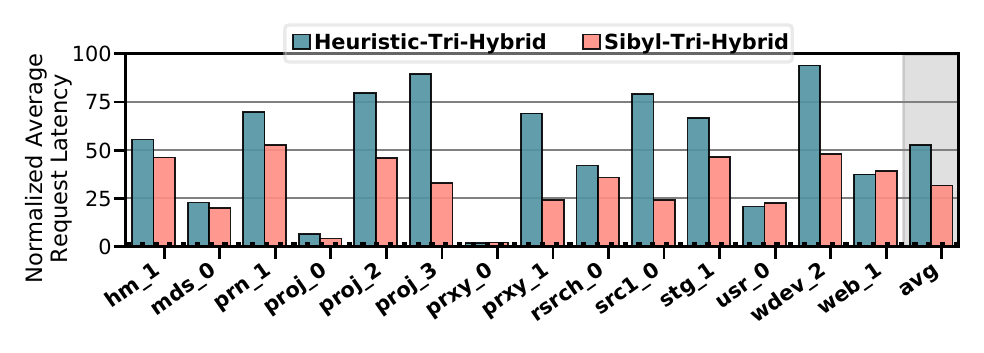}
  \vspace{-2em}
   \caption{\thold~configuration\label{fig:triybrid1}}
\end{subfigure}%

\begin{subfigure}[h]{.45\textwidth}
  \centering
  \includegraphics[width=\linewidth,trim={0.2cm 0.2cm 0.2cm 0.2cm},clip]{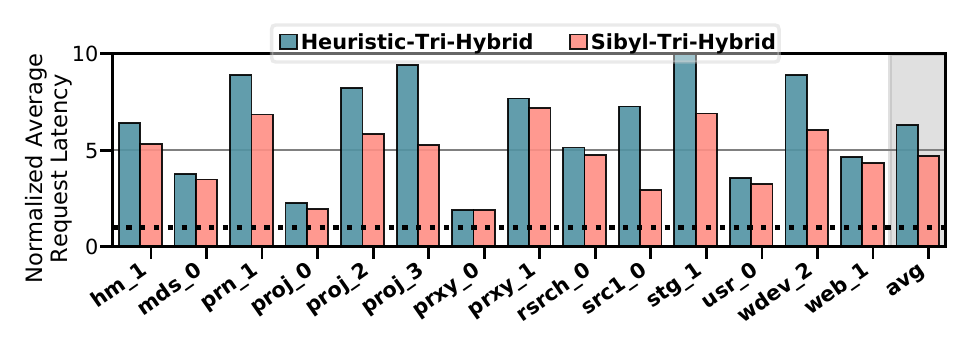}
    \vspace{-2em}
   \caption{\thnew ~configuration \label{fig:triybrid2}}
\end{subfigure}
  \caption{\gor{\gon{Average request} latency for \rcamfix{the tri-hybrid HSS} (normalized to \fast) }}
\label{fig:triybrid}
\end{figure}

We evaluate two \gca{different} tri-HSS configurations, \thold~and \thnew ~(Table \ref{tab:devices}), implemented as a single flat block device. The \thnew \ configuration has a low-end SSD (L$_{SSD}$), 
\gon{whose} performance \gon{is} lower than \rcamfix{the} \textsf{H} and \textsf{M} devices but higher than \rcamfix{the} \textsf{L} device. We restrict the available capacity of \textsf{H} and \textsf{M} 
\rcamfix{to 5\% and 10\%, respectively,}
of the working set size of \gon{a given workload}. This ensures data eviction from \textsf{H} and \textsf{M} devices once \gon{they are} full. We compare the performance of \namePaper on a tri-hybrid system with a \gon{state-of-the-art} heuristic-based policy~\cite{matsui2017design, matsui2017tri} that divides data into \emph{hot}, \emph{cold}, and \emph{frozen} and places \gon{them} \rcamfix{respectively} into \textsf{H}, \textsf{M}, and \textsf{L} devices.\footnote{\gon{\cde, \hps, \arc, and \kleio do \textit{not} consider more than two devices and \gonz{are not easily adaptable to a tri-hybrid HSS}.}} Figure~\ref{fig:triybrid} shows the performance \gon{of the} heuristic-based and \namePaper data placement policies. 

We observe that  \namePaper outperforms the heuristic-based policy by, on average, \gor{43.5\% (48.2\%) and 23.9\% (25.2\%) for \mbox{\thold} (\mbox{\thnew})}. \rcamfix{This is because Sibyl is much more dynamic and adaptive to  \gonz{the storage} system configuration \gonz{due to its RL-based \rncam{decision-making}} than the baseline heuristic-based policy, which is rigid in its decision-making.}
To extend \namePaper for three storage devices, we had to only (1) add a new action in \namePaper's action space, and (2) add the remaining capacity in the \textsf{M} device as a state feature. \gon{We conclude that \namePaper provides \gca{ease of extensibility} \rcamfix{to new storage \gonz{system} configurations, which} reduces the system architect's burden in designing sophisticated data placement mechanisms.}  

\vspace{-0.1cm}
 
\section{Explainability Analysis}
 \label{sec:explanability}
\begin{figure}[h]
  \centering
    \includegraphics[width=1\linewidth,trim={0.2cm 0.2cm 0.2cm 0.2cm},clip]{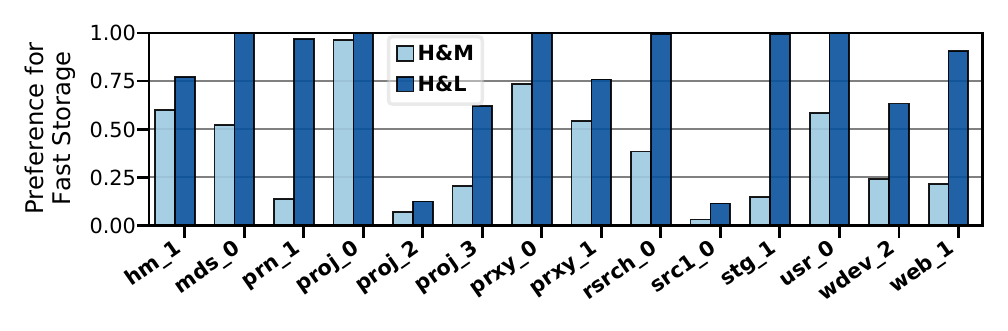}
  \caption{\namePaper's preference for the fast {storage device} under different \gon{HSS} configurations}
\label{fig:explain}
\vspace{-0.1cm}
 \end{figure}

We perform an explainability analysis to understand our results further and explain Sibyl’s decisions. 
We extract \namePaper's actions for different workloads under \hmssd and \hlssd~ \gon{HSS configurations and analyze the page placements for each workload}. 
Figure~\ref{fig:explain} shows \gon{\namePaper's preference for \rcamfix{the} fast storage \rcamfix{device} over \rcamfix{the} slow storage \rcamfix{device}, measured as the} ratio of the  number of fast storage placements to the \gont{sum of the number of placements in \rcamfix{both} fast and slow storage devices} \gon{(i.e., $\text{Preference=}\frac{\# \text{fast placements}}{\# \text{fast}+\# \text{slow placements}}$)}.

  \begin{figure*}[h]
\centering
\begin{subfigure}[h]{.47\textwidth}
  \centering
  \includegraphics[width=\linewidth]{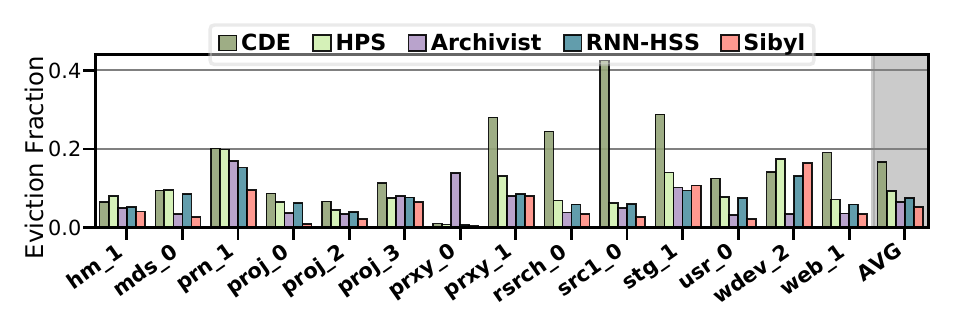}
   \vspace{-0.8cm}
  \caption{\textsf{H\&M} \gon{HSS configuration}
§  \label{fig:evict_1}}
\end{subfigure}%
\begin{subfigure}[h]{.47\textwidth}
  \centering
  \vspace{-0.2cm}
  \includegraphics[width=\linewidth,trim={0cm 0cm 0cm 0cm},clip]{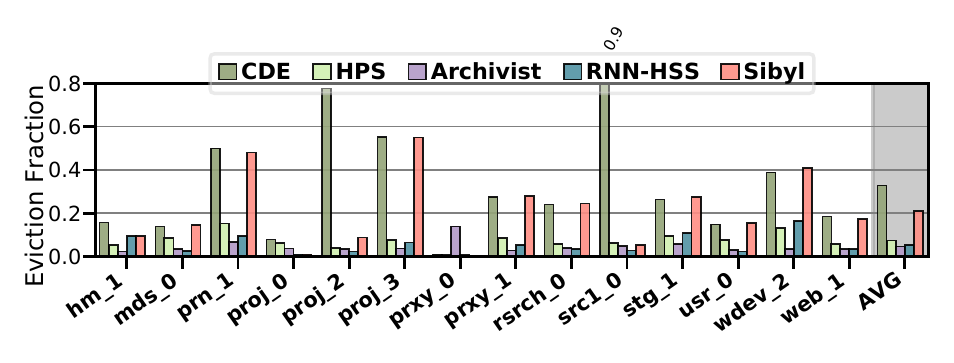}
   \vspace{-0.8cm}
   \caption{\textsf{H\&L}  \gon{HSS configuration}
  \label{fig:evict_2}}
\end{subfigure}
\caption[Two numerical solutions]{Comparison of evictions from the fast storage \gon{to the slow storage} (normalized to the total number of \gon{storage requests}) \label{fig:evict}}
\vspace{-0.7cm}
\end{figure*}

We make the following four observations. 
First, in the \js{\textsf{H\&L}} configuration, where the latency difference is large {between the two storage devices}, \namePaper~ \gon{prefers} to place \gon{more} data in the fast \gon{storage} device. \namePaper learns that despite the eviction penalty, the benefit of serving more requests from the fast \gon{storage} device is significant. On the other hand, in the \js{\textsf{H\&M}} device configuration, where the latency difference between two devices is smaller compared to \rakesh{\textsf{H\&L}}, \namePaper places only performance-critical pages in the faster \rcamfix{storage device} to avoid the eviction penalty.

Second, \rcamfix{in the} {\textsf{H\&M}} configuration, \namePaper~shows less preference to place pages from  \texttt{mds\_0}, \texttt{prn\_1}, \texttt{proj\_2}, \texttt{proj\_3}, \texttt{src1\_0}, \texttt{stg\_1}, and \texttt{web\_1}  in \juan{the} fast storage \gon{device}. These workloads \gonz{are} cold and sequential (Table~\ref{tab:workload}) \gonz{and thus} \gon{are less suitable for} the fast storage \gon{device}. Therefore, \gon{for such workloads}, \namePaper shows more preference for  \gon{the} slow storage \gon{device}. \rcamfix{In contrast,} for hot and random workloads (\texttt{prxy\_0} and \texttt{prxy\_1}), \namePaper shows \rcamfix{more} preference to place \rcamfix{pages} in the fast \gon{storage} device.


Third, for \texttt{rsrch\_0}, \texttt{wdev\_2}, and \texttt{web\_1}, \namePaper~ \gonzz{places} \gonff{$\le$40\%} \gon{of pages} in the fast storage \gon{device}. Such requests have random access patterns, while pages with cold and sequential accesses are \gonzz{placed in the} slow storage. 

\gont{Fourth},  in the \js{\textsf{H\&L}} setting, \namePaper shows more preference to \gonzz{place} requests in the fast \gon{storage} device, except for \texttt{proj\_2} and \texttt{src1\_0}  workloads. {We observe that these two workloads \gonff{are} \gon{highly random with} a low average access \gonz{count} (Table~\ref{tab:workload}). Therefore,  aggressive placement in \gonzz{the}  fast storage is not beneficial for  long-term performance.  }


\js{We also measure} the number of evictions \gonff{(as a \gonz{fraction} of all storage requests)} that occur while using \namePaper~and other baseline policies, as shown in  Figure~\ref{fig:evict}. 
We make two observations. First, \gon{in \gonff{the} \hmssd HSS configuration, \namePaper~\gonff{leads to} 68.4\%, 43.2\%, 19.7\%, and 29.3\% \gonff{fewer} evictions from the fast storage than \cde, \hps, \arc, and \kleio, respectively. }
Second, \cde places more data in the fast \gon{storage}, which leads to a \js{large} number of evictions in both \gon{HSS} configurations. However, if the latency difference between the two devices is large (e.g., \js{\textsf{H\&L}} configuration), \gon{\cde} provides higher performance than other baseline policies (see Figure~\ref{fig:perf}(b)). 
Therefore, in the \js{\textsf{H\&L} HSS configuration}, we observe \gonzz{that}~\namePaper follows a similar policy, leading to more evictions compared to other baselines. 



\section{Overhead Analysis}
\label{sec:overhead_analysis}

\subsection{\gont{Inference and Training Latencies}} 
The input layer \gon{of the} \gonz{training}  and inference network\gont{s} consists of \js{six} neurons, \gont{equal} to the number of features \js{listed} in Table~\ref{tab:state}. \gont{Each feature is} normalized to transform the \gont{value} range of different features to a common scale. { The size of one state entry is 40 bits (32 bits for state features and 8 bits for the counter used for tracking \gonzz{the remaining capacity in the} fast storage device).  We make use of two hidden layers with 20 and 30 neurons each. The final output layer has neurons equivalent to our action space, \gca{ i.e., two for \gont{dual-HSS} configurations and three for \gont{the} tri-HSS configurations}. 

\head{Inference \gonff{latency}} Our inference network has 52 \gont{inference} neurons (20+30+2)  with 780 \gon{weights} (6$\times$20+20$\times$30+30$\times$2). 
{As a result, {\namePaper} requires 780 MAC operations per inference (1$\times$6$\times$20+1$\times$20$\times$30 +1$\times$30$\times$2). On our evaluated \gont{CPU}, we can perform these operations in $\sim$10ns, which is several orders of magnitude smaller than the I/O read latency of even a high-end SSD ($\sim$10us)~\cite{inteloptane, samsung2017znand}.} 
\gca{\gont{\namePaper's inference} computation can also be performed in the SSD controller.}

\gor{{\noindent{\head{Training \gonff{latency}}}  For each training step, \mbox{\namePaper} needs to compute 1,597,440 MAC operations, where each batch requires 128$\times$6$\times$20+ 128$\times$20$\times$30+128$\times$30$\times$2 MAC operations. 
This computation takes $\sim$2us on our evaluated \gont{CPU}. 
\gonff{This} training latency does not affect the \gonff{benefits} of \mbox{\namePaper}
\rcamfix{because (1)}
training occurs asynchronously \rcamfix{with} inference, \gonzz{and}
\rcamfix{(2)} training latency is \gonff{$\sim$5$\times$} smaller than the I/O read latency of even a high-end SSD.  }}

\noindent\gonff{We conclude that Sibyl's performance benefits come at small latency overheads that are easily realizable in existing CPUs.}


\vspace{-0.35cm}
\subsection{Area Overhead} 

\noindent{\head{Storage cost}} We use a half-precision floating-point format for the weights \gon{of the \gonz{training} and the inference network\rcamfix{s}}. With 780 16-bit weights, \rcamfix{each} neural network requires 12.2 KiB of memory. 
Since we use 
\rcamfix{the same} network architecture for the \gonz{two} networks, we need 24.4 KiB of memory. In total, with an experience buffer of 100 KiB (\cref{subsec:detail_design}), \namePaper require\rcamfix{s} \gonzz{124.4} KiB of \gont{DRAM} overhead, \js{which is negligible compared to {the} memory size of modern computing systems}.  } 

\noindent{\head{Metadata cost}} HSSs need to maintain the address mapping information for the underlying storage devices~\cite{tsukada2021metadata}.  {\namePaper} \gonff{requires} 40 bits to store  \gont{state information} (\gonff{i.e., the per-page state features;} see Table~\ref{tab:state}). This overhead is $\sim$0.1\% of the total storage capacity when using a 4-KiB data placement granularity (5-byte per 4-KiB data). 

\noindent\gon{We conclude that \namePaper~\gon{has \gonzz{a} very} modest cost in \gon{terms of} \gon{storage capacity} overhead \gon{in \gonz{main memory (DRAM)}}.}

\vspace{0.9cm}
\section{Discussion}
\label{sec:discussion}

\gsa{\head{Cost of generality} 
\gon{We identify two main limitations of using RL for data placement. First, currently, RL is largely a \emph{black-box policy}.  Our explainability analysis (\cref{sec:explanability}) tries to provide intuition into Sibyl’s internal mechanism. However, providing rigorous explainability to reinforcement learning-based mechanisms is an active field of research~\cite{verma2018programmatically,liu2018toward,juozapaitis2019explainable,madumal2020explainable,sequeira2020interestingness,puiutta2020explainable}, a problem that is beyond the scope of this paper. Perfectly finding worst-case workloads against an RL policy is, therefore, very difficult, in fact, impossible, given the state-of-the-art in reinforcement learning. There are many dynamic decisions that the agent performs, which cannot be easily explained or modeled in human-understandable terms.}   
Second, \gonff{\namePaper} requires \emph{engineering effort} to (1) \gonff{thoroughly} tune the RL hyper-parameters, and (2) implement and integrate \namePaper components into the host OS's storage management layer. 
This second limitation is not specific to Sibyl and applies to any ML-based storage management technique. \gonff{As quantified in \cref{sec:overhead_analysis}, \namePaper's storage and latency overheads are small. }   }

\head{\ho{\namePaper's implications}} \gor{\gont{\namePaper} (1) provides performance improvements on a wide variety of workloads and system configurations (our evaluations \gonff{in} \cref{sec:results} show that \mbox{\namePaper} outperforms all evaluated state-of-the-art data placement policies \gonff{under all system configurations}),  (2) provides extensibility by \rcamfix{reducing} the designer burden when extending data placement policies to multiple devices \gonff{and different storage configurations}, and (3) \gonff{enables reducing } the fast storage \gonx{device} \gonff{size} by taking \gonff{better} advantage of \rcamfix{the} fast-yet-small storage device and large-yet-slow storage device to deliver high storage capacity at low latency.}

\head{Adding more features and optimization objectives}
\gsa{An RL-based approach simplifies adding new features (such as bandwidth utilization) \gont{in} \gonzz{the} RL state and optimization objectives (such as endurance) using \gonzz{the} RL reward function. This flexibility allows an RL-based mechanism to self-optimize and adapt its decision-making policy to achieve an objective without \gont{the designer} explicitly defining \emph{how} to achieve it. We demonstrate and evaluate example implementations of Sibyl using a reward scheme that is a function of  request latency and eviction latency. We \gont{find} that \gonx{request} latency in the reward structure best encapsulates system conditions since latency could vary for each \gon{storage} request \gont{based on complex system conditions}. To optimize for a different device-level objective, one needs to define a new reward function with appropriate state features, e.g., to optimize for endurance, one might use the number of writes to an endurance-critical device in the reward function. Another interesting research direction would be to perform multi-objective optimization, \gont{e.g., optimizing for \gonx{both}} performance and energy. We leave \gont{the study of different objectives and features to} future \gont{work}.
}

\head{Necessity of the reward} \gsa{RL training is highly dependent upon the quality of the reward function and state features. Using an incorrect reward or improper state features could lead to severe performance degradation. }Creating the right reward is a human-driven effort that could benefit from design insights. We tried two other reward structures to achieve our objective to \grakesh{\rcamfix{improve}  \gonff{system} performance}:
\begin{itemize}[leftmargin=*, noitemsep, topsep=0pt]
\item \textbf{Hit rate of the fast storage device:} Maximizing the hit rate of \ghpca{the} fast \gon{storage} device is another \gonff{potentially plausible objective}. However,  
\rcamfix{if we use} \gonff{the} hit rate as \gonzz{a} reward, \namePaper (1) tries to aggressively place data in \gonzz{the} fast storage \rcamfix{device}, \gon{which leads to unnecessary evictions}, \gonzz{and} (2)  cannot capture the asymmetry in the latencies  present in \gonx{modern} storage devices \gonff{(e.g., due to  \mbox{read/write} latencies, latency of garbage collection, queuing delays, error handling latencies, \gonzz{and} write buffer state)}.

\item \textbf{High negative reward for eviction:} We also tried a negative reward for eviction and \gonz{a} zero \rcamfix{reward} in other cases. We observe that such a reward structure \gon{provides suboptimal performance because \gonff{\namePaper places more pages in the slow device to avoid evictions. Thus, \gonx{with such a reward structure,} \namePaper is not able to effectively utilize the fast storage.}} 

\end{itemize}
  \noindent We conclude that our chosen  reward structure works well for a wide variety of workloads \gonff{\cref{sec:results}}, as \gonff{\gonx{reinforced} by} our generality studies using unseen workloads in \mbox{\cref{subsection:perf_unseen}}.   

\head{Managing hybrid main memory using RL} \gca{The key idea of \namePaper can be adapted for managing hybrid main memory architectures. However, managing data placement at different levels of the memory hierarchy
has its own set of challenges~\cite{li2017utility_HMM,agarwal2015page_HMM,agarwal2017thermostat_HMM,goglin2016exposing,ham2013disintegrated,lin2016memif,malladi2016dramscale,pavlovic2013data,pena2014toward,qureshi2009scalable,yoon2012row,meza2013case, meza2012enabling, ren2015thynvm} that \namePaper would need to adapt \rcamfix{to}, such as the low latency decision-making \gonff{and control} requirements in  main memory. 
Even with the \gonx{use} of hybrid main memories, many systems \rcamfix{continue to} benefit from using hybrid storage devices \gonff{due to much lower cost\gonx{-}per\gonx{-}bit of storage, which accommodates increasingly larger datasets.} 
 Therefore,  we focus on hybrid storage systems and leave it \rcamfix{to} future \rcamfix{work} to \rcamfix{\gonff{study} RL to manage} hybrid main memories. 
}

\vspace{-0.1cm}
\section{Related Work}

\gagan{To our knowledge, this is the first work to propose \rcamfix{a reinforcement learning}-based data placement technique for hybrid storage systems. \rcamfix{\namePaper can} continuously learn from and adapt to the \rcamfix{running} application and \rcamfix{the} storage \rcamfix{configuration and} device characteristics. }\gca{We \rcamfix{briefly} discuss closely-related prior works that propose data management techniques for hybrid memory/storage systems and RL-based system optimizations.}

\head{Heuristic-based data placement}
Many prior works~\cite{matsui2017design,sun2013high,heuristics_hyrbid_hystor_sc_2011,vasilakis2020hybrid2,lv2013probabilistic,li2014greendm,guerra2011cost,elnably2012efficient,heuristics_usenix_2014,li2017utility_HMM,agarwal2015page_HMM,agarwal2017thermostat_HMM,ham2013disintegrated_HMM, bu2012optimization, krish2016efficient, lin2011hot, tai2015sla, zhang2010automated, wu2012data, iliadis2015exaplan, lv2013hotness, matsui2017tri, feng2014hdstore,salkhordeh2015operating, pavlovic2013data, meza2013case, yang2017autotiering, chiachenchou2015batman, kim2011hybridstore, wang2019panthera, ramos2011page, liu2019hierarchical,luo2020optimal,doudali2021cori} propose heuristic-based techniques to perform data placement. 
These techniques rely on statically-chosen design features that \rcamfix{usually} favor certain workload\gonzz{s} and/or device characteristics, \rcamfix{leading to \gonx{relatively} rigid policies}.
In \cref{sec:motivation_limitations} and \cref{sec:results}, we  show \gca{that} \namePaper outperforms two \rcamfix{state-of-the-art} works, CDE~\cite{matsui2017design} and HPS~\cite{meswani2015heterogeneous}.

\head{ML-based data placement}
\rcamfix{Several works~\cite{doudali2019kleio,ren2019archivist,cheng2019optimizing,shetti2019machine,sen2019machine} propose ML-based techniques for data placement in hybrid memory/storage systems. These  works 1) are based on supervised learning techniques that require frequent \rcamfix{and very costly} retraining to adapt to changing workload and device characteristics, and 2) have not been evaluated on a real system.} \gonn{\rcamfix{We evaluate} \kleio, which is  inspired by the state-of-the-art data placement technique in hybrid main memory~\cite{doudali2019kleio}. It uses sophisticated recurrent neural networks (RNNs) for data placement \rcamfix{and} shows promising results \rcamfix{compared to heuristic-based techniques}. However, it has \gonx{two major} limitations \gonx{that make it impractical or difficult to implement}: \juangg{it} (1) trains  an RNN for each page, which \rcamfix{leads to large} computation\gonx{, storage, and training time \gonzz{overheads}}, and (2) requires offline application profiling.}
\gagan{Our evaluation (ref. \cref{subsection:Evaluation_perf}) shows that \namePaper outperforms \rcamfix{two state-of-the-art ML-based data placement techniques}, \kleio~\cite{doudali2019kleio} and \arc~\cite{ren2019archivist}, \rcamfix{across a wide variety of workloads.}}
\head{{\rcamfix{RL-based techniques in storage systems}}}
Recent works \gon{(e.g., \cite{liu2019learning,yoo2020reinforcement,wang2020reinforcement,rl_GC_TECS, kang2018dynamic})} propose the use of RL-based approaches for managing different aspects of storage systems. These works cater to use cases and objectives that are \gonx{very} different \rcamfix{from} \namePaper's.
Specifically, {Liu}~\etal~\cite{liu2019learning} (1) \rcamfix{propose} data placement in cloud systems and \emph{not} \gf{hybrid storage \rcamfix{systems}, (2) consider devices with unlimited capacity, sidestepping the capacity limitations, (3) \emph{emulate} a data center network rather than \gon{use} a real system \gon{for design and evaluation}, and (4) focus only on data-analytics workloads. }
{Yoo}~\etal~\cite{yoo2020reinforcement} do \emph{not} focus on data placement; they instead deal with dynamic storage resizing 
based on workload characteristics using a trace-based simulator.  Wang~\etal~\cite{wang2020reinforcement} (1) focus on cloud \gon{systems} to predict the data storage consumption, and 
(2) do \emph{not} consider hybrid storage systems. 
\namePaper is the first RL-based mechanism \gon{for data placement in hybrid storage systems.} 
 

\head{RL-based system optimizations}
Past works~\ghpca{\cite{lin2020deepNOC,ipek2008self,liu2020imitation,rl_NOC_AIDArc_2018,rl_voltage_scaling_TC_2018,mirhoseini2021chip,mutlu2021intelligent_DATE, pythia, peled2015semantic, martinez2009dynamic, multi_scheduler_HPCA_2012, jain2016machine,zheng2020agile, pd2015q}} propose RL-based methods for various system optimizations, such as memory scheduling~\cite{ipek2008self, multi_scheduler_HPCA_2012}, data prefetching~\cite{pythia,peled2015semantic}, cache replacement~\cite{liu2020imitation}, and network-on-chip arbitration~\cite{rl_NOC_AIDArc_2018,lin2020deepNOC}. \gca{Along with \namePaper\rcamfix{, designed} for efficient data placement in hybrid storage systems, \rcamfix{this body of work} demonstrate\rcamfix{s} that RL is a promising approach to designing high-performance\rcamfix{, and highly-adaptive} self-optimizing computing systems. }


\vspace{-0.1cm}
\section{Conclusion}

We introduce \namePaper, the first reinforcement learning-based mechanism for  data placement in hybrid storage systems.
Our extensive \gagan{real-system} {evaluation} demonstrate\rcamfix{s} that \namePaper~\gon{provides adaptivity and extensibility by} continuously learning from and autonomously adapting to the workload characteristics\rcamfix{, storage configuration and device characteristics,} and system-level feedback to maximize the overall long-term performance of a hybrid storage system. {We interpret \gca{\namePaper}'s policy through our explainability analysis} and {conclude that \namePaper provides an \rcamfix{effective} and robust approach \gonx{to} data placement in current and future hybrid storage systems.} 
\rcamfix{We hope that \namePaper and our open-sourced implementation of it~\cite{sibylLink} inspire future work and ideas in self-optimizing storage and memory systems.}

\begin{acks}
We thank anonymous reviewers \gonn{of ISCA 2022, HPCA 2022, and MICRO 2022} for their feedback and comments.
We thank the SAFARI Research Group members for valuable feedback and the stimulating intellectual environment they provide. We acknowledge the generous gifts of our industrial partners, especially  Google, Huawei, Intel, Microsoft, VMware. \gon{This research was partially supported by the Semiconductor Research Corporation and \gonn{the ETH Future Computing Laboratory}.}

\end{acks}


{
\balance
\bibliographystyle{IEEEtran}
\bibliography{ms}

\begin{thebibliography}{100}
\providecommand{\url}[1]{#1}
\csname url@samestyle\endcsname
\providecommand{\newblock}{\relax}
\providecommand{\bibinfo}[2]{#2}
\providecommand{\BIBentrySTDinterwordspacing}{\spaceskip=0pt\relax}
\providecommand{\BIBentryALTinterwordstretchfactor}{4}
\providecommand{\BIBentryALTinterwordspacing}{\spaceskip=\fontdimen2\font plus
\BIBentryALTinterwordstretchfactor\fontdimen3\font minus
  \fontdimen4\font\relax}
\providecommand{\BIBforeignlanguage}[2]{{%
\expandafter\ifx\csname l@#1\endcsname\relax
\typeout{** WARNING: IEEEtran.bst: No hyphenation pattern has been}%
\typeout{** loaded for the language `#1'. Using the pattern for}%
\typeout{** the default language instead.}%
\else
\language=\csname l@#1\endcsname
\fi
#2}}
\providecommand{\BIBdecl}{\relax}
\BIBdecl

\bibitem{meza2013case}
J.~Meza, Y.~Luo, S.~Khan, J.~Zhao, Y.~Xie, and O.~Mutlu, ``{A Case for
  Efficient Hardware/Software Cooperative Management of Storage and Memory},''
  in \emph{WEED}, 2013.

\bibitem{bailey2013exploring}
K.~A. Bailey, P.~Hornyack, L.~Ceze, S.~D. Gribble, and H.~M. Levy, ``{Exploring
  Storage Class Memory with Key Value Stores},'' in \emph{SOSP}, 2013.

\bibitem{smullen2010accelerating}
C.~W. Smullen, J.~Coffman, and S.~Gurumurthi, ``{Accelerating Enterprise
  Solid-State Disks With Non-Volatile Merge Caching},'' in \emph{IGSC}, 2010.

\bibitem{lu2012pram}
N.~Lu, I.-S. Choi, S.-H. Ko, and S.-D. Kim, ``{A PRAM Based Block Updating
  Management for Hybrid Solid State Disk},'' in \emph{ELEX}, 2012.

\bibitem{tarihi2015hybrid}
M.~Tarihi, H.~Asadi, A.~Haghdoost, M.~Arjomand, and H.~Sarbazi-Azad, ``{A
  Hybrid Non-Volatile Cache Design for Solid-State Drives Using Comprehensive
  I/O Characterization},'' in \emph{TC}, 2015.

\bibitem{xiao2016hs}
W.~Xiao, H.~Dong, L.~Ma, Z.~Liu, and Q.~Zhang, ``{HS-BAS: A Hybrid Storage
  System Based on Band Awareness of Shingled Write Disk},'' in \emph{ICCD},
  2016.

\bibitem{wang2017larger}
C.~Wang, D.~Wang, Y.~Chai, C.~Wang, and D.~Sun, ``{Larger, Cheaper, but Faster:
  SSD-SMR Hybrid Storage Boosted by a New SMR-Oriented Cache Framework},'' in
  \emph{MSST}, 2017.

\bibitem{lu2016design}
Z.-W. Lu and G.~Zhou, ``{Design and Implementation of Hybrid Shingled Recording
  RAID System},'' in \emph{PiCom}, 2016.

\bibitem{luo2015design}
D.~Luo, J.~Wan, Y.~Zhu, N.~Zhao, F.~Li, and C.~Xie, ``{Design and
  Implementation of a Hybrid Shingled Write Disk System },'' in \emph{TPDS},
  2015.

\bibitem{srinivasan2010flashcache}
M.~Srinivasan, P.~Saab, and V.~Tkachenko, ``Flashcache,'' in \emph{Facebook},
  2010.

\bibitem{reinsel2013breaking}
D.~Reinsel and J.~Rydning, ``{Breaking the 15K-rpm HDD Performance Barrier with
  Solid State Hybrid Drives},'' in \emph{IDC}, 2013.

\bibitem{lee2014mining}
S.~Lee, Y.~Won, and S.~Hong, ``{Mining-Based File Caching in a Hybrid Storage
  System},'' in \emph{JISE}, 2014.

\bibitem{felter2011reliability}
W.~Felter, A.~Hylick, and J.~Carter, ``{Reliability-Aware Energy Management for
  Hybrid Storage Systems},'' in \emph{MSST}, 2011.

\bibitem{bu2012optimization}
K.~Bu, M.~Wang, H.~Nie, W.~Huang, and B.~Li, ``{The Optimization of the
  Hierarchical Storage System Based on the Hybrid SSD Technology},'' in
  \emph{ISDEA}, 2012.

\bibitem{canim2010ssd}
M.~Canim, G.~A. Mihaila, B.~Bhattacharjee, K.~A. Ross, and C.~A. Lang, ``{SSD
  Bufferpool Extensions for Database Systems},'' in \emph{VLDB}, 2010.

\bibitem{bisson2007reducing}
T.~Bisson and S.~A. Brandt, ``{Reducing Hybrid Disk Write Latency with
  Flash-Backed I/O Requests},'' in \emph{MASCOTS}, 2007.

\bibitem{saxena2012flashtier}
M.~Saxena, M.~M. Swift, and Y.~Zhang, ``{FlashTier: A Lightweight, Consistent
  and Durable Storage Cache},'' in \emph{EuroSys}, 2012.

\bibitem{krish2016efficient}
K.~Krish, B.~Wadhwa, M.~S. Iqbal, M.~M. Rafique, and A.~R. Butt, ``{On
  Efficient Hierarchical Storage for Big Data Processing},'' in \emph{CCGrid},
  2016.

\bibitem{zhao2016towards}
D.~Zhao, K.~Qiao, and I.~Raicu, ``{Towards Cost-Effective and High-Performance
  Caching Middleware for Distributed Systems},'' in \emph{IJBDI}, 2016.

\bibitem{lin2011hot}
L.~Lin, Y.~Zhu, J.~Yue, Z.~Cai, and B.~Segee, ``{Hot Random Off-Loading: A
  Hybrid Storage System with Dynamic Data Migration},'' in \emph{MASCOTS},
  2011.

\bibitem{chen2015duplication}
X.~Chen, W.~Chen, Z.~Lu, P.~Long, S.~Yang, and Z.~Wang, ``{A Duplication-Aware
  SSD-Based Cache Architecture for Primary Storage in Virtualization
  Environment},'' in \emph{ISJ}, 2015.

\bibitem{niu2018hybrid}
J.~Niu, J.~Xu, and L.~Xie, ``{Hybrid Storage Systems: A Survey of Architectures
  and Algorithms},'' in \emph{IEEE Access}, 2018.

\bibitem{oh2015enabling}
Y.~Oh, E.~Lee, C.~Hyun, J.~Choi, D.~Lee, and S.~H. Noh, ``{Enabling
  Cost-Effective Flash Based Caching with an Array of Commodity SSDs},'' in
  \emph{Middleware}, 2015.

\bibitem{liu2013molar}
Y.~Liu, X.~Ge, X.~Huang, and D.~H. Du, ``{MOLAR: A Cost-Efficient,
  High-Performance SSD-Based Hybrid Storage Cache},'' in \emph{CLUSTER}, 2013.

\bibitem{tai2015sla}
J.~Tai, B.~Sheng, Y.~Yao, and N.~Mi, ``{SLA-Aware Data Migration in a Shared
  Hybrid Storage Cluster},'' in \emph{CC}, 2015.

\bibitem{huang2016improving}
S.~Huang, Q.~Wei, D.~Feng, J.~Chen, and C.~Chen, ``{Improving Flash-Based Disk
  Cache with Lazy Adaptive Replacement},'' in \emph{TOS}, 2016.

\bibitem{kgil2006flashcache}
T.~Kgil and T.~Mudge, ``{FlashCache: A NAND Flash Memory File Cache for Low
  Power Web Servers},'' in \emph{CASES}, 2006.

\bibitem{kgil2008improving}
T.~Kgil, D.~Roberts, and T.~Mudge, ``{Improving NAND Flash Based Disk
  Caches},'' in \emph{ISCA}, 2008.

\bibitem{oh2012caching}
Y.~Oh, J.~Choi, D.~Lee, and S.~H. Noh, ``{Caching Less For Better Performance:
  Balancing Cache Size and Update Cost of Flash Memory Cache in Hybrid Storage
  Systems},'' in \emph{FAST}, 2012.

\bibitem{yang2013hec}
J.~Yang, N.~Plasson, G.~Gillis, N.~Talagala, S.~Sundararaman, and R.~Wood,
  ``{HEC: Improving Endurance of High Performance Flash-Based Cache Devices},''
  in \emph{SYSTOR}, 2013.

\bibitem{ou2014edm}
J.~Ou, J.~Shu, Y.~Lu, L.~Yi, and W.~Wang, ``{EDM: An Endurance-Aware Data
  Migration Scheme for Load Balancing in SSD Storage Clusters},'' in
  \emph{IPDPS}, 2014.

\bibitem{appuswamy2013cache}
R.~Appuswamy, D.~C. van Moolenbroek, and A.~S. Tanenbaum, ``{Cache, Cache
  Everywhere, Flushing All Hits Down the Sink: On Exclusivity in Multilevel,
  Hybrid Caches},'' in \emph{MSST}, 2013.

\bibitem{cheng2015amc}
Y.~Cheng, W.~Chen, Z.~Wang, X.~Yu, and Y.~Xiang, ``{AMC: An Adaptive
  Multi-Level Cache Algorithm in Hybrid Storage Systems},'' in \emph{CCPE},
  2015.

\bibitem{chai2015wec}
Y.~Chai, Z.~Du, X.~Qin, and D.~A. Bader, ``{WEC: Improving Durability of SSD
  Cache Drives by Caching Write-Efficient Data},'' in \emph{TC}, 2015.

\bibitem{dai2015etd}
N.~Dai, Y.~Chai, Y.~Liang, and C.~Wang, ``{ETD-Cache: An Expiration-Time Driven
  Cache Scheme to Make SSD-Based Read Cache Endurable and Cost-Efficient},'' in
  \emph{CF}, 2015.

\bibitem{ye2015regional}
F.~Ye, J.~Chen, X.~Fang, J.~Li, and D.~Feng, ``{A Regional Popularity-Aware
  Cache Replacement Algorithm to Improve the Performance and Lifetime of
  SSD-Based Disk Cache},'' in \emph{NAS}, 2015.

\bibitem{chang2015profit}
H.-P. Chang, S.-Y. Liao, D.-W. Chang, and G.-W. Chen, ``{Profit Data Caching
  and Hybrid Disk-Aware Completely Fair Queuing Scheduling Algorithms For
  Hybrid Disks},'' in \emph{SPE}, 2015.

\bibitem{saxena2014design}
M.~Saxena and M.~M. Swift, ``{Design and Prototype of a Solid-State Cache},''
  in \emph{TOS}, 2014.

\bibitem{li2014enabling}
Y.~Li, L.~Guo, A.~Supratak, and Y.~Guo, ``{Enabling Performance as a Service
  For a Cloud Storage System},'' in \emph{CLOUD}, 2014.

\bibitem{zong2014faststor}
Z.~Zong, R.~Fares, B.~Romoser, and J.~Wood, ``{FastStor: Data-Mining-Based
  Multilayer Prefetching for Hybrid Storage Systems},'' in \emph{CC}, 2014.

\bibitem{do2011turbocharging}
J.~Do, D.~Zhang, J.~M. Patel, D.~J. DeWitt, J.~F. Naughton, and A.~Halverson,
  ``{Turbocharging DBMS Buffer Pool Using SSDs},'' in \emph{SIGMOD}, 2011.

\bibitem{lee2015effective}
D.~Lee, C.~Min, and Y.~I. Eom, ``{Effective SSD Caching For High-Performance
  Home Cloud Server},'' in \emph{ICCE}, 2015.

\bibitem{baek2016fully}
S.~H. Baek and K.-W. Park, ``{A Fully Persistent and Consistent Read/Write
  Cache Using Flash-Based General SSDs for Desktop Workloads},'' in
  \emph{ICEIS}, 2016.

\bibitem{liu2010raf}
Y.~Liu, J.~Huang, C.~Xie, and Q.~Cao, ``{RAF: A Random Access First Cache
  Management to Improve SSD-Based Disk Cache},'' in \emph{NAS}, 2010.

\bibitem{liang2016elastic}
Y.~Liang, Y.~Chai, N.~Bao, H.~Chen, and Y.~Liu, ``{Elastic Queue: A Universal
  SSD Lifetime Extension Plug-in for Cache Replacement Algorithms},'' in
  \emph{SYSTOR}, 2016.

\bibitem{yadgar2011management}
G.~Yadgar, M.~Factor, K.~Li, and A.~Schuster, ``{Management of Multilevel,
  Multiclient Cache Hierarchies with Application Hints},'' in \emph{TOCS},
  2011.

\bibitem{zhang2012multi}
Z.~Zhang, Y.~Kim, X.~Ma, G.~Shipman, and Y.~Zhou, ``{Multi-level Hybrid Cache:
  Impact and Feasibility},'' in \emph{ORNL Tech. Rep}, 2012.

\bibitem{klonatos2011azor}
Y.~Klonatos, T.~Makatos, M.~Marazakis, M.~D. Flouris, and A.~Bilas, ``{Azor:
  Using Two-Level Block Selection to Improve SSD-Based I/O Caches},'' in
  \emph{NAS}, 2011.

\bibitem{matsui2017design}
C.~Matsui, C.~Sun, and K.~Takeuchi, ``{Design of Hybrid SSDs With Storage Class
  Memory and NAND Flash Memory},'' in \emph{Proc. IEEE}, 2017.

\bibitem{sun2013high}
C.~Sun, K.~Miyaji, K.~Johguchi, and K.~Takeuchi, ``{A High Performance and
  Energy-Efficient Cold Data Eviction Algorithm for 3D-TSV Hybrid ReRAM/MLC
  NAND SSD},'' in \emph{CAS}, 2013.

\bibitem{heuristics_hyrbid_hystor_sc_2011}
F.~Chen, D.~A. Koufaty, and X.~Zhang, ``{Hystor: Making the Best Use of Solid
  State Drives in High Performance Storage Systems},'' in \emph{SC}, 2011.

\bibitem{vasilakis2020hybrid2}
E.~Vasilakis, V.~Papaefstathiou, P.~Trancoso, and I.~Sourdis, ``{Hybrid2:
  Combining Caching and Migration in Hybrid Memory Systems},'' in \emph{HPCA},
  2020.

\bibitem{lv2013probabilistic}
Y.~Lv, X.~Chen, G.~Sun, and B.~Cui, ``{A Probabilistic Data Replacement
  Strategy for Flash-Based Hybrid Storage System},'' in \emph{APWeb}, 2013.

\bibitem{li2014greendm}
Z.~Li, ``{GreenDM: A Versatile Tiering Hybrid Drive for the Trade-Off
  Evaluation of Performance, Energy, and Endurance},'' Ph.D. dissertation,
  Stony Brook University, NY, 2014.

\bibitem{guerra2011cost}
J.~Guerra, H.~Pucha, J.~S. Glider, W.~Belluomini, and R.~Rangaswami, ``{Cost
  Effective Storage Using Extent Based Dynamic Tiering},'' in \emph{FAST},
  2011.

\bibitem{elnably2012efficient}
A.~Elnably, H.~Wang, A.~Gulati, and P.~J. Varman, ``{Efficient QoS for
  Multi-Tiered Storage Systems},'' in \emph{HotStorage}, 2012.

\bibitem{heuristics_usenix_2014}
H.~Wang and P.~Varman, ``{Balancing Fairness and Efficiency in Tiered Storage
  Systems with Bottleneck-Aware Allocation},'' in \emph{FAST}, 2014.

\bibitem{doudali2019kleio}
T.~D. Doudali, S.~Blagodurov, A.~Vishnu, S.~Gurumurthi, and A.~Gavrilovska,
  ``{Kleio: A Hybrid Memory Page Scheduler with Machine Intelligence},'' in
  \emph{HPDC}, 2019.

\bibitem{ren2019archivist}
J.~Ren, X.~Chen, Y.~Tan, D.~Liu, M.~Duan, L.~Liang, and L.~Qiao, ``{Archivist:
  A Machine Learning Assisted Data Placement Mechanism for Hybrid Storage
  Systems},'' in \emph{ICCD}, 2019.

\bibitem{cheng2019optimizing}
P.~Cheng, Y.~Lu, Y.~Du, Z.~Chen, and Y.~Liu, ``{Optimizing Data Placement on
  Hierarchical Storage Architecture via Machine Learning},'' in \emph{NPC},
  2019.

\bibitem{raghavan2014tiera}
A.~Raghavan, A.~Chandra, and J.~B. Weissman, ``{Tiera: Towards Flexible
  Multi-Tiered Cloud Storage Instances},'' in \emph{Middleware}, 2014.

\bibitem{salkhordeh2015operating}
R.~Salkhordeh, H.~Asadi, and S.~Ebrahimi, ``{Operating System Level Data
  Tiering Using Online Workload Characterization},'' in \emph{JSC}, 2015.

\bibitem{hui2012hash}
J.~Hui, X.~Ge, X.~Huang, Y.~Liu, and Q.~Ran, ``{E-HASH: An Energy-Efficient
  Hybrid Storage System Composed of One SSD and Multiple HDDs},'' in
  \emph{ICSI}, 2012.

\bibitem{xue2014storage}
J.~Xue, F.~Yan, A.~Riska, and E.~Smirni, ``{Storage Workload Isolation via Tier
  Warming},'' in \emph{ICAC}, 2014.

\bibitem{zhang2010automated}
G.~Zhang, L.~Chiu, C.~Dickey, L.~Liu, P.~Muench, and S.~Seshadri, ``{Automated
  Lookahead Data Migration in SSD-enabled Multi-tiered Storage Systems},'' in
  \emph{MSST}, 2010.

\bibitem{zhao2010fdtm}
X.~Zhao, Z.~Li, and L.~Zeng, ``{FDTM: Block Level Data Migration Policy in
  Tiered Storage System},'' in \emph{NPC}, 2010.

\bibitem{shi2013optimal}
H.~Shi, R.~V. Arumugam, C.~H. Foh, and K.~K. Khaing, ``{Optimal Disk Storage
  Allocation for Multitier Storage System},'' in \emph{TMAG}, 2013.

\bibitem{wu2012data}
X.~Wu and A.~N. Reddy, ``{Data Organization in a Hybrid Storage System},'' in
  \emph{ICNC}, 2012.

\bibitem{ma2014providing}
S.~Ma, H.~Chen, Y.~Shen, H.~Lu, B.~Wei, and P.~He, ``{Providing Hybrid Block
  Storage for Virtual Machines using Object-based Storage},'' in \emph{ICPADS},
  2014.

\bibitem{iliadis2015exaplan}
I.~Iliadis, J.~Jelitto, Y.~Kim, S.~Sarafijanovic, and V.~Venkatesan,
  ``{ExaPlan: Queueing-Based Data Placement and Provisioning for Large Tiered
  Storage Systems},'' in \emph{MASCOTS}, 2015.

\bibitem{wu2009managing}
X.~Wu and A.~N. Reddy, ``{Managing Storage Space in a Flash and Disk Hybrid
  Storage System},'' in \emph{MASCOTS}, 2009.

\bibitem{wu2010exploiting}
X.~Wu and A.~N. Reddy, ``{Exploiting Concurrency to Improve Latency and
  throughput in a Hybrid Storage System},'' in \emph{MASCOTS}, 2010.

\bibitem{park2011hot}
D.~Park and D.~H. Du, ``{Hot Data Identification for Flash-based Storage
  Systems Using Multiple Bloom Filters},'' in \emph{MSST}, 2011.

\bibitem{lv2013hotness}
Y.~Lv, B.~Cui, X.~Chen, and J.~Li, ``{Hotness-Aware Buffer Management For
  Flash-Based Hybrid Storage Systems},'' in \emph{CIKM}, 2013.

\bibitem{montgomery2014extent}
D.~Montgomery, ``{Extent Migration For Tiered Storage Architecture},'' in
  \emph{USPTO}, 2014.

\bibitem{matsui2017tri}
C.~Matsui, T.~Yamada, Y.~Sugiyama, Y.~Yamaga, and K.~Takeuchi, ``{Tri-Hybrid
  SSD with Storage Class Memory (SCM) and MLC/TLC NAND Flash Memories},''
  \emph{Proc. IEEE}, 2017.

\bibitem{wiki:Sibyl}
Wikipedia, ``{Sibyl}, \url{https://en.wikipedia.org/wiki/Sibyl}.''

\bibitem{sutton_2018}
R.~S. Sutton and A.~G. Barto, \emph{{Reinforcement Learning: An Introduction}},
  2018.

\bibitem{minsky1961steps}
M.~Minsky, ``{Steps Toward Artificial Intelligence},'' in \emph{Proc. IRE},
  1961.

\bibitem{cai2015data}
Y.~Cai, Y.~Luo, E.~F. Haratsch, K.~Mai, and O.~Mutlu, ``Data {R}etention in
  {MLC} {NAND} {F}lash {M}emory: {C}haracterization, {O}ptimization, and
  {R}ecovery,'' in \emph{HPCA}, 2015.

\bibitem{grupp2009characterizing}
L.~M. Grupp, A.~M. Caulfield, J.~Coburn, S.~Swanson, E.~Yaakobi, P.~H. Siegel,
  and J.~K. Wolf, ``{Characterizing Flash Memory: Anomalies, Observations, and
  Applications},'' in \emph{MICRO}, 2009.

\bibitem{jung2012nandflashsim}
M.~Jung, E.~H. Wilson, D.~Donofrio, J.~Shalf, and M.~T. Kandemir,
  ``{NANDFlashSim: Intrinsic Latency Variation Aware NAND Flash Memory System
  Modeling and Simulation at Microarchitecture Level},'' in \emph{MSST}, 2012.

\bibitem{cai2014neighbor}
Y.~Cai, G.~Yalcin, O.~Mutlu, E.~F. Haratsch, O.~Unsal, A.~Cristal, and K.~Mai,
  ``Neighbor-{C}ell {A}ssisted {E}rror {C}orrection for {MLC} {NAND} {F}lash
  {M}emories,'' in \emph{SIGMETRICS}, 2014.

\bibitem{cai2017error}
Y.~Cai, S.~Ghose, E.~F. Haratsch, Y.~Luo, and O.~Mutlu, ``Error
  {C}haracterization, {M}itigation, and {R}ecovery in {F}lash-{M}emory-{B}ased
  {S}olid-{S}tate {D}rives,'' in \emph{Proc. IEEE}, 2017.

\bibitem{cui2017dlv}
J.~Cui, Y.~Zhang, W.~Wu, J.~Yang, Y.~Wang, and J.~Huang, ``{DLV: Exploiting
  Device Level Latency Variations for Performance Improvement on Flash Memory
  Storage Systems},'' in \emph{TCAD}, 2017.

\bibitem{Cai2018}
Y.~Cai, S.~Ghose, E.~F. Haratsch, Y.~Luo, and O.~Mutlu, ``Reliability {I}ssues
  in {F}lash-{M}emory-{B}ased {S}olid-{S}tate {D}rives: {E}xperimental
  {A}nalysis, {M}itigation, {R}ecovery,'' in \emph{Inside Solid State Drives
  (SSDs)}, 2018.

\bibitem{park2021reducing}
J.~Park, M.~Kim, M.~Chun, L.~Orosa, J.~Kim, and O.~Mutlu, ``Reducing
  {S}olid-{S}tate {D}rive {R}ead {L}atency by {O}ptimizing {R}ead-{R}etry,'' in
  \emph{ASPLOS}, 2021.

\bibitem{ipek2008self}
E.~Ipek, O.~Mutlu, J.~F. Mart{\'\i}nez, and R.~Caruana, ``{Self-Optimizing
  Memory Controllers: A Reinforcement Learning Approach},'' in \emph{ISCA},
  2008.

\bibitem{pythia}
R.~Bera, K.~Kanellopoulos, A.~Nori, T.~Shahroodi, S.~Subramoney, and O.~Mutlu,
  ``{Pythia: A Customizable Hardware Prefetching Framework Using Online
  Reinforcement Learning},'' in \emph{MICRO}, 2021.

\bibitem{zell1994simulation}
A.~Zell, \emph{Simulation neuronaler netze}.\hskip 1em plus 0.5em minus
  0.4em\relax Addison-Wesley Bonn, 1994.

\bibitem{MSR}
``{MSR Cambridge Traces.}, \url{http://iotta.snia.org/traces/388}.''

\bibitem{tarasov2016filebench}
V.~Tarasov, E.~Zadok, and S.~Shepler, ``{Filebench: A Flexible Framework for
  File System Benchmarking},'' 2016.

\bibitem{sibylLink}
CMU-SAFARI, ``{Sibyl}, \url{https://github.com/CMU-SAFARI/Sibyl}.''

\bibitem{inteloptane}
Intel, ``{Intel Optane SSD DC P4801X Series},
  \url{https://ark.intel.com/content/www/us/en/ark/products/149365/intel-optane-ssd-dc-p4801x-series-100gb-2-5in-pcie-x4-3d-xpoint.html}.''

\bibitem{samsung2017znand}
Samsung, ``{Ultra-Low Latency with Samsung Z-NAND SSD},
  \url{https://www.samsung.com/semiconductor/global.semi.static/Ultra-Low_Latency_with_Samsung_Z-NAND_SSD-0.pdf}.''

\bibitem{intels4510}
Intel, ``{Intel SSD D3-S4510 Series},
  \url{https://www.intel.com/content/www/us/en/products/memory-storage/solid-state-drives/data-center-ssds/d3-series/d3-s4510-series/d3-s4510-1-92tb-2-5inch-3d2.html}.''

\bibitem{intelqlc}
Intel, ``{Intel SSD 660p Series},
  \url{https://www.intel.com/content/www/us/en/products/docs/memory-storage/solid-state-drives/consumer-ssds/660p-series-brief.html}.''

\bibitem{seagate}
Seagate, ``{Seagate Barracuda Datasheet},
  \url{https://www.seagate.com/www-content/datasheets/pdfs/3-5-barracuda-3tbDS1900-10-1710US-en_US.pdf}.''

\bibitem{adatasu630}
ADATA, ``{ADATA Ultimate Series: SU630},
  \url{https://shop.adata.com/adata-ultimate-series-su630-960gb-sata-iii-internal-2-5-solid-state-drive/}.''

\bibitem{feng2014hdstore}
Z.~Feng, Z.~Feng, X.~Wang, G.~Rao, Y.~Wei, and Z.~Li, ``{HDStore: An SSD/HDD
  Hybrid Distributed Storage Scheme for Large-Scale Data},'' in \emph{WAIM},
  2014.

\bibitem{micheloni2018hybrid}
R.~Micheloni, L.~Crippa, and M.~Picca, ``{Hybrid Storage Systems},'' in
  \emph{Inside Solid State Drives (SSDs)}, 2018.

\bibitem{kim2014evaluating}
H.~Kim, S.~Seshadri, C.~L. Dickey, and L.~Chiu, ``Evaluating {P}hase {C}hange
  {M}emory for {E}nterprise {S}torage {S}ystems: A {S}tudy of {C}aching and
  {T}iering {A}pproaches,'' in \emph{FAST}, 2014.

\bibitem{tsuchida201064mb}
K.~Tsuchida, T.~Inaba, K.~Fujita, Y.~Ueda, T.~Shimizu, Y.~Asao, T.~Kajiyama,
  M.~Iwayama, K.~Sugiura, S.~Ikegawa \emph{et~al.}, ``A {64Mb} {MRAM} with
  {C}lamped-{R}eference and {A}dequate-{R}eference {S}chemes,'' in
  \emph{ISSCC}, 2010.

\bibitem{kawahara20128}
A.~Kawahara, R.~Azuma, Y.~Ikeda, K.~Kawai, Y.~Katoh, Y.~Hayakawa, K.~Tsuji,
  S.~Yoneda, A.~Himeno, K.~Shimakawa \emph{et~al.}, ``{An {8Mb} Multi-Layered
  Cross-Point ReRAM Macro with 443MB/s Write Throughput},'' in \emph{ISSCC},
  2012.

\bibitem{choi201220nm}
Y.~Choi, I.~Song, M.-H. Park, H.~Chung, S.~Chang, B.~Cho, J.~Kim, Y.~Oh,
  D.~Kwon, J.~Sunwoo \emph{et~al.}, ``{A 20nm 1.8V 8Gb PRAM with 40MB/s Program
  Bandwidth},'' in \emph{ISSCC}, 2012.

\bibitem{oh2013hybrid}
Y.~Oh, E.~Lee, J.~Choi, D.~Lee, and S.~H. Noh, ``{Hybrid Solid State Drives for
  Improved Performance and Enhanced Lifetime},'' in \emph{MSST}, 2013.

\bibitem{lee2010high}
H.~G. Lee, ``{High-Performance NAND and PRAM Hybrid Storage Design for Consumer
  Electronics},'' in \emph{TCE}, 2010.

\bibitem{okamoto2015application}
S.~Okamoto, C.~Sun, S.~Hachiya, T.~Yamada, Y.~Saito, T.~O. Iwasaki, and
  K.~Takeuchi, ``{Application Driven SCM \& NAND Flash Hybrid SSD Design for
  Data-Centric Computing System},'' in \emph{IMW}, 2015.

\bibitem{gal2005algorithms}
E.~Gal and S.~Toledo, ``{Algorithms and Data Structures for Flash Memories},''
  in \emph{CSUR}, 2005.

\bibitem{nvme}
N.~Express, ``{Everything You Need to Know About the NVMe 2.0 Specifications
  and New Technical Proposals},'' \url{https://nvmexpress.org/}.

\bibitem{sata}
S.~A.~W. Group, ``{Serial ATA: High Speed Serialized AT Attachment},''
  \url{https://web.archive.org/web/20161009182351/http://www.ece.umd.edu/courses/enee759h.S2003/references/serialata10a.pdf}.

\bibitem{linux}
{Linux Source Code (v5.10.16)},
  \url{https://elixir.bootlin.com/linux/latest/source}.

\bibitem{meswani2015heterogeneous}
M.~R. Meswani, S.~Blagodurov, D.~Roberts, J.~Slice, M.~Ignatowski, and G.~H.
  Loh, ``{Heterogeneous Memory Architectures: A HW/SW Approach for Mixing
  Die-stacked and Off-package Memories},'' in \emph{HPCA}, 2015.

\bibitem{cheong2018flash}
W.~{Cheong}, C.~{Yoon}, S.~{Woo}, K.~{Han}, D.~{Kim}, C.~{Lee}, Y.~{Choi},
  S.~{Kim}, D.~{Kang}, G.~{Yu}, J.~{Kim}, J.~{Park}, K.~{Song}, K.~{Park},
  S.~{Cho}, H.~{Oh}, D.~D.~G. {Lee}, J.~{Choi}, and J.~{Jeong}, ``{A Flash
  Memory Controller for 15$\mu$s Ultra-Low-Latency SSD Using High-Speed 3D NAND
  Flash with 3$\mu$s Read Time},'' in \emph{ISSCC}, 2018.

\bibitem{micron3dxpoint}
Micron, ``{3D XPoint Technology},
  \url{https://www.micron.com/products/advanced-solutions/3d-xpoint-technology}.''

\bibitem{hady2017platform}
F.~T. Hady, A.~Foong, B.~Veal, and D.~Williams, ``{Platform Storage Performance
  With 3D XPoint Technology},'' in \emph{Proc. IEEE}, 2017.

\bibitem{intelp4610}
Intel, ``{Intel SSD DC P4610 Series},
  \url{https://ark.intel.com/content/www/us/en/ark/products/140103/intel-ssd-dc-p4610-series-1-6tb-2-5in-pcie-3-1-x4-3d2-tlc.html}.''

\bibitem{watkins1989learning}
C.~J. C.~H. Watkins, ``{Learning From Delayed Rewards},'' Ph.D. dissertation,
  1989.

\bibitem{qlearning_ML_1992}
C.~J. Watkins and P.~Dayan, ``Q-learning,'' in \emph{ML}, 1992.

\bibitem{rummery1994line}
G.~A. Rummery and M.~Niranjan, \emph{{On-line Q-Learning Using Connectionist
  Systems}}.\hskip 1em plus 0.5em minus 0.4em\relax Citeseer, 1994.

\bibitem{silver2016mastering}
D.~Silver, A.~Huang, C.~J. Maddison, A.~Guez, L.~Sifre, G.~Van Den~Driessche,
  J.~Schrittwieser, I.~Antonoglou, V.~Panneershelvam, M.~Lanctot \emph{et~al.},
  ``{Mastering the Game of Go With Deep Neural Networks and Tree Search},'' in
  \emph{Nature}, 2016.

\bibitem{silver2017mastering}
D.~Silver, J.~Schrittwieser, K.~Simonyan, I.~Antonoglou, A.~Huang, A.~Guez,
  T.~Hubert, L.~Baker, M.~Lai, A.~Bolton \emph{et~al.}, ``{Mastering the Game
  of Go Without Human Knowledge},'' in \emph{Nature}, 2017.

\bibitem{mnih2013playing}
V.~Mnih, K.~Kavukcuoglu, D.~Silver, A.~Graves, I.~Antonoglou, D.~Wierstra, and
  M.~Riedmiller, ``{Playing Atari with Deep Reinforcement Learning},'' in
  \emph{NIPS}, 2013.

\bibitem{liang2016deep}
S.~Liang and R.~Srikant, ``{Why Deep Neural Networks for Function
  Approximation?}'' in \emph{arXiv}, 2016.

\bibitem{sutton1999policy}
R.~S. Sutton, D.~McAllester, S.~Singh, and Y.~Mansour, ``{Policy Gradient
  Methods for Reinforcement Learning With Function Approximation},'' in
  \emph{NIPS}, 1999.

\bibitem{baird1995residual}
L.~Baird, ``{Residual Algorithms: Reinforcement Learning With Function
  Approximation},'' in \emph{ML}, 1995.

\bibitem{kira1992featureselection}
K.~Kira and L.~A. Rendell, ``{A Practical Approach to Feature Selection},'' in
  \emph{ML}, 1992.

\bibitem{silver2021reward}
D.~Silver, S.~Singh, D.~Precup, and R.~S. Sutton, ``{Reward is Enough},''
  \emph{AI}, 2021.

\bibitem{zhong2009program}
Y.~Zhong, X.~Shen, and C.~Ding, ``{Program Locality Analysis Using Reuse
  Distance},'' in \emph{TOPLAS}, 2009.

\bibitem{bottou2003stochastic}
L.~Bottou, ``{Stochastic Learning},'' in \emph{Advanced Lectures on Machine
  Learning}, 2003.

\bibitem{C51}
M.~G. Bellemare, W.~Dabney, and R.~Munos, ``{A Distributional Perspective on
  Reinforcement Learning},'' in \emph{arXiv}, 2017.

\bibitem{harrold2022data}
D.~J. Harrold, J.~Cao, and Z.~Fan, ``{Data-Driven Battery Operation For Energy
  Arbitrage Using Rainbow Deep Reinforcement Learning},'' in \emph{Energy},
  2021.

\bibitem{dqn}
V.~Mnih, K.~Kavukcuoglu, D.~Silver, A.~A. Rusu, J.~Veness, M.~G. Bellemare,
  A.~Graves, M.~Riedmiller, A.~K. Fidjeland, G.~Ostrovski, S.~Petersen,
  C.~Beattie, A.~Sadik, I.~Antonoglou, H.~King, D.~Kumaran, D.~Wierstra,
  S.~Legg, and D.~Hassabis, ``{Human-Level Control Through Deep Reinforcement
  Learning},'' in \emph{Nature}, 2015.

\bibitem{tokic2011value}
M.~Tokic and G.~Palm, ``{Value-Difference Based Exploration: Adaptive Control
  between Epsilon-Greedy and Softmax},'' in \emph{AAAI}, 2011.

\bibitem{bebis1994feed}
G.~Bebis and M.~Georgiopoulos, ``{Feed-Forward Neural Networks},'' in
  \emph{IEEE Potentials}, 1994.

\bibitem{de1993backpropagation}
J.~De~Villiers and E.~Barnard, ``{Backpropagation Neural Nets with One and Two
  Hidden Layers},'' in \emph{IEEE Trans. Neural Netw. Learn. Sys}, 1993.

\bibitem{ramachandran2017searching}
P.~Ramachandran, B.~Zoph, and Q.~V. Le, ``{Searching for Activation
  Functions},'' in \emph{arXiv}, 2017.

\bibitem{agarap2018relu}
A.~F. Agarap, ``{Deep Learning using Rectified Linear Units (ReLU)},'' in
  \emph{arXiv}, 2018.

\bibitem{paine2020hyperparameter}
T.~L. Paine, C.~Paduraru, A.~Michi, C.~Gulcehre, K.~Zolna, A.~Novikov, Z.~Wang,
  and N.~de~Freitas, ``{Hyperparameter Selection For Offline Reinforcement
  Learning},'' in \emph{arXiv}, 2020.

\bibitem{NAPEL}
G.~Singh, J.~G{\'o}mez-Luna, G.~Mariani, G.~F. Oliveira, S.~Corda, S.~Stuijk,
  O.~Mutlu, and H.~Corporaal, ``{NAPEL: Near-Memory Computing Application
  Performance Prediction via Ensemble Learning},'' in \emph{DAC}, 2019.

\bibitem{arlot2010survey}
S.~Arlot and A.~Celisse, ``{A Survey of Cross-Validation Procedures for Model
  Selection},'' \emph{SS}, 2010.

\bibitem{montgomery2017design}
D.~C. Montgomery, ``{Design and Analysis of Experiments},'' 2017.

\bibitem{linuxmint}
{Linux Mint 20.1 ``Ulyssa''}, \url{https://linuxmint.com/edition.php?id=284}.

\bibitem{tweedie1998journaling}
S.~C. Tweedie \emph{et~al.}, ``{Journaling the Linux ext2fs Filesystem},'' in
  \emph{The Fourth Annual Linux Expo}, 1998.

\bibitem{TFagents}
S.~Guadarrama, A.~Korattikara, O.~Ramirez, P.~Castro, E.~Holly, S.~Fishman,
  K.~Wang, E.~Gonina, N.~Wu, E.~Kokiopoulou, L.~Sbaiz, J.~Smith, G.~Bartók,
  J.~Berent, C.~Harris, V.~Vanhoucke, and E.~Brevdo, ``{TF-Agents: A Library
  for Reinforcement Learning in TensorFlow},
  \url{https://github.com/tensorflow/agents},'' 2018.

\bibitem{amdryzen}
AMD, ``{AMD Ryzen™ 7 PRO 2700 Processor},
  \url{https://www.amd.com/en/products/cpu/amd-ryzen-7-2700}.''

\bibitem{cooper2010benchmarking}
B.~F. Cooper, A.~Silberstein, E.~Tam, R.~Ramakrishnan, and R.~Sears,
  ``{Benchmarking Cloud Serving Systems With YCSB},'' in \emph{SOCC}, 2010.

\bibitem{tsukada2021metadata}
S.~Tsukada, H.~Takayashiki, M.~Sato, K.~Komatsu, and H.~Kobayashi, ``{A
  Metadata Prefetching Mechanism for Hybrid Memory Architectures},'' in
  \emph{COOL CHIPS}, 2021.

\bibitem{verma2018programmatically}
A.~Verma, V.~Murali, R.~Singh, P.~Kohli, and S.~Chaudhuri, ``{Programmatically
  Interpretable Reinforcement Learning},'' in \emph{ICML}, 2018.

\bibitem{liu2018toward}
G.~Liu, O.~Schulte, W.~Zhu, and Q.~Li, ``{Toward Interpretable Deep
  Reinforcement Learning With Linear Model u-Trees},'' in \emph{ECML PKDD},
  2018.

\bibitem{juozapaitis2019explainable}
Z.~Juozapaitis, A.~Koul, A.~Fern, M.~Erwig, and F.~Doshi-Velez, ``{Explainable
  Reinforcement Learning via Reward Decomposition},'' in \emph{IJCAI-ECAI},
  2019.

\bibitem{madumal2020explainable}
P.~Madumal, T.~Miller, L.~Sonenberg, and F.~Vetere, ``{Explainable
  Reinforcement Learning Through a Causal Lens},'' in \emph{AAAI}, 2020.

\bibitem{sequeira2020interestingness}
P.~Sequeira and M.~Gervasio, ``{Interestingness Elements For Explainable
  Reinforcement Learning: Understanding Agents' Capabilities and
  Limitations},'' in \emph{AI}, 2020.

\bibitem{puiutta2020explainable}
E.~Puiutta and E.~Veith, ``{Explainable Reinforcement Learning: A Survey},'' in
  \emph{CD-MAKE}, 2020.

\bibitem{li2017utility_HMM}
Y.~Li, S.~Ghose, J.~Choi, J.~Sun, H.~Wang, and O.~Mutlu, ``{Utility-Based
  Hybrid Memory Management},'' in \emph{CLUSTER}, 2017.

\bibitem{agarwal2015page_HMM}
N.~Agarwal, D.~Nellans, M.~Stephenson, M.~O'Connor, and S.~W. Keckler, ``{Page
  Placement Strategies for GPUs Within Heterogeneous Memory Systems},'' in
  \emph{ASPLOS}, 2015.

\bibitem{agarwal2017thermostat_HMM}
N.~Agarwal and T.~F. Wenisch, ``{Thermostat: Application-Transparent Page
  Management for Two-Tiered Main Memory},'' in \emph{ASPLOS}, 2017.

\bibitem{goglin2016exposing}
B.~Goglin, ``{Exposing the Locality of Heterogeneous Memory Architectures to
  HPC Applications},'' in \emph{MEMSYS}, 2016.

\bibitem{ham2013disintegrated}
T.~J. Ham, B.~K. Chelepalli, N.~Xue, and B.~C. Lee, ``{Disintegrated Control
  for Energy-Efficient and Heterogeneous Memory Systems},'' in \emph{HPCA},
  2013.

\bibitem{lin2016memif}
F.~X. Lin and X.~Liu, ``{memif: Towards Programming Heterogeneous Memory
  Asynchronously},'' in \emph{ASPLOS}, 2016.

\bibitem{malladi2016dramscale}
K.~T. Malladi, U.~Kang, M.~Awasthi, and H.~Zheng, ``{DRAMScale: Mechanisms to
  Increase DRAM Capacity},'' in \emph{MEMSYS}, 2016.

\bibitem{pavlovic2013data}
M.~Pavlovic, N.~Puzovic, and A.~Ramirez, ``{Data Placement in HPC Architectures
  With Heterogeneous Off-Chip Memory},'' in \emph{ICCD}, 2013.

\bibitem{pena2014toward}
A.~J. Pena and P.~Balaji, ``{Toward the Efficient Use of Multiple Explicitly
  Managed Memory Subsystems},'' in \emph{CLUSTER}, 2014.

\bibitem{qureshi2009scalable}
M.~K. Qureshi, V.~Srinivasan, and J.~A. Rivers, ``{Scalable High Performance
  Main Memory System Using Phase-Change Memory Technology},'' in \emph{ISCA},
  2009.

\bibitem{yoon2012row}
H.~Yoon, J.~Meza, R.~Ausavarungnirun, R.~A. Harding, and O.~Mutlu, ``{Row
  Buffer Locality Aware Caching Policies for Hybrid Memories},'' in
  \emph{ICCD}, 2012.

\bibitem{meza2012enabling}
J.~Meza, J.~Chang, H.~Yoon, O.~Mutlu, and P.~Ranganathan, ``{Enabling Efficient
  and Scalable Hybrid Memories Using Fine-Granularity DRAM Cache Management},''
  in \emph{IEEE CAL}, 2012.

\bibitem{ren2015thynvm}
J.~Ren, J.~Zhao, S.~Khan, J.~Choi, Y.~Wu, and O.~Mutlu, ``{ThyNVM: Enabling
  Software-Transparent Crash Consistency in Persistent Memory Systems},'' in
  \emph{MICRO}, 2015.

\bibitem{ham2013disintegrated_HMM}
T.~J. Ham, B.~K. Chelepalli, N.~Xue, and B.~C. Lee, ``{Disintegrated Control
  for Energy-Efficient and Heterogeneous Memory Systems},'' in \emph{HPCA},
  2013.

\bibitem{yang2017autotiering}
Z.~Yang, M.~Hoseinzadeh, A.~Andrews, C.~Mayers, D.~T. Evans, R.~T. Bolt,
  J.~Bhimani, N.~Mi, and S.~Swanson, ``{AutoTiering: Automatic Data Placement
  Manager in Multi-Tier All-Flash Datacenter},'' in \emph{IPCCC}, 2017.

\bibitem{chiachenchou2015batman}
C.~Chou, A.~Jaleel, and M.~K. Qureshi, ``{BATMAN: Maximizing Bandwidth
  Utilization of Hybrid Memory Systems}.''\hskip 1em plus 0.5em minus
  0.4em\relax Citeseer, 2015.

\bibitem{kim2011hybridstore}
Y.~Kim, A.~Gupta, B.~Urgaonkar, P.~Berman, and A.~Sivasubramaniam,
  ``{HybridStore: A Cost-Efficient, High-Performance Storage System Combining
  SSDs and HDDs},'' in \emph{MASCOTS}, 2011.

\bibitem{wang2019panthera}
C.~Wang, H.~Cui, T.~Cao, J.~Zigman, H.~Volos, O.~Mutlu, F.~Lv, X.~Feng, and
  G.~H. Xu, ``{Panthera: Holistic Memory Management for Big Data Processing
  over Hybrid Memories},'' in \emph{PLDI}, 2019.

\bibitem{ramos2011page}
L.~E. Ramos, E.~Gorbatov, and R.~Bianchini, ``{Page Placement in Hybrid Memory
  Systems},'' in \emph{ICS}, 2011.

\bibitem{liu2019hierarchical}
L.~Liu, S.~Yang, L.~Peng, and X.~Li, ``{Hierarchical Hybrid Memory Management
  in OS for Tiered Memory Systems},'' in \emph{TPDS}, 2019.

\bibitem{luo2020optimal}
Y.~Luo, P.~Jin, and S.~Wan, ``{Optimal Data Placement for Data-Centric
  Algorithms on NVM-Based Hybrid Memory},'' in \emph{DSAA}, 2020.

\bibitem{doudali2021cori}
T.~D. Doudali, D.~Zahka, and A.~Gavrilovska, ``{Cori: Dancing to the Right Beat
  of Periodic Data Movements over Hybrid Memory Systems},'' in \emph{IPDPS},
  2021.

\bibitem{shetti2019machine}
M.~Shetti, B.~Li, and D.~Du, ``{Machine Learning-based Adaptive Migration
  Algorithm for Hybrid Storage Systems},'' in \emph{TOS}, 2019.

\bibitem{sen2019machine}
S.~Sen and N.~Imam, ``{Machine Learning Based Design Space Exploration for
  Hybrid Main-Memory Design},'' in \emph{MEMSYS}, 2019.

\bibitem{liu2019learning}
K.~Liu, J.~Peng, J.~Wang, B.~Yu, Z.~Liao, Z.~Huang, and J.~Pan, ``{A
  Learning-Based Data Placement Framework for Low Latency in Data Center
  Networks},'' in \emph{TCC}, 2019.

\bibitem{yoo2020reinforcement}
S.~Yoo and D.~Shin, ``{Reinforcement Learning-Based SLC Cache Technique for
  Enhancing SSD Write Performance},'' in \emph{USENIX HotStorage}, 2020.

\bibitem{wang2020reinforcement}
H.~Wang, H.~Shen, Q.~Liu, K.~Zheng, and J.~Xu, ``{A Reinforcement Learning
  Based System for Minimizing Cloud Storage Service Cost},'' in \emph{ICPP},
  2020.

\bibitem{rl_GC_TECS}
W.~Kang, D.~Shin, and S.~Yoo, ``{Reinforcement Learning-Assisted Garbage
  Collection to Mitigate Long-Tail Latency in SSD},'' in \emph{TECS}, 2017.

\bibitem{kang2018dynamic}
W.~Kang and S.~Yoo, ``{Dynamic Management of Key States for Reinforcement
  Learning-assisted Garbage Collection to Reduce Long Tail Latency in SSD},''
  in \emph{DAC}, 2018.

\bibitem{lin2020deepNOC}
T.-R. Lin, D.~Penney, M.~Pedram, and L.~Chen, ``{A Deep Reinforcement Learning
  Framework for Architectural Exploration: A Routerless NoC Case Study},'' in
  \emph{HPCA}, 2020.

\bibitem{liu2020imitation}
E.~Liu, M.~Hashemi, K.~Swersky, P.~Ranganathan, and J.~Ahn, ``{An Imitation
  Learning Approach for Cache Replacement},'' in \emph{ICML}, 2020.

\bibitem{rl_NOC_AIDArc_2018}
J.~Yin, Y.~Eckert, S.~Che, M.~Oskin, and G.~Loh, ``{Toward More Efficient NoC
  Arbitration: A Deep Reinforcement Learning Approach},'' in \emph{AIDArc},
  2018.

\bibitem{rl_voltage_scaling_TC_2018}
Q.~Fettes, M.~Clark, R.~Bunescu, A.~Karanth, and A.~Louri, ``{Dynamic Voltage
  and Frequency Scaling in NoCs With Supervised and Reinforcement Learning
  Techniques},'' in \emph{TC}, 2018.

\bibitem{mirhoseini2021chip}
A.~Mirhoseini, A.~Goldie, M.~Yazgan, J.~Jiang, E.~Songhori, S.~Wang, Y.-J. Lee,
  E.~Johnson, O.~Pathak, S.~Bae \emph{et~al.}, ``{Chip Placement With Deep
  Reinforcement Learning},'' \emph{Nature}, 2021.

\bibitem{mutlu2021intelligent_DATE}
O.~Mutlu, ``{Intelligent Architectures for Intelligent Computing Systems},'' in
  \emph{DATE}, 2021.

\bibitem{peled2015semantic}
L.~Peled, S.~Mannor, U.~Weiser, and Y.~Etsion, ``{Semantic Locality and
  Context-based Prefetching Using Reinforcement Learning},'' in \emph{ISCA},
  2015.

\bibitem{martinez2009dynamic}
J.~F. Martinez and E.~Ipek, ``{Dynamic Multicore Resource Management: A Machine
  Learning Approach},'' in \emph{IEEE Micro}, 2009.

\bibitem{multi_scheduler_HPCA_2012}
J.~Mukundan and J.~F. Martinez, ``{MORSE: Multi-Objective Reconfigurable
  Self-Optimizing Memory Scheduler},'' in \emph{HPCA}, 2012.

\bibitem{jain2016machine}
R.~Jain, P.~R. Panda, and S.~Subramoney, ``{Machine Learned Machines: Adaptive
  Co-optimization of Caches, Cores, and On-chip Network},'' in \emph{DATE},
  2016.

\bibitem{zheng2020agile}
H.~Zheng and A.~Louri, ``{Agile: A Learning-Enabled Power And
  Performance-Efficient Network-On-Chip Design},'' in \emph{TETCI}, 2020.

\bibitem{pd2015q}
S.~M. PD, H.~Yu, H.~Huang, and D.~Xu, ``{A Q-Learning Based Self-Adaptive I/O
  Communication for 2.5D Integrated Many-Core Microprocessor and Memory},'' in
  \emph{IEEE TOC}, 2015.

\end{thebibliography}
}


\end{document}